\documentclass[aps,prd,onecolumn,preprintnumbers,nofootinbib,superscriptaddress]{revtex4-2}

\usepackage{amssymb,amsfonts,graphicx}
\usepackage{hyperref}
\usepackage{physics}
\usepackage{xcolor}
\usepackage{multirow}
\usepackage{caption}
\usepackage{subcaption}
\usepackage{booktabs}
\usepackage{siunitx}
\usepackage{xspace}
\usepackage{amsmath}
\usepackage{colortbl}
\usepackage{enumitem}
\usepackage{array}
\usepackage{braket}
\usepackage{wrapfig}
\usepackage{float}
\usepackage{slashed}

\usepackage{xcolor}
\usepackage{ulem}
\usepackage{tikz}
\usepackage{tikz-feynman}
\usepackage{natbib}
\tikzfeynmanset{compat=1.1.0}
\hypersetup{
    colorlinks=true,
    linkcolor=blue,
    citecolor=magenta,
    urlcolor=blue
}

\begin{document}

\title{Search for New Physics through the Observables of Semileptonic $B_{c}^+\to D^{\ast+}\ell^{+}\ell^{-}$ Decay}

\author{Zohaib Aarfi}
\email{zohaib.phdphy22sns@student.nust.edu.pk}
\affiliation{Department of Physics and Astronomy, School of Natural Sciences, National University of Sciences and Technology (NUST), H-12, Islamabad 44000, Pakistan}
\author{Qazi Maaz Us Salam}
\email{qazimaaz92@gmail.com}
\affiliation{Department of Physics, Lahore University of Management
Sciences (LUMS), Opposite Sector U, D.H.A, Lahore 54792, Pakistan}
\affiliation{National Centre for Physics, Shahdra Valley Road, Islamabad 45320,Pakistan}
\author{Faisal Munir Bhutta}
\email{faisal.munir@sns.nust.edu.pk}
\affiliation{Department of Physics and Astronomy, School of Natural Sciences, National University of Sciences and Technology (NUST), H-12, Islamabad 44000, Pakistan}
\author{Ishtiaq Ahmed}
\email{ishtiaq.ahmed@ncp.edu.pk}
\affiliation{National Centre for Physics, Shahdra Valley Road, Islamabad 45320,Pakistan}
\author{M. Ali Paracha}
\email{aliparacha@sns.nust.edu.pk (corresponding author)}
\affiliation{Department of Physics and Astronomy, School of Natural Sciences, National University of Sciences and Technology (NUST), H-12, Islamabad 44000, Pakistan}


\begin{abstract}

Motivated by the current flavor anomalies and the comparatively less explored nature of the $b\to d$ sector, we present a model-independent study of the rare semileptonic $B_c^{+}\to D^{\ast+}\ell^{+}\ell^{-}$ ($\ell=\mu,\tau$) decay, to investigate its sensitivity to New Physics effects. The analysis incorporates penguin-box contributions, long-distance resonance effects, and the sizable weak-annihilation amplitudes. Using the $B_c\to D^\ast$ transition form factors calculated in the covariant confined quark model and the weak-annihilation form factors obtained within the Bethe--Salpeter approach, we analyze the differential branching fraction, forward-backward asymmetry, longitudinal helicity fraction, and a comprehensive set of normalized angular observables in several one- and two-dimensional New Physics scenarios and compare our results with the Standard Model predictions. Notably, the branching fraction, forward-backward asymmetry, and the normalized angular coefficients such as, $\langle I_{2c}\rangle$, $\langle I_{3}\rangle$, $\langle I_{5}\rangle$, and $\langle I_{6s} \rangle$, show clear sensitivity to New Physics effects. Our results indicate that the decay $B_c^{+}\to D^{\ast+}\ell^{+}\ell^{-}$ serves as a complementary probe of the flavor structure of New Physics and can therefore be investigated in future measurements at LHCb and other high-luminosity flavor experiments.

\end{abstract}

\maketitle
\newpage

\section{Introduction}
In the framework of the Standard Model (SM), the rare semileptonic $B_{c}^+\to D^{\ast+}\ell^{+}\ell^{-}$, $(\ell=\mu,\tau)$ decays are mediated through flavor changing neutral current (FCNC) $b\to d$ transition. The dominant contribution to these processes originates from an internal top quark loop, where the large mass of the top quark  enhances the virtual effects. The branching ratio for this transition is highly suppressed, reaching the order $\mathcal{O}(10^{-8})$, primarily as a consequence of a loop-induced FCNC and additionally due to the small magnitude of the CKM matrix element $V_{td}~$\cite{Cabibbo:1963yz,Kobayashi:1973fv}. While this suppression makes these decays sensitive probes of New Physics (NP), they are also theoretically interesting because the CP violation in $b\to d$ transition is significantly larger than the $b\to s$ transition.

Interest in the FCNC rare decays, driven by $b\to s\ell^+\ell^-$, has increased significantly following the observation of flavor anomalies. While high-precision measurements from LHCb collaboration show that the lepton flavor universality (LFU) ratios $R_{K^{(\ast)}}$ are aligned with SM predictions~\cite{LHCb:2022qnv}, the angular observable $P^{\prime}_{5}$ continues to exhibit a departure from the SM with approximate significance of $2.5\sigma-3.0\sigma$ \cite{LHCb:2020lmf}, and the branching fraction of $B_{s}\to\phi\mu^{+}\mu^{-}$ decay deviates by about $3.6\sigma$ from the SM prediction \cite{LHCb:2021xxq}. These anomalies suggest that the flavor sector may still be influenced by NP and motivate researchers to explore NP through decays mediated by the $b\to d$ transition, as a complementary test. Furthermore, investigating $b\to d$ transitions allows us to test whether potential NP effects originate from new sources of flavor symmetry breaking. Since these transitions are more Cabibbo-suppressed than $b\to s$ processes, they offer a unique window where NP contributions might be more distinguishable.

Experimentally, several measurements and upper bounds for rare $b\to d$ channels have already been reported by the  LHCb, BaBaR and Belle collaborations, including 
$B^{+}\to\pi^{+}\mu^{+}\mu^{-}$,
$B^{0}\to\pi^{+}\pi^{-}\mu^{+}\mu^{-}$,
and 
$B^{0}\to\rho^{0}\mu^{+}\mu^{-}$~\cite{LHCb:2026xvw,LHCb:2015hsa,BaBar:2013qaj,Belle:2008tjs}.  Recently, Belle has extended these searches to the modes 
$B\to(\pi,\eta,\rho,\omega)\ell^{+}\ell^{-}$ and has set upper limits at the level of $(3.8-47)\times10^{-8}$~\cite{Belle:2024cis}. Furthermore, evidence for the decay 
$B_s\to \bar K^{*}\mu^{+}\mu^{-}$, having a branching fraction of 
$(2.9\pm1.1)\times10^{-8}$~\cite{LHCb:2018rym}, is observed by the LHCb. The upcoming data from the current and future collider experiments are expected to improve the sensitivity to these suppressed modes and enable precision measurements on the observables associated with these decays~\cite{Belle2WhitePaper2018}. 

In this context, the charm $B_{c}$ meson provides a unique laboratory for testing the SM. Its decay dynamics provides complementary information to the $B_{u,d,s}$ systems~\cite{Ishaq:2013toa,Huang:2018rys,Munir:2015gsp,MunirBhutta:2020ber,Bhutta:2024zwj,Das:2018orb,Mohapatra:2021izl,Rajeev:2020aut}. Furthermore, owing to the large production cross-section of $B_{c}$ meson at the LHC, several hadronic and semileptonic $B_{c}$ decay channels have been established~\cite{LHCb:2012ag,Dutta:2019wxo,Mohapatra:2021ynn,Zaki:2023mcw,Li:2023mrj,Salam:2024nfv}, providing fertile ground for testing the SM parameters as well as NP. These developments motivate us to study NP through semileptonic $B_{c}\to D^{(\ast)}_{s,d}\ell^{+}\ell^{-}$ decays, which offer a rich set of physical observables to analyze the NP effects. One of the important features of $B_{c}\to D^{(\ast)}_{s,d}\ell^{+}\ell^{-}$ decays, that distinguishes them from other $B_{u,s,d}$ decays, is the significant role of the weak annihilation (WA) contribution, since in $B_{c}$ decays the penguin-box (PB) and WA contributions enter on equal footing due to their comparable CKM matrix element values. In the literature the decays $B_{c}\to D^{\ast}_{s,d}\ell^{+}\ell^{-}$ have been studied both in the frameworks of the SM and in different NP models~\cite{Geng:2001vy,Azizi:2008vv,Azizi:2008vy,Faessler:2002ut,Choi:2010ha,Wang:2014yia,Yilmaz:2012ah,Ebert:2010dv,Maji:2020zlq,Maji:2020wer,Paracha:2011gt,Ahmed:2011fy,Ahmed:2011sa}.

In this work, we investigate the decays $B_c^+ \to D^{\ast+}(\to P_1P_2)\ell^+\ell^-$, with $P_1P_2=D^+\pi^0,D^0\pi^+$, and $\ell=\mu,\tau$, in a model-independent NP framework. This study serves as a natural extension of our previous work \cite{Aarfi:2026hgi}, where we analyzed the impact of WA, PB, and long-distance (LD) contributions within the SM. The impact of NP can be probed using the differential branching ratio, forward-backward (FB) asymmetry, longitudinal helicity fraction of the final state $D^{\ast}$ meson, and angular coefficients $I_{i}$'s. The PB form factors which we use to analyze the physical observables were computed in the framework of  covariant confined quark model (CCQM)~\cite{Ivanov:2024iat}, and the form factors associated with WA amplitude were computed numerically adopting the Bethe Salpeter (BS) approach~\cite{Ju:2013oba}. The decay is analyzed both in low and high $q^{2}$ ranges for muons as the final state leptons, whereas for the case of final state tauons, only high $q^{2}$ range is applicable as the allowed kinematic range. To investigate NP signatures, we utilize the fitted values of NP Wilson Coefficients (WCs) obtained from global fit analyses of $b\to d$ transition~\cite{Bause:2022rrs}. Furthermore, we also present the predictions of the physical observables for the allowed ranges of the NP WCs, as well as correlation plots between observables in both one-dimensional (1D) and two-dimensional (2D) NP scenarios.

The paper is organized as follows. In Section II, we present the formalism to analyze $B_{c}^+\to D^{\ast+}\ell^{+}\ell^{-}$ decay, describing the effective Hamiltonian and the decay amplitude. The hadronic matrix elements associated with PB and WA amplitudes are parametrized in terms of transition form factors. Next, we give the four-fold angular distribution of the $B_{c}^+\to D^{\ast+}(\to P_{1}P_{2})\ell^{+}\ell^{-}$ decays, containing the angular coefficients $I_{i}$'s. Finally, the section ends with the presentation of the expressions of the helicity amplitudes, which contain the SM and NP WCs, as well as the transition form factors associated with both PB and WA amplitudes. In section-III, we present the formulas of the physical observables, which is followed by the phenomenological analysis of $B_{c}^+\to D^{\ast+}(\to P_{1}P_{2})\ell^{+}\ell^{-}$ decays both in the SM and the selected 1D and 2D NP scenarios. Finally, in section-IV, we conclude the results of our study.

\section{ Formalism to Analyze \texorpdfstring{$B_{c}^+\to D^{\ast+}\ell^{+}\ell^{-}$}{Bc to D*ll} Decay}

The semileptonic  $B_{c}^+\to D^{\ast+}\ell^{+}\ell^{-}$ decay receives contributions from the PB and WA topologies. Additionally, the decay is also affected by LD resonance effects which primarily arise from intermediate vector meson states decaying into a lepton pair. At quark level, the decay is driven by $b\to d\ell^{+}\ell^{-}$ transition and its low energy effective Hamiltonian can be written as~\cite{Bause:2022rrs,Farooq:2024owx},

\begin{eqnarray}
\mathcal{H}_{\text{eff}}
=
-\frac{4G_{F}}{\sqrt{2}}V_{tb}V^{\ast}_{td}
\left[
\mathcal{H}^{(t)}_{\text{eff}}
+
\lambda_{u}^{(d)}
\mathcal{H}^{(u)}_{\text{eff}}
\right]
+\text{h.c.},
\label{Heff}
\end{eqnarray}
where, $G_{F}$ refers to the Fermi coupling constant, $V_{tb}V^{\ast}_{td}$ are the CKM matrix elements, and
\begin{eqnarray}
\lambda_{u}^{(d)}=\frac{V_{ub}V^{\ast}_{ud}}{V_{tb}V^{\ast}_{td}}.
\end{eqnarray}

The  form of $\mathcal{H}^{(t)}_{\text{eff}}$ and $\mathcal{H}^{(u)}_{\text{eff}}$ in terms of WCs $C_{i}$'s and local quark operators $O_{i}$'s can be written as,
\begin{eqnarray}
\mathcal{H}^{(t)}_{\text{eff}}
=
C_{1}O^{c}_{1}
+
C_{2}O^{c}_{2}
+
\sum_{i=3}^{6}C_{i}O_{i}
+
\sum_{i=7,9,10}
\left[
(C_{i}+C_{i}^{NP})O_{i}
+
C_{i}^{\prime\,NP}O_{i}^{\prime}
\right],
\label{Heff1}
\end{eqnarray}
and
\begin{eqnarray}
\mathcal{H}^{(u)}_{\text{eff}}
=
C_{1}(O^{c}_{1}-O^{u}_{1})
+
C_{2}(O^{c}_{2}-O^{u}_{2}).
\label{Heff2}
\end{eqnarray}

With the effective Hamiltonian in Eq.~(\ref{Heff}), the PB amplitude for the decay is described as

\begin{eqnarray}
\mathcal{M}^{PB}
\left(B_{c}^+\to D^{\ast+}\ell^{+}\ell^{-}\right)
=
\frac{G_{F}\alpha}{2\sqrt{2}\pi}
V_{tb}V^{\ast}_{td}
\Big\{
T^{1,PB}_{\mu}(\bar{\ell}\gamma^{\mu}\ell)
+
T^{2,PB}_{\mu}(\bar{\ell}\gamma^{\mu}\gamma_{5}\ell)
\Big\},
\label{Amp1}
\end{eqnarray}
where
\begin{eqnarray}
T^{1,PB}_{\mu}
&=&
(C_{9}^{eff}+C_{9}^{NP})
\langle D^{\ast}(k,\epsilon)|
\bar d\gamma_{\mu}(1-\gamma_{5})b
|B_{c}(p)\rangle
\notag\\
&+&
C_{9}^{\prime\,NP}
\langle D^{\ast}(k,\epsilon)|
\bar d\gamma_{\mu}(1+\gamma_{5})b
|B_{c}(p)\rangle
\notag\\
&-&
\frac{2m_b}{q^2}
(C_{7}^{eff}+C_{7}^{NP})
\langle D^{\ast}(k,\epsilon)|
\bar d i\sigma_{\mu\nu}q^{\nu}(1+\gamma_{5})b
|B_{c}(p)\rangle,
\label{Amp1a}
\end{eqnarray}

and

\begin{eqnarray}
T^{2,PB}_{\mu}
&=&
(C_{10}+C_{10}^{NP})
\langle D^{\ast}(k,\epsilon)|
\bar d\gamma_{\mu}(1-\gamma_{5})b
|B_{c}(p)\rangle
\notag\\
&+&
C_{10}^{\prime\,NP}
\langle D^{\ast}(k,\epsilon)|
\bar d\gamma_{\mu}(1+\gamma_{5})b
|B_{c}(p)\rangle.
\label{Amp1b}
\end{eqnarray}

The explicit forms of the effective WCs and local quark operators, including the LD resonance effects and NP contributions are presented in~\cite{Aarfi:2026hgi,Bobeth:1999mk,Beneke:2001at,Asatrian:2001de,Asatryan:2001zw,Greub:2008cy,Du:2015tda,Aarfi:2025qcp}.

The hadronic matrix elements appearing in Eqs.~(\ref{Amp1a}) and (\ref{Amp1b}) are parametrized in terms of Lorentz-invariant form factors. The matrix elements for the vector and axial-vector quark currents are defined as
\begin{align}
\langle D^\ast(k,\epsilon)|
\bar d\gamma_{\mu}b
|B_c(p)\rangle
&=
\frac{2\epsilon_{\mu\nu\alpha\beta}}
{m_{B_c}+m_{D^\ast}}
\epsilon^{\ast\nu}p^{\alpha}k^{\beta}
V(q^{2}),
\label{2.13a}
\\
\langle D^\ast(k,\epsilon)|
\bar d\gamma_{\mu}\gamma_{5}b
|B_c(p)\rangle
&=
i(m_{B_c}+m_{D^\ast})
g_{\mu\nu}\epsilon^{\ast\nu}A_{1}(q^{2})
\nonumber\\
&-
i\frac{P_{\mu}(\epsilon^{\ast}\cdot q)}
{m_{B_c}+m_{D^\ast}}
A_{2}(q^{2})
\nonumber\\
&-
i\frac{2m_{D^\ast}}{q^{2}}
q_{\mu}(\epsilon^{\ast}\cdot q)
[A_{3}(q^{2})-A_{0}(q^{2})],
\label{2.13b}
\end{align}
where
\begin{eqnarray}
A_{3}(q^{2})
=
\frac{m_{B_c}+m_{D^\ast}}{2m_{D^\ast}}A_{1}(q^{2})
-
\frac{m_{B_c}-m_{D^\ast}}{2m_{D^\ast}}A_{2}(q^{2}),
\end{eqnarray}
with $A_{3}(0)=A_{0}(0)$.

The tensor-current matrix elements are written as
\begin{align}
\langle D^\ast(k,\epsilon)|
\bar d i\sigma_{\mu\nu}q^{\nu}b
|B_c(p)\rangle
&=
-2\epsilon_{\mu\nu\alpha\beta}
\epsilon^{\ast\nu}p^{\alpha}k^{\beta}
T_{1}(q^{2}),
\label{FF11}
\\
\langle D^\ast(k,\epsilon)|
\bar d i\sigma_{\mu\nu}q^{\nu}\gamma_{5}b
|B_c(p)\rangle
&=
i\Big[
(m^{2}_{B_c}-m^{2}_{D^\ast})
g_{\mu\nu}\epsilon^{\ast\nu}
-
(\epsilon^{\ast}\cdot q)P_{\mu}
\Big]
T_{2}(q^{2})
\nonumber\\
&+
i(\epsilon^{\ast}\cdot q)
\left[
q_{\mu}
-
\frac{q^{2}}{m^{2}_{B_c}-m^{2}_{D^\ast}}
P_{\mu}
\right]
T_{3}(q^{2}).
\label{F3}
\end{align}

Using the naive factorization approach, the WA amplitude of the $B_{c}^+\to D^{\ast+}\ell^{+}\ell^{-}$ decay is expressed as~\cite{Ju:2013oba},
\begin{equation}
\mathcal{M}_{Ann}
=
V_{cb}V_{cd}^{*}
\frac{i\alpha}{q^2}
\frac{G_F}{2\sqrt{2}\pi}
\left(
\frac{C_1}{N_c}+C_2
\right)
\mathcal{T}_{ann}^{\mu}
\,
\bar{\ell}\gamma_{\mu}\ell,
\end{equation}
where
\begin{equation}
\mathcal{T}_{ann}^{\mu}
=
\mathcal{T}_{1}^{\mu}
+
\mathcal{T}_{2}^{\mu}
+
\mathcal{T}_{3}^{\mu}
+
\mathcal{T}_{4}^{\mu}.
\end{equation}

The explicit expressions for the hadronic matrix elements 
$\mathcal{T}_{i}^{\mu}$ can be found in Ref.~\cite{Ju:2013oba}. 
The corresponding WA form factors are obtained within the BS framework~\cite{Ju:2013oba,Ju:2014oha}. In terms of form factors, the WA matrix elements can be parametrized as,
\begin{align}
T^{\mu}_{ann}(B_{c}\to D^{\ast})
=
(m_{B_c}-m_{D^\ast})
\Big[
&T_{1ann}m_{B_c}^{2}\epsilon^{\mu}
+
T_{2ann}(\epsilon\cdot q)p^{\mu}
+
T_{3ann}(\epsilon\cdot q)k^{\mu}
\nonumber\\
&
+\frac{i}{2}V_{ann}
\epsilon^{\mu\nu\alpha\beta}
\epsilon_{\nu}p_{\alpha}k_{\beta}
\Big].
\label{WAFF}
\end{align}
To analyze the physical observables, we adopt the four-fold differential decay distribution which can be written in terms of the dilepton invariant mass squared $q^{2}$, and the angular variables $\theta_{\ell}$, $\theta_{V}$, and $\phi$ as \cite{Aarfi:2026hgi,Aarfi:2025qcp},
\begin{align}
\frac{d^4\Gamma}{dq^2\,d\cos\theta_{\ell}\,d\cos\theta_V\,d\phi}
=
\frac{9}{32\pi}
\text{Br}(D^\ast\to P_1P_2)
\Big[
&I_{1s}\sin^{2}\theta_V
+
I_{1c}\cos^{2}\theta_V
\nonumber\\
&
+
(I_{2s}\sin^{2}\theta_V
+
I_{2c}\cos^{2}\theta_V)
\cos2\theta_{\ell}
\nonumber\\
&
+
(I_{6s}\sin^{2}\theta_V
+
I_{6c}\cos^{2}\theta_V)
\cos\theta_{\ell}
\nonumber\\
&
+
(I_3\cos2\phi+I_9\sin2\phi)
\sin^{2}\theta_V\sin^{2}\theta_{\ell}
\nonumber\\
&
+
(I_4\cos\phi+I_8\sin\phi)
\sin2\theta_V\sin2\theta_{\ell}
\nonumber\\
&
+
(I_5\cos\phi+I_7\sin\phi)
\sin2\theta_V\sin\theta_{\ell}
\Big].
\label{fullad}
\end{align}

The expressions of angular coefficients $I_i$ containing the helicity amplitudes remain the same as those in the previous analyses \cite{Aarfi:2026hgi,Aarfi:2025qcp}, and are not repeated here for brevity. In terms of the SM and NP WCs and the WA effects, the helicity amplitudes can be expressed as follows:

\begin{align}
H^{1,D^\ast}_{\pm}
&= -i(m_{B_c}^{2}-m_{D^\ast}^{2})
\Bigg[
(C_{9}^{\text{eff}}+C_{9}^{NP}-C_{9}^{\prime NP})
\frac{A_1}{m_{B_c}-m_{D^\ast}}
+
\frac{2m_b}{q^{2}}(C_{7}^{\text{eff}}+C_{7}^{NP})T_2
\Bigg] \nonumber\\
&\quad \pm i\sqrt{\lambda}
\Bigg[
(C_{9}^{\text{eff}}+C_{9}^{NP}+C_{9}^{\prime NP})
\frac{V}{m_{B_c}+m_{D^\ast}}
+
\frac{2m_b}{q^{2}}(C_{7}^{\text{eff}}+C_{7}^{NP})T_1
\Bigg] \nonumber\\
&\quad - iR_{PBAnn}
\Big[
m_{B_c}^{2}(m_{B_c}-m_{D^\ast})T_{1\text{ann}}
\pm \tfrac{\sqrt{\lambda}}{2}(m_{B_c}-m_{D^\ast})V_{\text{ann}}
\Big],
\\[6pt]
H^{2,D^\ast}_{\pm}
&= -i(C_{10}+C_{10}^{NP}-C_{10}^{\prime NP})
(m_{B_c}+m_{D^\ast})A_1 \nonumber\\
&\quad \pm i\sqrt{\lambda}(C_{10}+C_{10}^{NP}+C_{10}^{\prime NP})
\frac{V}{m_{B_c}+m_{D^\ast}},
\\[6pt]
H^{1,D^\ast}_{0}
&= -\frac{i}{2m_{D^\ast}\sqrt{q^{2}}}
\Bigg[
(C_{9}^{\text{eff}}+C_{9}^{NP}-C_{9}^{\prime NP})
\Big(
(m_{B_c}^{2}-m_{D^\ast}^{2}-q^{2})(m_{B_c}+m_{D^\ast})A_1
- \frac{\lambda}{m_{B_c}+m_{D^\ast}}A_2
\Big) \nonumber\\
&\quad + 2m_b(C_{7}^{\text{eff}}+C_{7}^{NP})
\Big(
(m_{B_c}^{2}+3m_{D^\ast}^{2}-q^{2})T_2
- \frac{\lambda}{m_{B_c}^{2}-m_{D^\ast}^{2}}T_3
\Big)
\Bigg] \nonumber\\
&\quad - \frac{iR_{PBAnn}}{4\sqrt{q^{2}}m_{D^\ast}(m_{B_c}+m_{D^\ast})}
(m_{D^\ast}^{2}-m_{B_c}^{2})
\Big[
2m_{B_c}^{2}T_{1\text{ann}}(q^{2}-m_{B_c}^{2}+m_{D^\ast}^{2})
- \lambda(T_{2\text{ann}}+T_{1\text{ann}})
\Big],
\\[6pt]
H^{2,D^\ast}_{0}
&= -\frac{i}{2m_{D^\ast}\sqrt{q^{2}}}
(C_{10}+C_{10}^{NP}-C_{10}^{\prime NP})
\Big[
(m_{B_c}^{2}-m_{D^\ast}^{2}-q^{2})(m_{B_c}+m_{D^\ast})A_1
- \frac{\lambda}{m_{B_c}+m_{D^\ast}}A_2
\Big],
\\[6pt]
H^{1,D^\ast}_{t}
&= -i\sqrt{\frac{\lambda}{q^{2}}}
(C_{9}^{\text{eff}}+C_{9}^{NP}-C_{9}^{\prime NP})A_0 \nonumber\\
&\quad - \frac{i\sqrt{\lambda}R_{PBAnn}}{4\sqrt{q^{2}}m_{D^\ast}}
(m_{B_c}-m_{D^\ast})
\Big[
2m_{D^\ast}^{2}T_{1\text{ann}}
+ (q^{2}+m_{B_c}^{2}-m_{D^\ast}^{2})T_{2\text{ann}}
- (q^{2}-m_{B_c}^{2}+m_{D^\ast}^{2})T_{3\text{ann}}
\Big],
\\[6pt]
H^{2,D^\ast}_{t}
&= -i\sqrt{\frac{\lambda}{q^{2}}}
(C_{10}+C_{10}^{NP}-C_{10}^{\prime NP})A_0,
\end{align}

where
\begin{eqnarray}
R_{PBAnn}
=
\left(
\frac{V_{cb}V_{cd}^{\ast}}
{V_{tb}V_{td}^{\ast}}
\right)
\left(
\frac{C_1}{3}+C_2
\right).
\end{eqnarray}

\section{Numerical inputs and Phenomenological Analysis}\label{NA}

In this section, we investigate the NP effects through the observables associated with  $B_{c}^+\to D^{\ast+}(\to P_{1}P_{2})\ell^{+}\ell^{-}$ decays for $\ell=\mu,\tau$. The observables considered in this work for the analysis are the differential branching fraction 
$d\text{Br}/dq^{2}$, the lepton forward-backward asymmetry $A_{\rm FB}$, the longitudinal helicity fraction $f_{L}$, and the normalized angular coefficients $\langle I_{i}\rangle$. The expressions of these observables in terms of angular coefficients $I_{i}$ are as follows.

\begin{itemize}

\item Differential branching fraction:
\begin{equation}
\frac{d \text{Br}(B_c^+\to D^{\ast+}(\to P_1P_2)\ell^{+}\ell^{-})}{dq^{2}}
=\tau_{B_c} \text{Br}(D^{\ast+}\to P_1P_2)
\frac{\left(
3I_{1c}
+
6I_{1s}
-
I_{2c}
-
2I_{2s}
\right)}{4}
.
\label{Br1}
\end{equation}
Ignoring the cascade decay $D^{\ast+}\to P_1P_2$, 
\begin{equation}
\frac{d \text{Br}(B_c^+\to D^{\ast+}\ell^{+}\ell^{-})}{dq^{2}}
=\tau_{B_c}
\frac{\left(
3I_{1c}
+
6I_{1s}
-
I_{2c}
-
2I_{2s}
\right)}{4}
.
\label{Br2}
\end{equation}

\item Lepton forward-backward asymmetry:
\begin{equation}
A_{\rm FB}(q^{2})
=
\frac{6I_{6s}}
{2(3I_{1c}+6I_{1s}-I_{2c}-2I_{2s})}.
\label{AFB1}
\end{equation}

\item Longitudinal helicity fraction:
\begin{equation}
f_{L}(q^{2})
=
\frac{3I_{1c}-I_{2c}}
{3I_{1c}+6I_{1s}-I_{2c}-2I_{2s}}.
\label{fL1}
\end{equation}

\item Normalized angular coefficients:
\begin{equation}
\langle I_i\rangle
=
\frac{
\text{Br}(D^{\ast+}\to P_1P_2)\,I_i
}{
d\Gamma(B_c^+\to D^{\ast+}(\to P_1P_2)\ell^{+}\ell^{-})/dq^{2}
}.
\label{angobsorig}
\end{equation}

\end{itemize}


\begin{table}[ht]
\centering
\caption{Input parameters used in the study~\cite{ParticleDataGroup:2024cfk}.}
\label{input}
\begin{tabular}{|c|c|}
\hline
$m_{B_c}=6.2749$ GeV & $m_{\mu}=0.106$ GeV \\
$m_d=0.0048$ GeV & $m_b=4.18$ GeV \\
$m_{\tau}=1.776$ GeV & $m_{D^\ast}=2.010$ GeV \\
$\alpha^{-1}=137$ & $G_F=1.166\times10^{-5}$ GeV$^{-2}$ \\
$|V_{tb}V_{td}^{\ast}|=8.1\times10^{-3}$ & $|V_{cb}|=0.041$ \\
$\tau_{B_c}=0.507\times10^{-12}$ sec &
$\text{Br}(D^{\ast +}\to D^0\pi^+)=67.7\%$
\\
\hline
\end{tabular}
\end{table}
The required input parameters are listed in Table~\ref{input}. To determine the effects of NP, we use several model-independent 1D and 2D NP scenarios involving the WCs 
$C_{7}^{NP}$, 
$C_{9}^{NP}$, 
$C_{10}^{NP}$, 
$C_{9}^{\prime NP}$, and 
$C_{10}^{\prime NP}$. 
The best-fit values and the corresponding $1\sigma$ and $2\sigma$ ranges of the NP WCs are adopted from global fit analysis \cite{Bause:2022rrs}, and their numerical values are summarized in Table~\ref{tableNPWCs}.

\begin{table*}[ht!]
 \centering
 \caption{Best-fit values and the corresponding $1\sigma$ and $2\sigma$ ranges of the NP WCs in different 1D and 2D NP scenarios~\cite{Bause:2022rrs}.}
\label{tableNPWCs}
 \renewcommand*{\arraystretch}{1.6}
 \resizebox{0.95\textwidth}{!}{
 \begin{tabular}{|c|c|c|c|c|c|c|c|}
  \hline
  \hline
  Scenario  & Fit Parameters & Best Fit & $1\sigma$ & $2\sigma$ & $\chi^2_{H_i, min}$ & $\text{Pull}_{H_i}$ & p-value(\%)\\
  \hline
  \hline

   SI & $C^{NP}_9$ & $-1.37$ & $[-2.97, -0.47]$ & [$-7.65$, $0.26$] & $1.23$ & $1.63$ & $94$ \\
   
   SII & $C^{NP}_{10}$ & $0.96$ & $[0.31, 1.74]$ & $[-0.27, 2.88]$ & $1.55$ & $1.53$ & $90$ \\
   
   SIII & $C^{NP}_9=-C^{NP}_{10}$ & $-0.54$ & $[-0.90, -0.20]$ & $[-1.29, 0.13]$ & $1.32$ & $1.60$ & $93$  \\
   
   SIV & $C^{NP}_9=-C^{NP}_{10}=-C^{' NP}_9=-C^{' NP}_{10}$ & $-0.58$ & $[-1.06, -0.20]$ & $[-4.04, 0.12]$ & $1.28$ & $1.61$ & $93$ \\
  
  \hline  
  
  SV & ($C^{NP}_7$,\,$C^{NP}_9$) & $(0.11, -1.55)$ & ($[-0.05, 0.34]$, $[-3.05, -0.61]$) & ($[-0.18, 1.46]$, $[-10.07, 0.18]$) & $0.77$ & $1.25$ & $94$ \\
  
  SVI & ($C^{NP}_9$,\,$C^{' NP}_9$) & $(-2.22, 1.18)$ & ($[-6.55, -0.63]$, $[-2.99, 2.89]$) & ($[-7.58, 0.23]$, $[-3.92, 3.81]$) & $0.87$ & $1.22$ & $92$ \\
  
  SVII & ($C^{NP}_9$,\,$C^{' NP}_{10}$) & $(-1.83, -0.38)$ & ($[-6.58, -0.6]$, $[-1.2, 0.32]$) & ($[-7.6, 0.25]$, $[-1.8, 0.99]$) & $0.95$ & $1.20$ & $91$ \\
  
  SVIII & $( C^{NP}_9=-C^{' NP}_9,\,C^{NP}_{10}=+C^{' NP}_{10} )$ & $(-1.73, 0.44)$ & ($[-3.34, -0.19]$, $[0.04, 0.95]$) & ($[-4.1, 0.51]$, $[-0.34, 4.52]$) & $0.88$ & $1.22$ & $92$ \\
  
  \hline
  \hline
  \end{tabular}
}
\end{table*}

It is worth mentioning that the NP scenarios SI, SII, and SIII emerge naturally in a variety of simple NP models featuring tree-level leptoquark or $Z^\prime$ boson exchanges. Specifically, SI and SII can be generated in $Z^\prime$ models, while SIII is realized in scalar or vector leptoquarks, and also in a $Z^\prime$ model having purely left-handed couplings~\cite{Crivellin:2018yvo,Bobeth:2016llm,DAlise:2024qmp}.

In the following subsections, we first discuss the predictions of the observables in different 1D NP scenarios and then extend our analysis to the 2D NP scenarios.

\subsection{Probing NP signatures using 1D NP scenarios\label{results-1D}}

\begin{figure}[H]
\centering
\includegraphics[width=2in,height=1.24in]{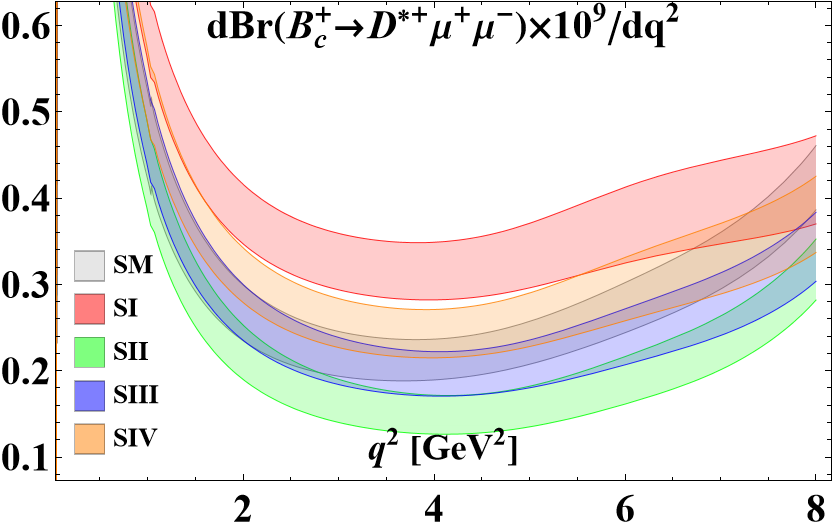}
\includegraphics[width=2in,height=1.24in]{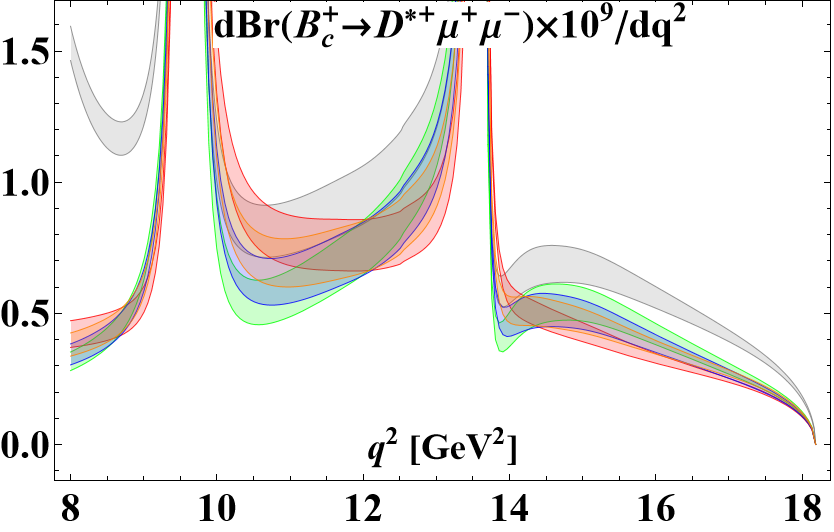}
\raisebox{0.0in}{\includegraphics[width=2in,height=1.24in]{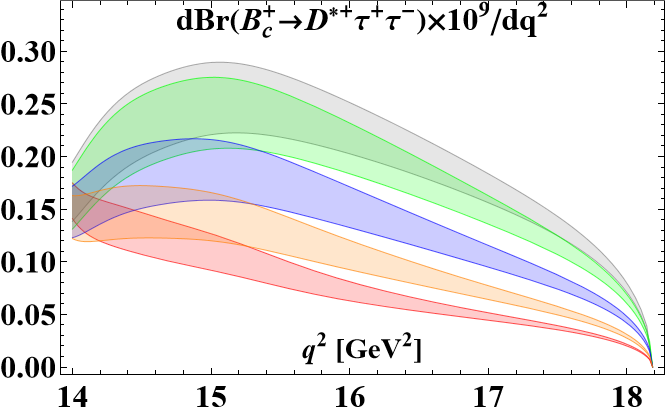}}
\includegraphics[width=2in,height=1.24in]{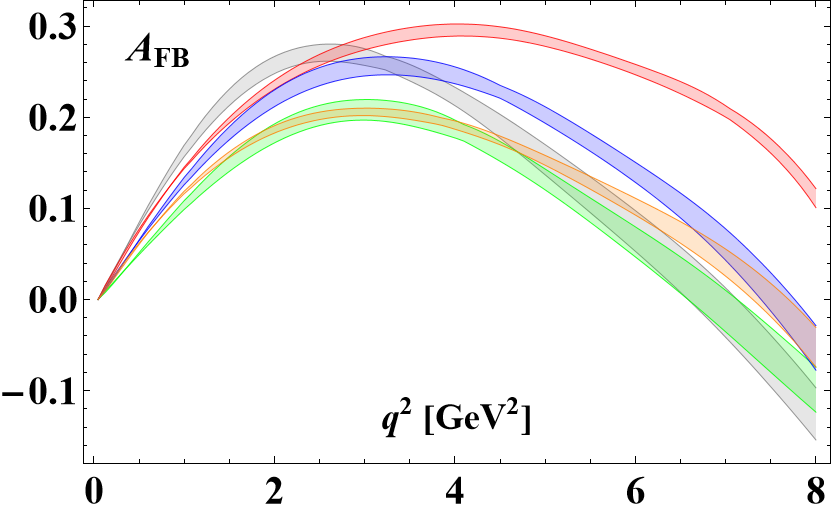}
\includegraphics[width=2in,height=1.24in]{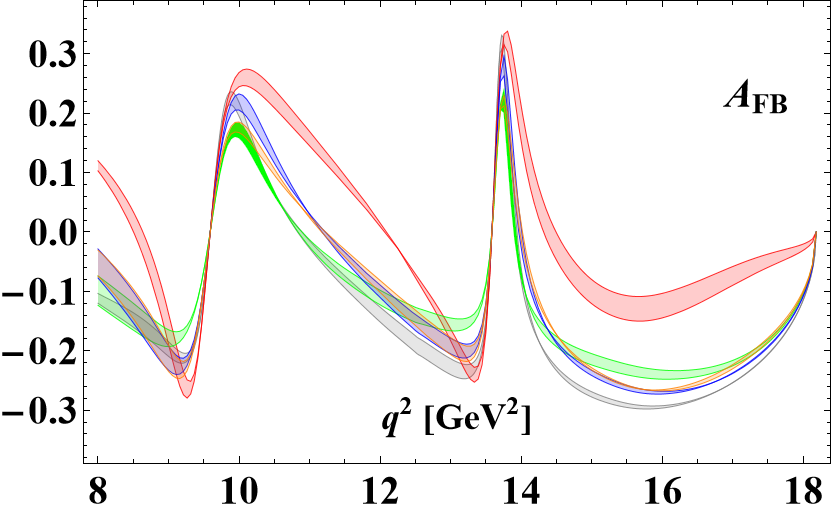}
\raisebox{0.0in}{\includegraphics[width=2in,height=1.24in]{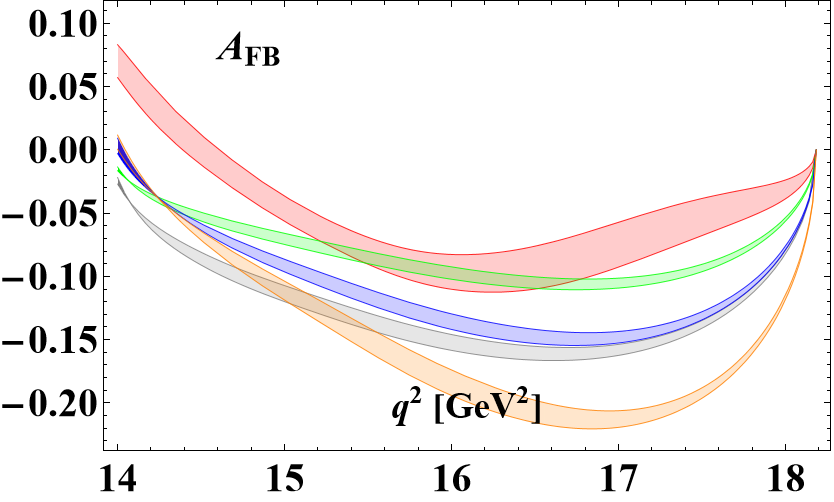}}
\includegraphics[width=2in,height=1.24in]{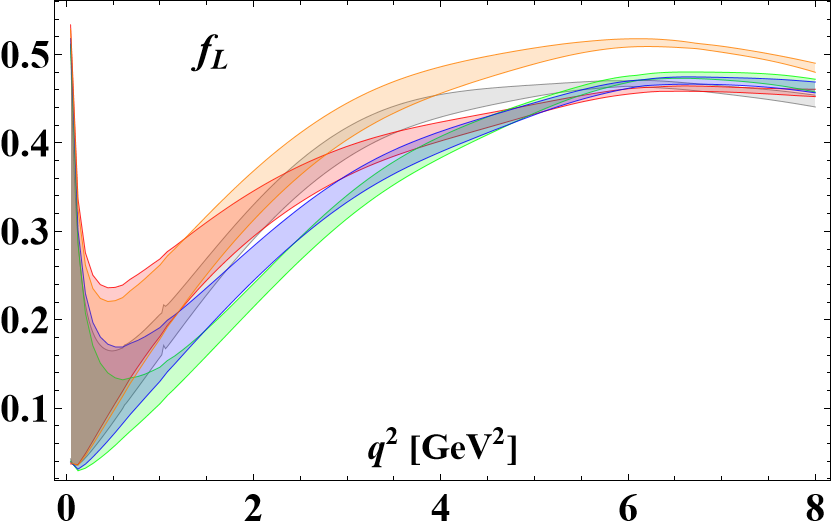}
\includegraphics[width=2in,height=1.24in]{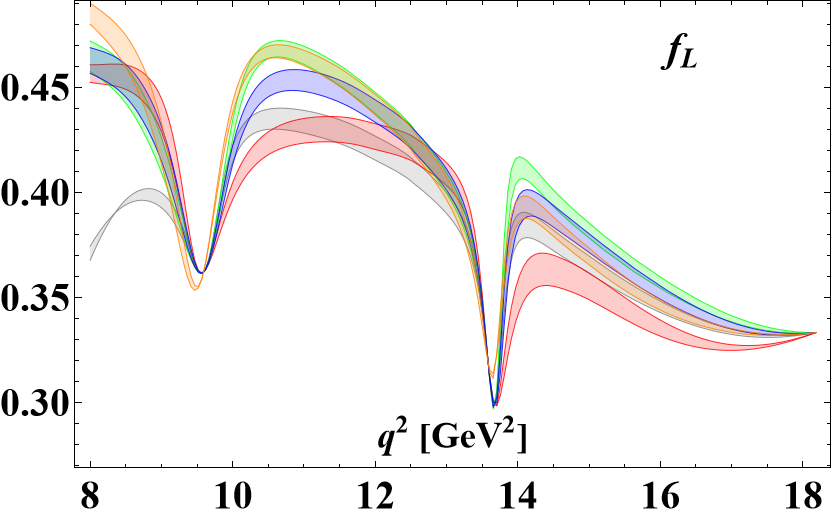}
\raisebox{0.0in}{\includegraphics[width=2in,height=1.24in]{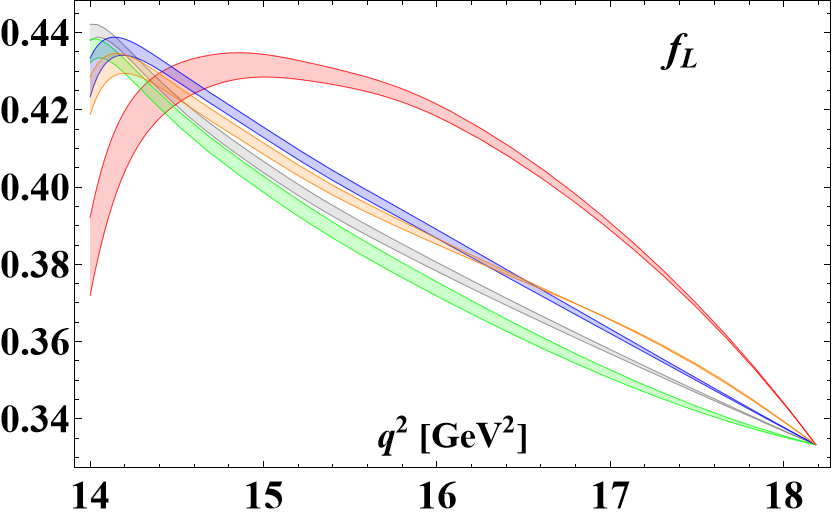}}
\caption{Predictions for the differential branching fraction
$d\text{Br}/dq^{2}$, the forward-backward asymmetry $A_{\rm FB}$, and the longitudinal helicity fraction $f_{L}$ of the decay
$B_c^{+}\to D^{\ast+}\ell^{+}\ell^{-}$ in the SM and the 1D NP scenarios SI--SIV. The first and second columns display these observables (from top to bottom) as functions of the dilepton invariant mass squared $q^{2}$ for the low- and high-$q^{2}$ ranges of the muon channel, respectively, whereas the third column shows the corresponding predictions for the high-$q^{2}$ range of the tau channel.}
\label{fig:1D_BrAFBfL}
\end{figure}

\begin{figure}[b]
\centering
\includegraphics[width=2in,height=1.24in]{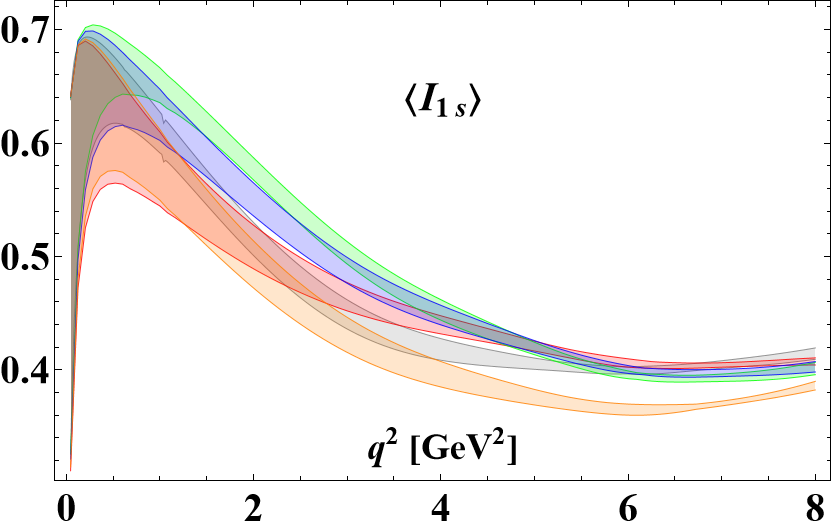}
\includegraphics[width=2in,height=1.24in]{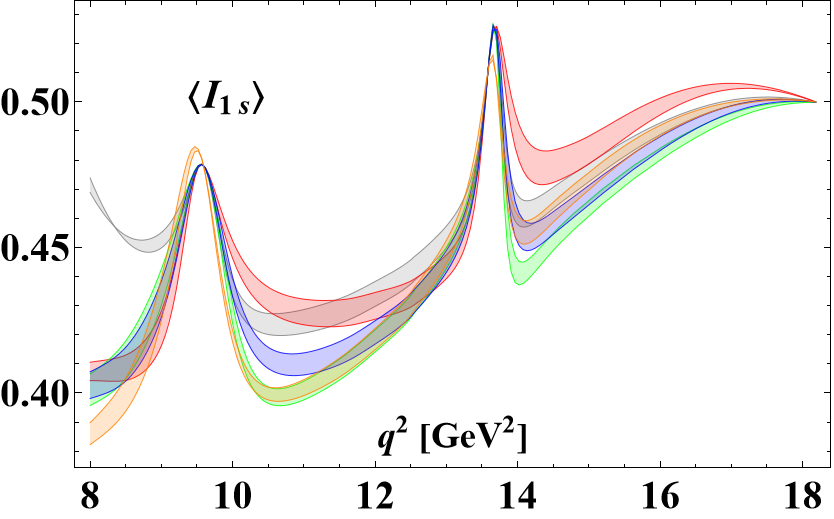}
\raisebox{0.0in}{\includegraphics[width=2in,height=1.24in]{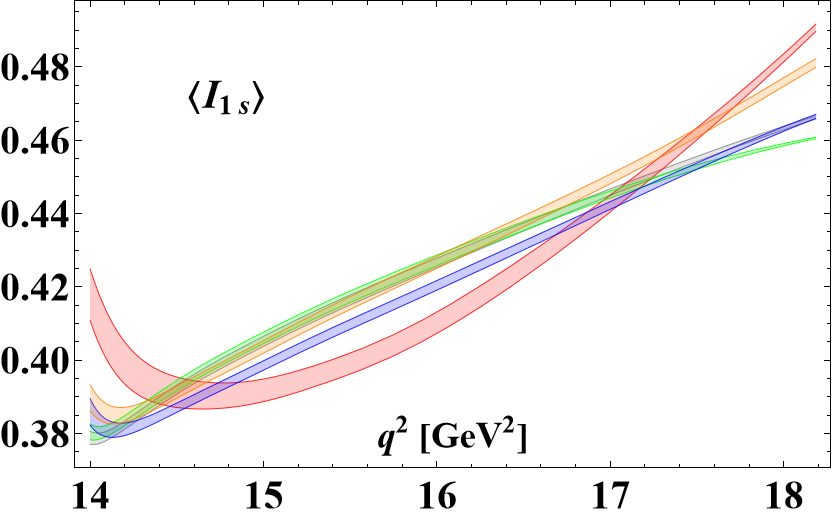}}
\\
\includegraphics[width=2in,height=1.24in]{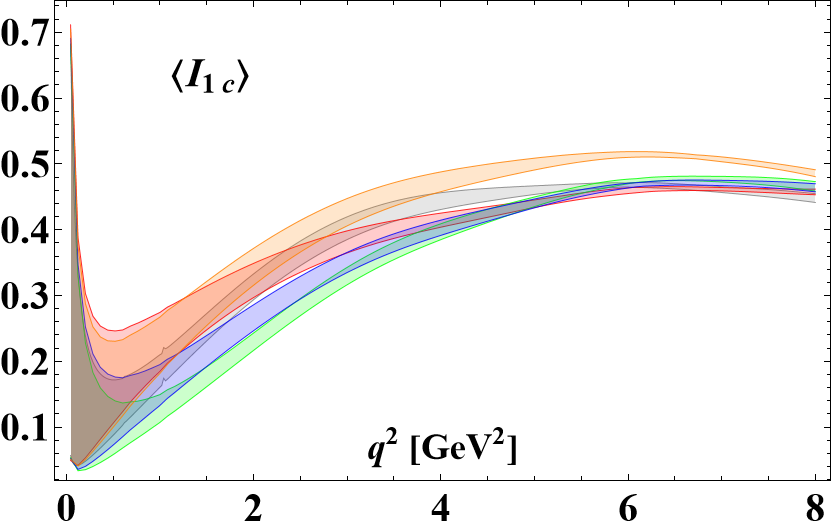}
\includegraphics[width=2in,height=1.24in]{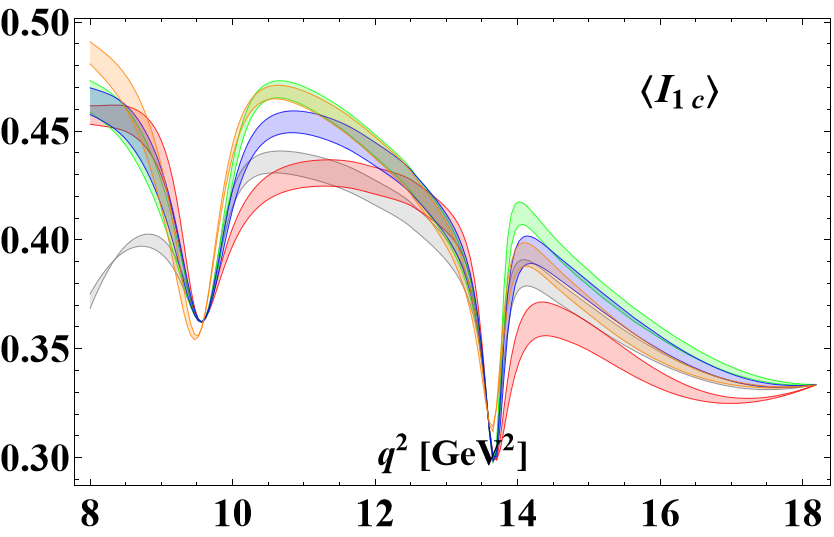}
\raisebox{0.00in}{\includegraphics[width=2in,height=1.24in]{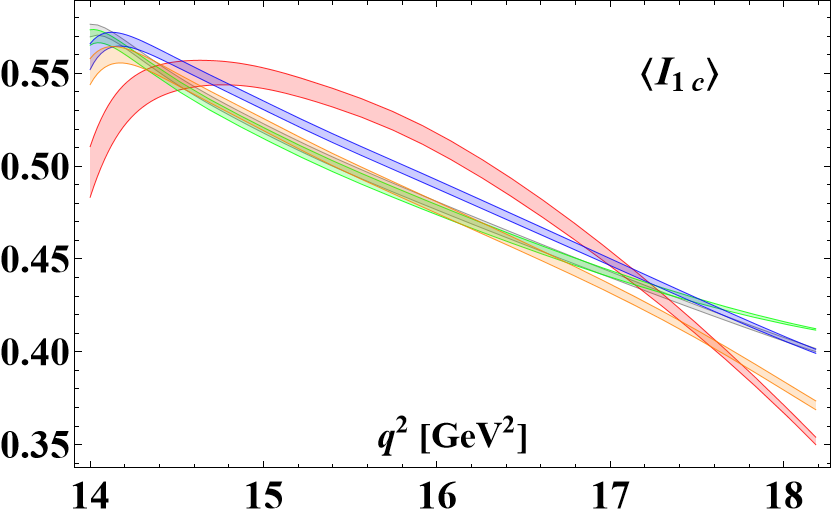}}
\\
\includegraphics[width=2in,height=1.24in]{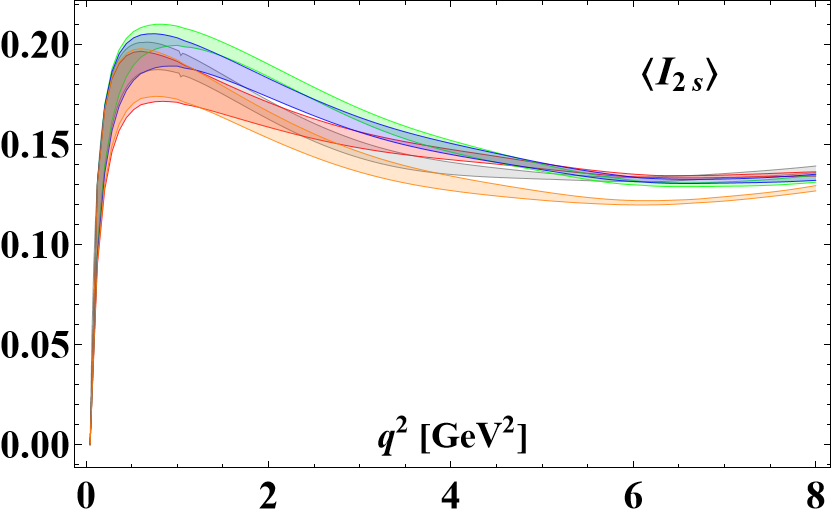}
\includegraphics[width=2in,height=1.24in]{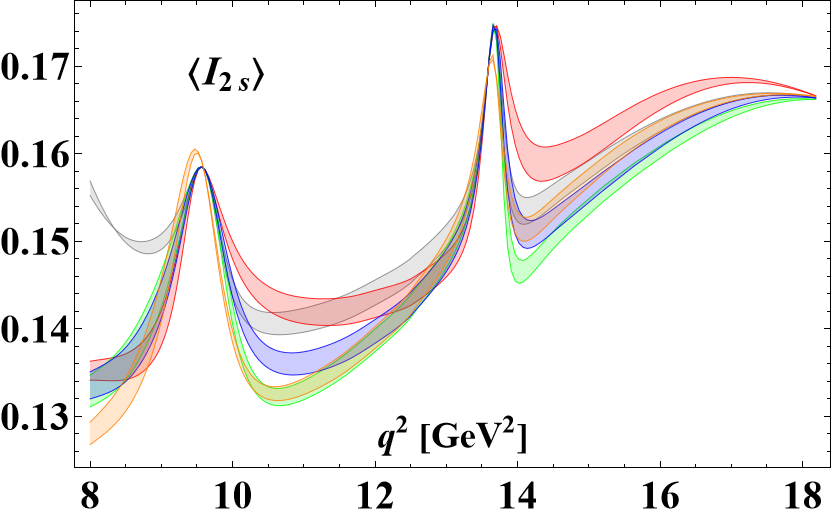}
\raisebox{0.0in}{\includegraphics[width=2in,height=1.24in]{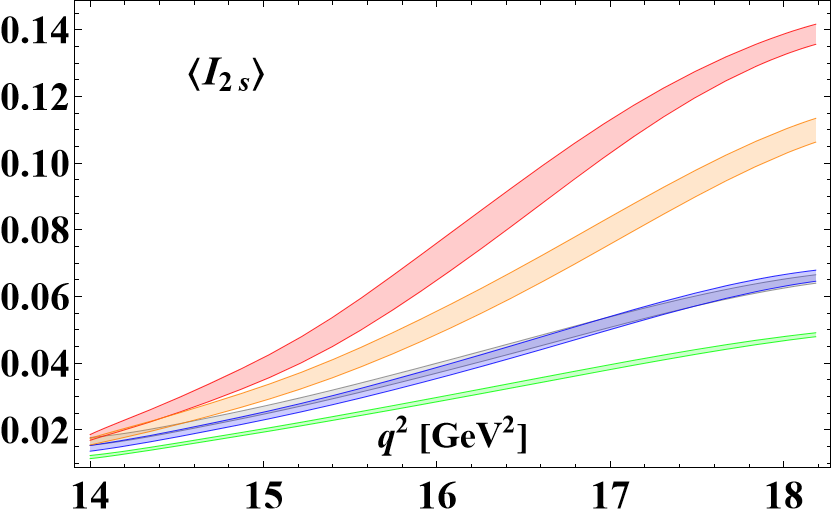}}
\\
\includegraphics[width=2in,height=1.24in]{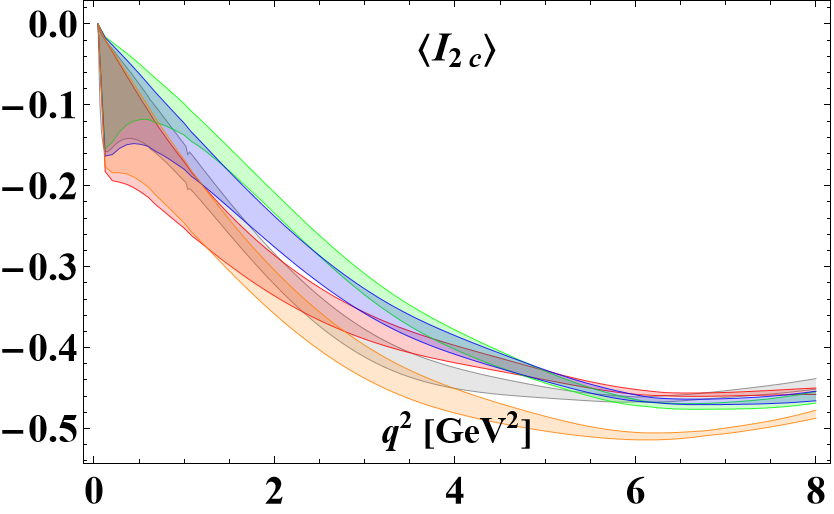}
\includegraphics[width=2in,height=1.24in]{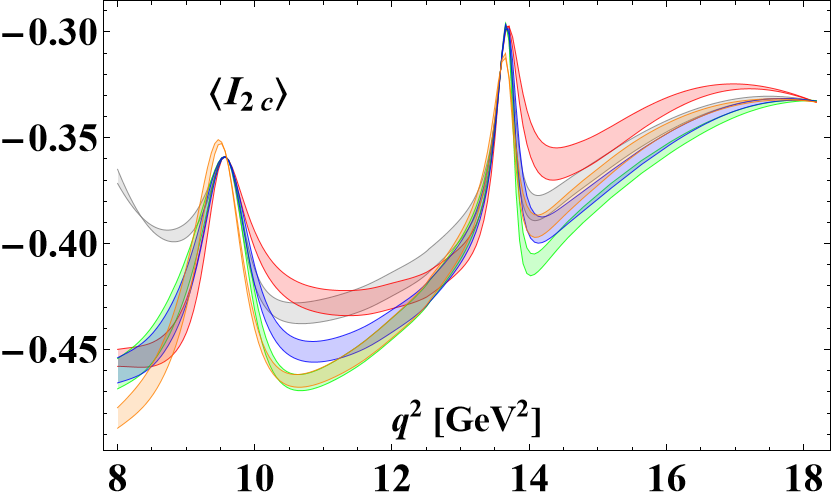}
\raisebox{0.0in}{\includegraphics[width=2in,height=1.24in]{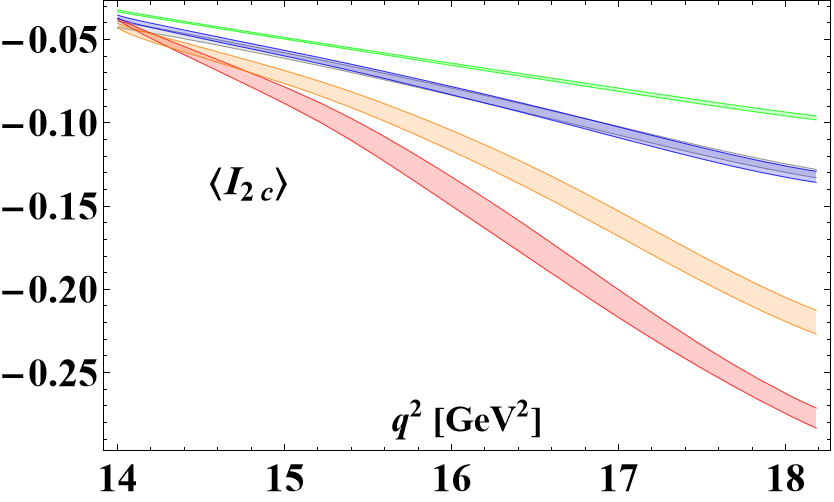}}
\caption{Predictions for the normalized angular coefficients $\langle I_{1s}\rangle$, $\langle I_{1c}\rangle$, $\langle I_{2s}\rangle$, and $\langle I_{2c}\rangle$ of the decay
$B_c^{+}\to D^{\ast+}\ell^{+}\ell^{-}$ in the SM and the 1D NP scenarios SI--SIV. The arrangement of the panels is the same as in Fig.~\ref{fig:1D_BrAFBfL}.}
\label{angularpnl11}
\end{figure}

\begin{figure}[H]
\centering
\includegraphics[width=2in,height=1.24in]{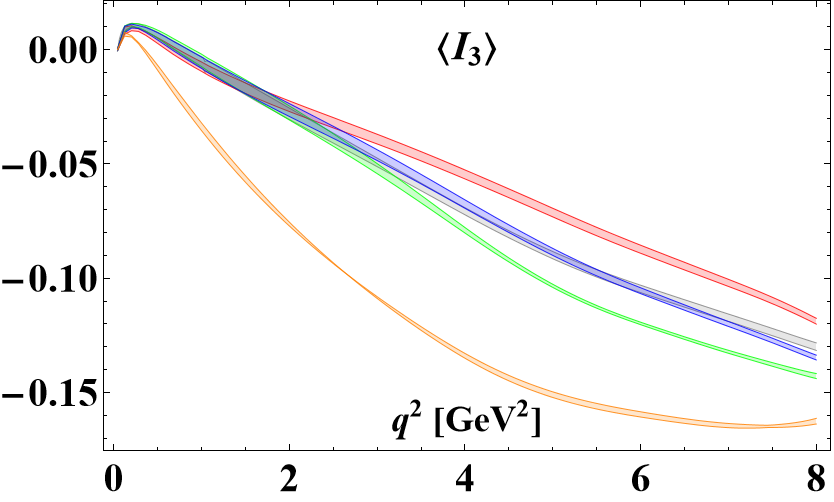}
\includegraphics[width=2in,height=1.24in]{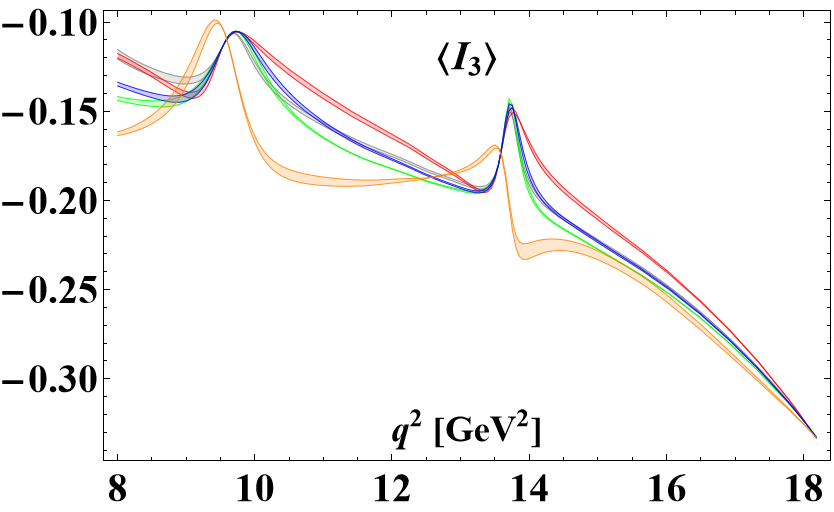}
\raisebox{0.0in}{\includegraphics[width=2in,height=1.24in]{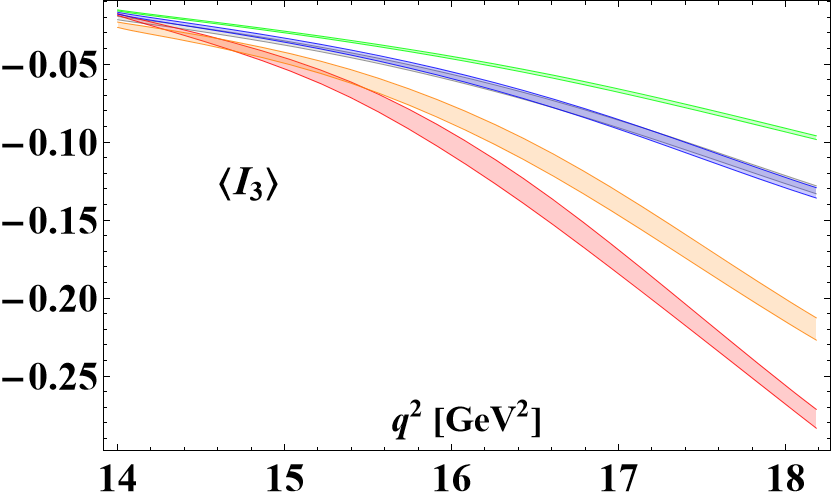}}
\\
\includegraphics[width=2in,height=1.24in]{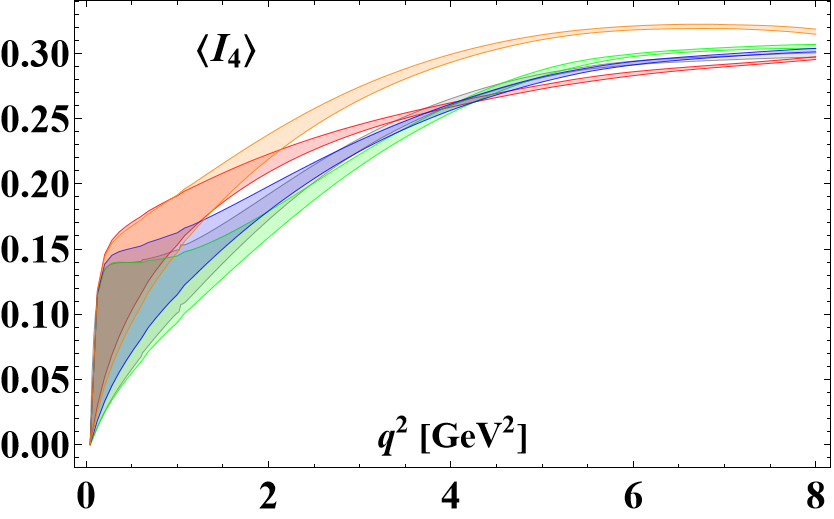}
\includegraphics[width=2in,height=1.24in]{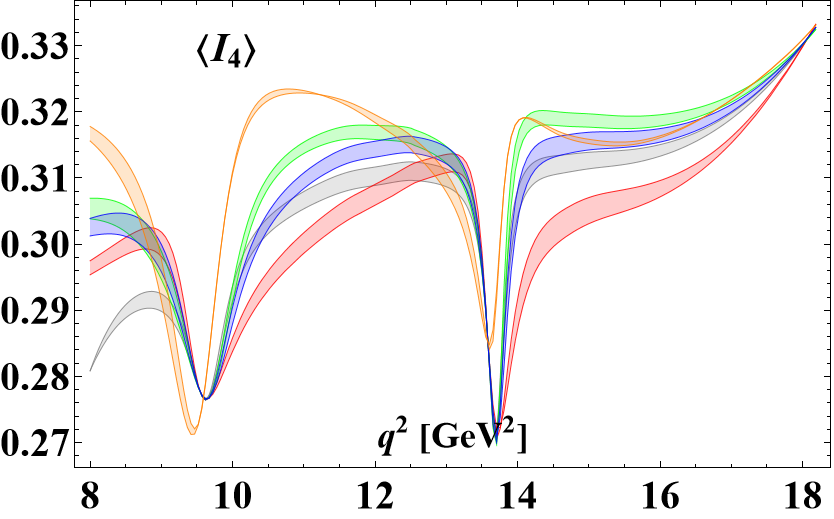}
\raisebox{0.00in}{\includegraphics[width=2in,height=1.24in]{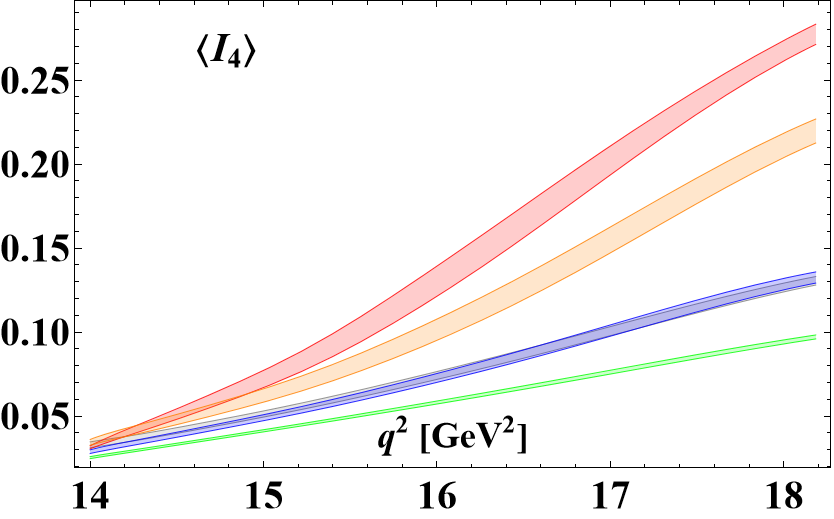}}
\\
\includegraphics[width=2in,height=1.24in]{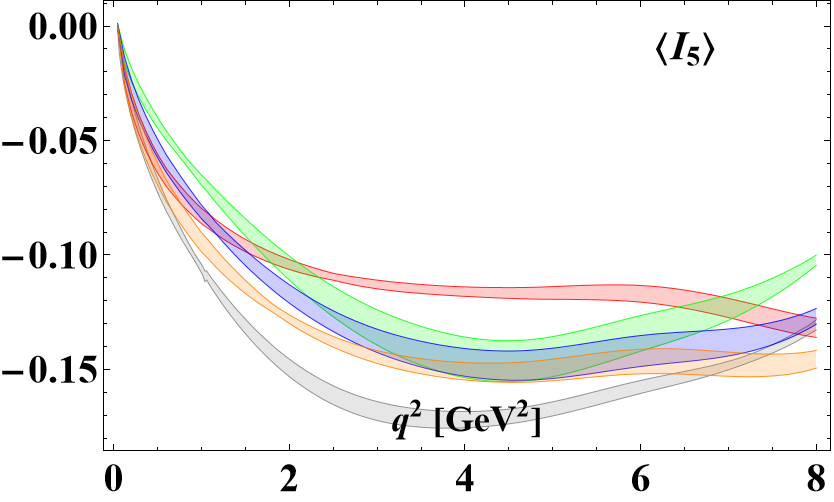}
\includegraphics[width=2in,height=1.24in]{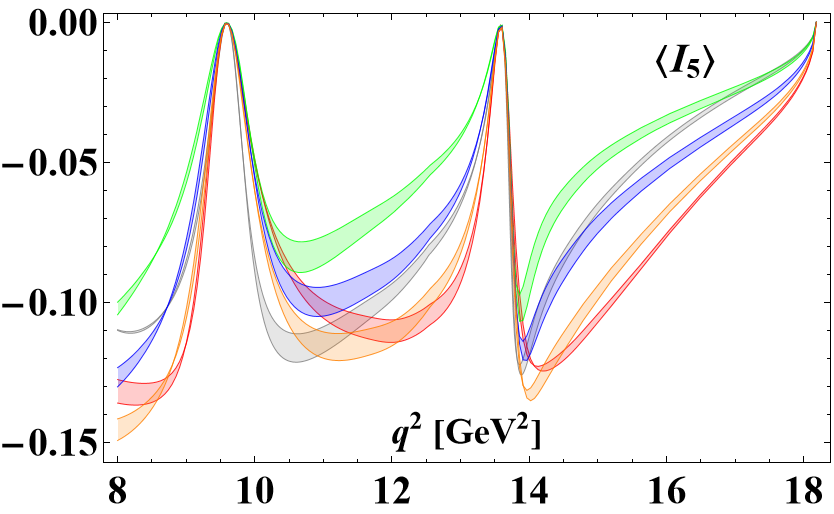}
\raisebox{0.0in}{\includegraphics[width=2in,height=1.24in]{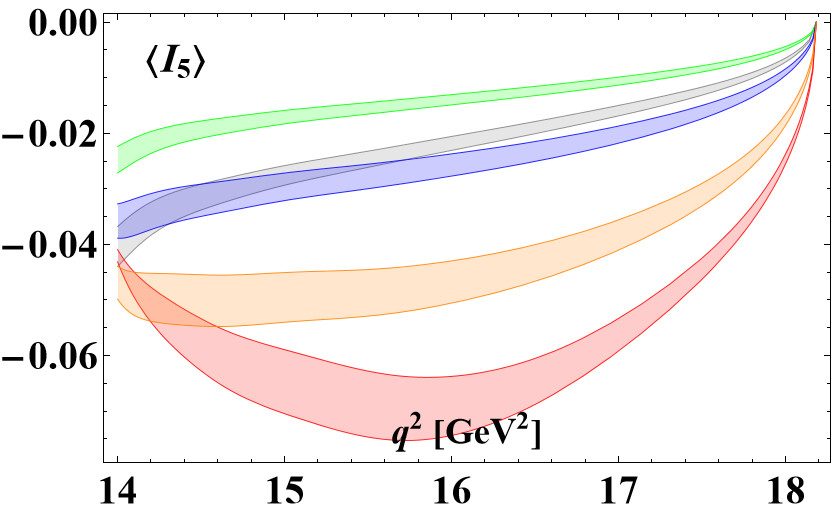}}
\\
\includegraphics[width=2in,height=1.24in]{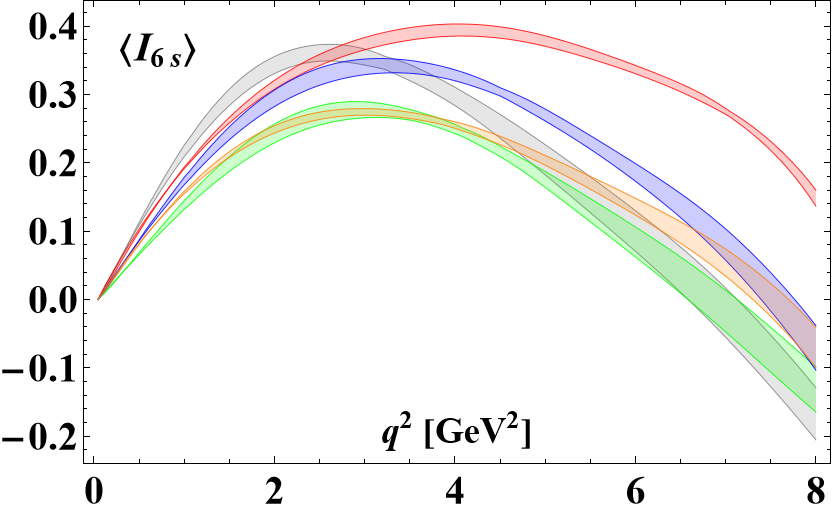}
\includegraphics[width=2in,height=1.24in]{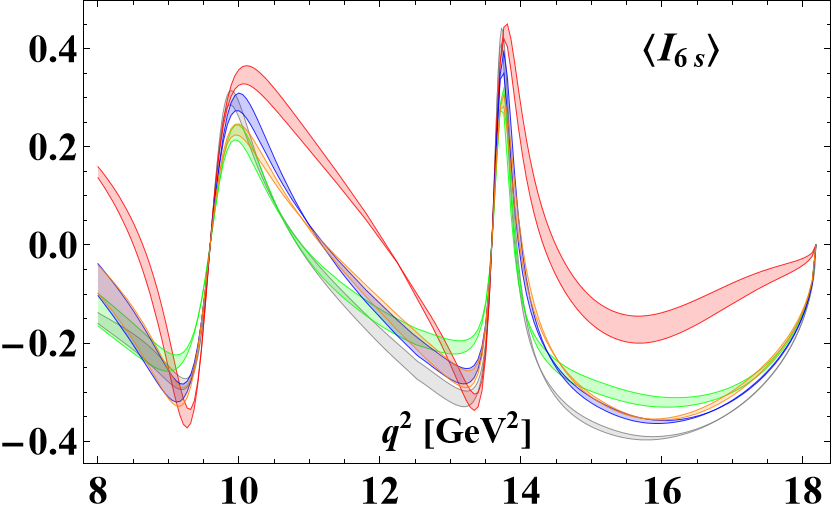}
\raisebox{0.0in}{\includegraphics[width=2in,height=1.24in]{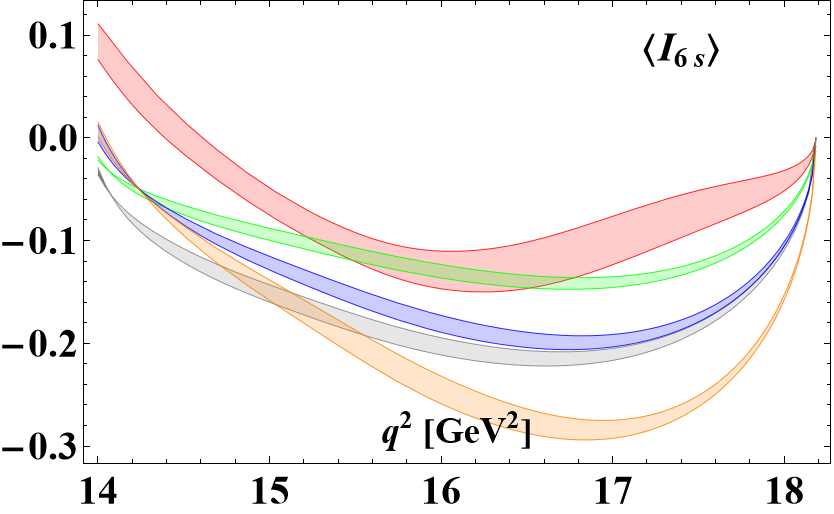}}    
\caption{Predictions for the normalized angular coefficients $\langle I_{3}\rangle$, $\langle I_{4}\rangle$, $\langle I_{5}\rangle$, and $\langle I_{6s}\rangle$ of the decay
$B_c^{+}\to D^{\ast+}\ell^{+}\ell^{-}$ in the SM and the 1D NP scenarios SI--SIV. The arrangement of the panels is the same as in Fig.~\ref{fig:1D_BrAFBfL}.}
\label{angularpnl22}
\end{figure}

In this subsection, we discuss the effects of the 1D NP scenarios SI--SIV on the differential branching fraction 
$d\text{Br}/dq^{2}$, lepton FB asymmetry $A_{\rm FB}$, longitudinal helicity fraction $f_{L}$, and the angular coefficients $\langle I_{i}\rangle$, for the decay 
$B_{c}^+\to D^{\ast+}\ell^{+}\ell^{-}$. 
The corresponding results are presented in Figs.~\ref{fig:1D_BrAFBfL}--~\ref{angularpnl22}. 
The plots are organized from left to right in three different kinematical ranges: the low-$q^{2}$ muon range 
($4m_\mu^2 \simeq 0.045~\mathrm{GeV}^2 \le q^{2}\le 8~{\rm GeV}^{2})$, the high-$q^{2}$ muon range 
$(8 \le q^{2}\le 18.18~{\rm GeV}^{2})$, and the high-$q^{2}$ tau range $(14 \le q^{2}\le 18.18~{\rm GeV}^{2})$. The bands in these figures correspond to the uncertainty in the predictions of the observables, coming mainly from the errors in the form factors. To enhance discrimination among different NP scenarios, the color bands are plotted using the lower limit $1\sigma$ value for SI--SIII and the upper limit $1\sigma$ value for SIV as given in Table \ref{tableNPWCs}. 
The scenarios are color-coded as follows:
\begin{itemize}
    \item SI ($C_{9}^{NP}$): red,
    \item SII ($C_{10}^{NP}$): green,
    \item SIII ($C_{9}^{NP}=-C_{10}^{NP}$): blue,
    \item SIV ($C_{9}^{NP}=-C_{10}^{NP}=-C_{9}^{\prime NP}=-C_{10}^{\prime NP}$): orange.
\end{itemize}

The first left panel of Fig.~\ref{fig:1D_BrAFBfL} shows the differential branching fraction 
$d\text{Br}/dq^{2}$. 
In the low-$q^{2}$ muon range, SI (red band) enhances the branching fraction compared to the SM prediction (gray band). The SIII and SIV scenarios remain close to the SM prediction, whereas SII suppresses the branching fraction.

In the middle plot of the first row, we show the predictions of the $d\text{Br}/dq^{2}$ for the remaining range $8 \leq q^2 \leq q_{\mathrm{max}}^2$, including the charmonium resonance range. In the high-$q^2$ muon range, it is observed that the NP predictions tend to lie below the SM prediction across much of the kinematic range, and significant overlap is seen among the NP scenarios. Therefore, the high-$q^2$ differential branching fraction shows sensitivity to NP effects, but the substantial overlap limits the ability to distinguish different NP scenarios. For the case of $\tau$ as the final state lepton, shown in the right-most panel of the first row, the branching fraction decreases with increasing $q^{2}$. The hierarchy among the NP scenarios differs from the muon case due to the significant impact of the lepton-mass effects in the helicity amplitudes. Among the considered scenarios, the SI scenario exhibits the largest deviation from the SM prediction, whereas the SII scenario shows the least pronounced deviation relative to the SM over most of the allowed kinematic range.

The second row of Fig.~\ref{fig:1D_BrAFBfL} displays the forward-backward asymmetry $A_{\rm FB}$. In the low-$q^2$ muon range, all scenarios predict positive values of $A_{FB}$ over a sizable part of the kinematic range. The shift in zero-crossing of $A_{\rm FB}$ depends on the NP scenario. From the trend shown in Fig.~\ref{fig:1D_BrAFBfL}, the zero crossing of $A_{\rm FB}$ in the SM is approximately at $q^2 \simeq 6.75~\mathrm{GeV}^2$, while the SII, SIII, and SIV scenarios exhibit the zero crossing of $A_{\rm FB}$ near $q^2 \simeq 6.73~\mathrm{GeV}^2$, $7.61~\mathrm{GeV}^2$, and $7.19~\mathrm{GeV}^2$, respectively. However, the SI scenario remains positive throughout the  low-$q^2$ interval. In the high $q^2$ muon range, we also incorporate the effects of the LD resonance along with the interference between the WA and PB amplitudes. The SI scenario depicts the notable deviation within the  plotted range, while the other scenarios shows a small deviation and partly overlap with each other. For the $\tau$ as final state lepton, it has been observed that $A_{\rm FB}$ is distinguishable for all the NP scenarios under consideration, when compared to that of SM. However scenarios SII, SIII and SIV approximately overlap with SM at $14 \le q^{2}\le 15~{\rm GeV}^{2}$. Additionally, the scenario SI shows a clear departure from the SM across the whole kinematic range for the $\tau$, and it further shows clear distinguishability from the other NP scenarios.

The longitudinal helicity fraction $f_L$ versus $q^{2}$ is shown in the third row of Fig.~\ref{fig:1D_BrAFBfL}. In the low-$q^2$ range for muons, $f_L$ increases with increasing $q^2$ values for all the considered 1D NP scenarios. At the lowest value of $q^2$, the NP bands remain close to each other. It has been observed that the SIV scenario becomes distinguishable around $q^{2}\approx 5$ $\text{GeV}^{2}$, while the other scenarios remain closer to the SM prediction. In the high-$q^2$  range for the muon, $f_L$ shows sharp variations around the charmonium resonance ranges. Between the resonances, the SII and SIII scenarios tend to increase $f_L$, while SI gives relatively smaller values. Above the $\psi(2S)$ range, all predictions decrease and gradually converge near the endpoint. Thus, $f_L$ provides moderate separation among the considered NP scenarios.
For the $\tau-$lepton, $f_{L}$ as a function of $q^{2}$ is depicted in extreme right side of third row of Fig.~\ref{fig:1D_BrAFBfL}. From the figure, it is evident that among all the NP scenarios under consideration, the scenario SI starts with the lowest value of $f_{L}$ and after crossing all the other NP scenarios and the SM at $q^{2}\approx 14.3$ $\text{GeV}^{2}$, it produces the most pronounced enhancement of the $f_{L}$ along with presenting a clear discrimination with the other NP scenarios. Contrary to SI, SIII, and SIV, scenario SII predicts lower values of the longitudinal helicity fraction compared to the SM predictions in most of the allowed kinematic range.

Overall, the 1D NP scenarios produce visible deviations from the SM results in these observables. From Fig.~\ref{fig:1D_BrAFBfL}, in the low-$q^{2}$ muon range, the differential branching fraction and $A_{FB}$ show the largest visible separation among the NP bands, whereas $f_L$ exhibits a comparatively larger overlap among the scenarios. Therefore, in the low-$q^{2}$ muon range, $d\text{Br}/dq^{2}$ and $A_{FB}$ are expected to be more suitable observables for distinguishing the 1D NP scenarios, whereas $f_L$ is less suitable. However, for the case of $\tau$ lepton, all three observables are sensitive to the significant deviation of NP from the SM predictions, in the allowed kinematic range. 


The predictions for the normalized angular coefficients 
$\langle I_{1s}\rangle$, 
$\langle I_{1c}\rangle$, 
$\langle I_{2s}\rangle$, and 
$\langle I_{2c}\rangle$ in the SM and the 1D NP scenarios are presented in Fig.~\ref{angularpnl11}, while the corresponding results for 
$\langle I_{3}\rangle$, 
$\langle I_{4}\rangle$, 
$\langle I_{5}\rangle$, and 
$\langle I_{6s}\rangle$ are shown in Fig.~\ref{angularpnl22}. 
The plots are arranged in the same order as discussed previously. For the angular coefficients 
$\langle I_{1s}\rangle$ and 
$\langle I_{2s}\rangle$, all NP scenarios produce relatively similar behaviors in the low-$q^{2}$ range, where the SII scenario generally predicts slightly larger values compared to the other scenarios. In the high-$q^{2}$ muon range, visible deviations among the scenarios emerge near the resonance ranges, where the interference between the weak annihilation and PB amplitudes becomes significant. 
For the $\tau$ channel, the spread among the NP scenarios increases considerably with increasing $q^{2}$, particularly for the SI and SIV scenarios.

The coefficients 
$\langle I_{1c}\rangle$ and 
$\langle I_{2c}\rangle$ exhibit a stronger dependence on the NP WCs. 
In the low-$q^{2}$ muon range, the SII and SIV scenarios lead to comparatively larger deviations from the SM prediction, whereas the SI and SIII scenarios remain relatively closer to the SM values. 
For the $\tau$ channel, the differences among the NP scenarios become more noticeable at higher $q^2$ values. This effect is especially visible in $\langle I_{2c}\rangle$, which becomes more negative in the SI and SIV scenarios compared with the SM prediction.
The observables 
$\langle I_{3}\rangle$ and 
$\langle I_{4}\rangle$ also show visible sensitivity to the NP scenarios. 
The coefficient 
$\langle I_{3}\rangle$ exhibits the largest deviations in the SI and SIV scenarios, particularly when the $\tau$ is considered as a final state lepton. 
On the other hand, 
$\langle I_{4}\rangle$ in the low-$q^{2}$ muon range increases steadily with $q^{2}$ and shows moderate deviations in different NP scenarios. In the high-$q^{2}$ muon range NP scenarios are relatively distinguishable, whereas for the $\tau$ lepton case, all NP scenarios are distinct for $q^{2}\ge 15$ $\text{GeV}^2$, and can be clearly discriminated from each other.

\begin{figure}[H]
\centering
\includegraphics[width=2.8in,height=1.2in]{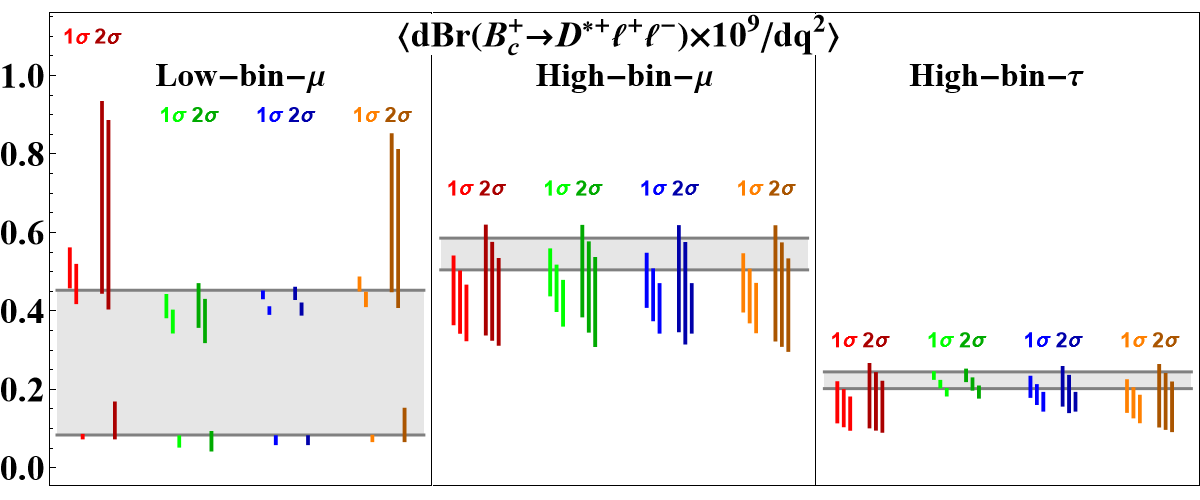}
\includegraphics[width=2.8in,height=1.2in]{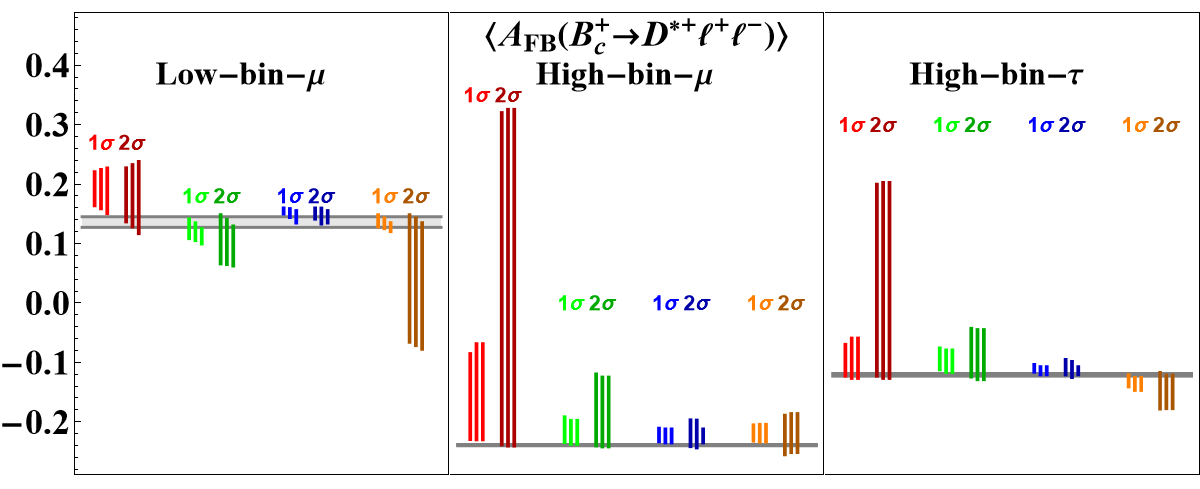}

\includegraphics[width=2.8in,height=1.2in]{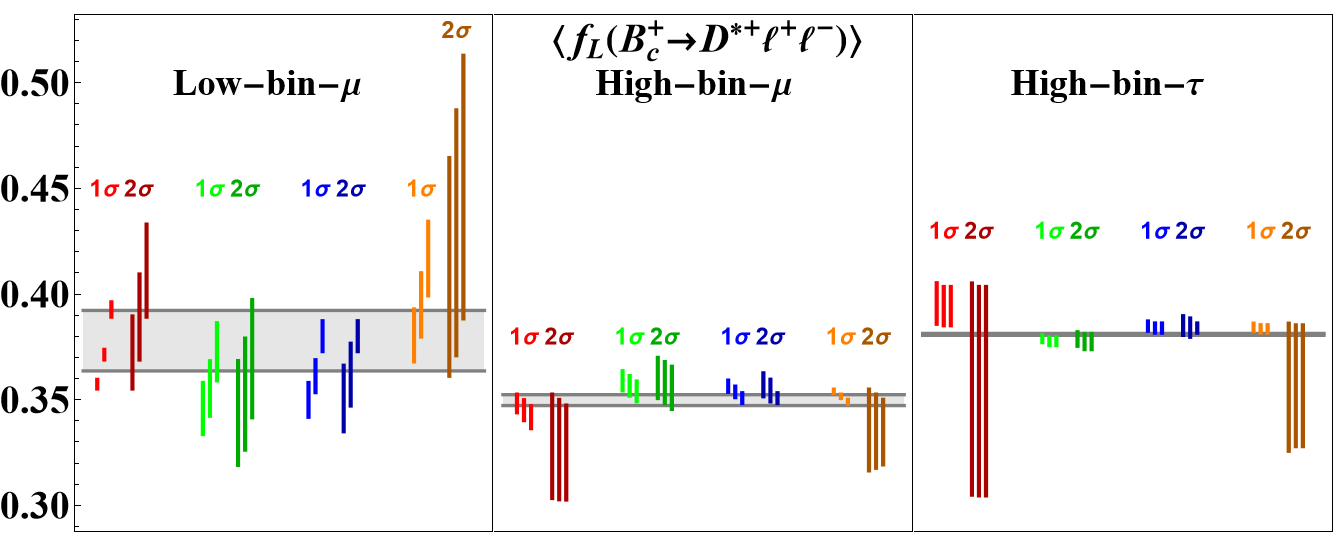}
\includegraphics[width=2.8in,height=1.2in]{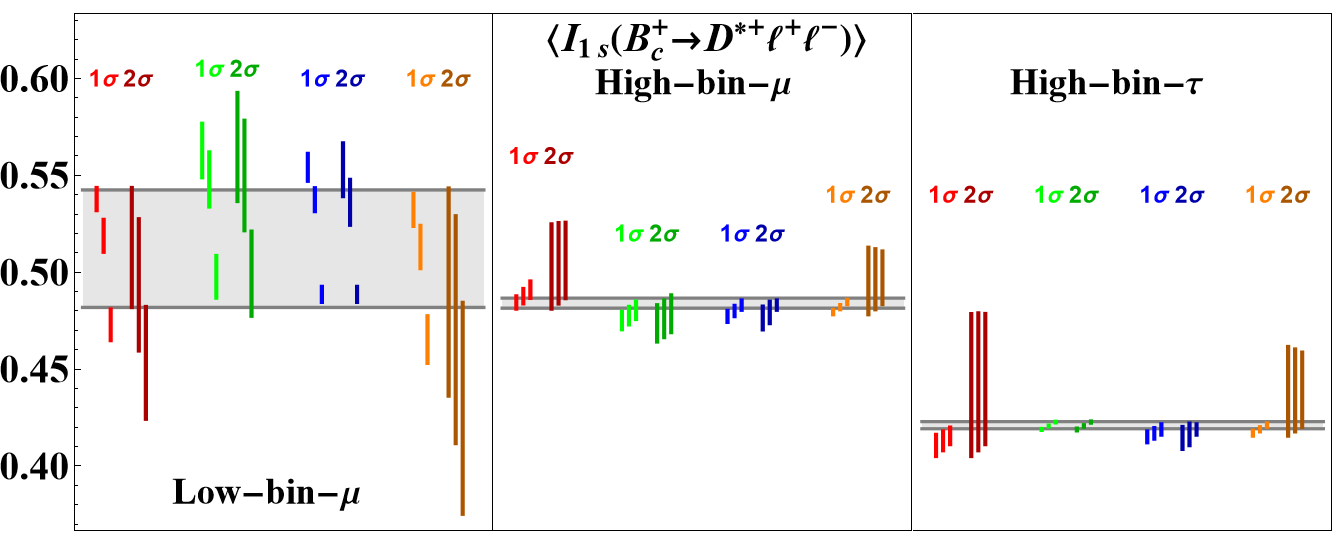}

\includegraphics[width=2.8in,height=1.2in]{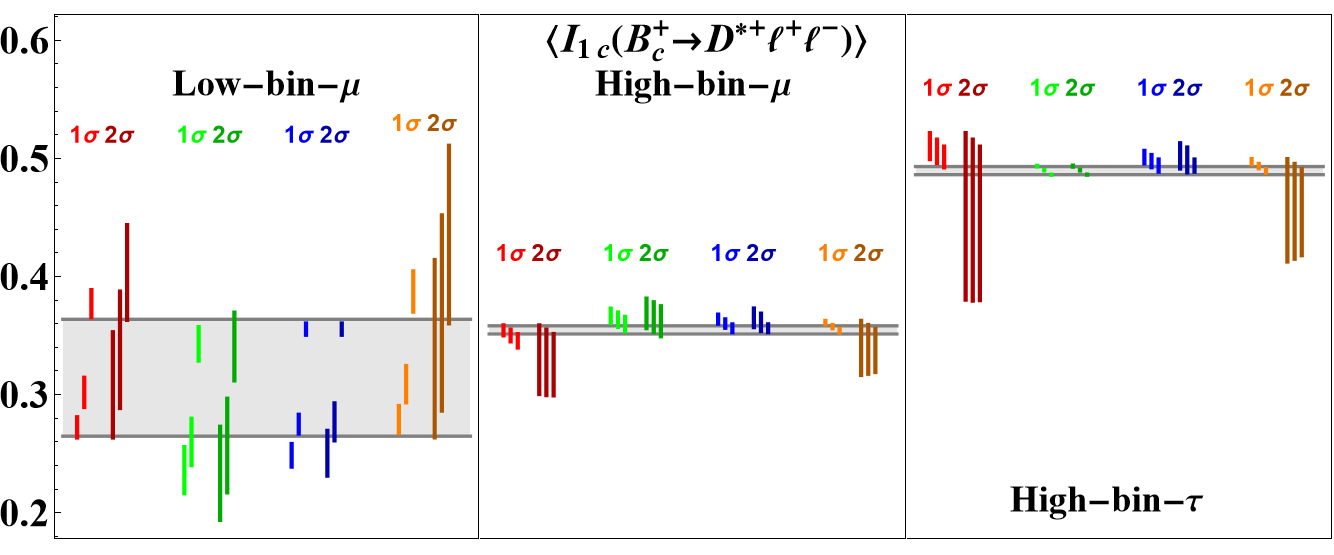}
\includegraphics[width=2.8in,height=1.2in]{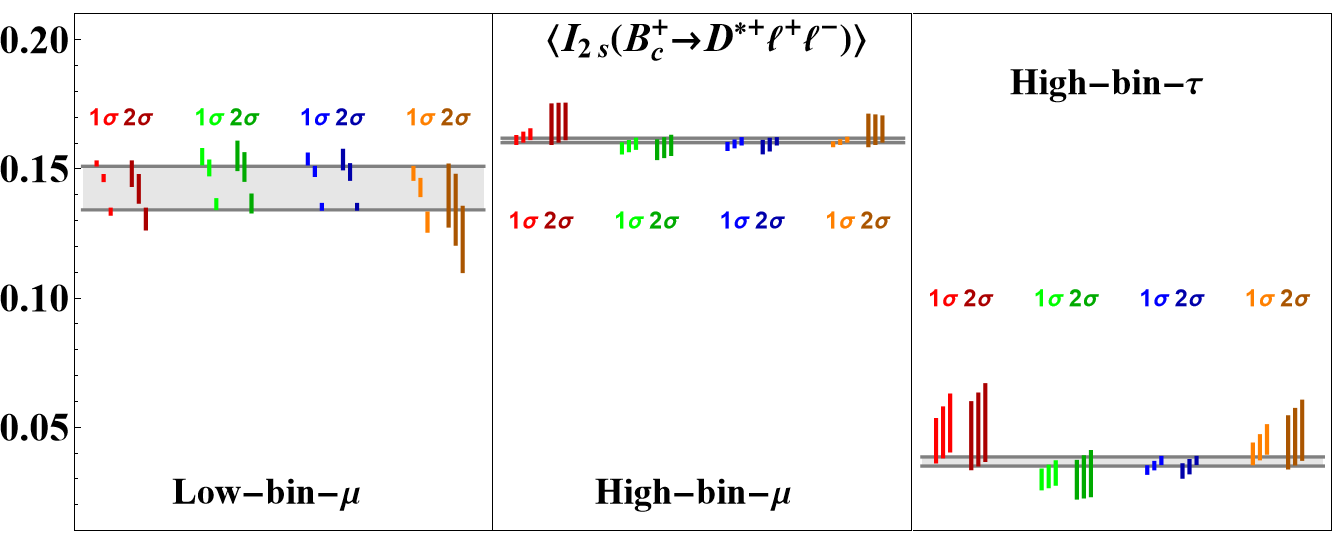}

\includegraphics[width=2.8in,height=1.2in]{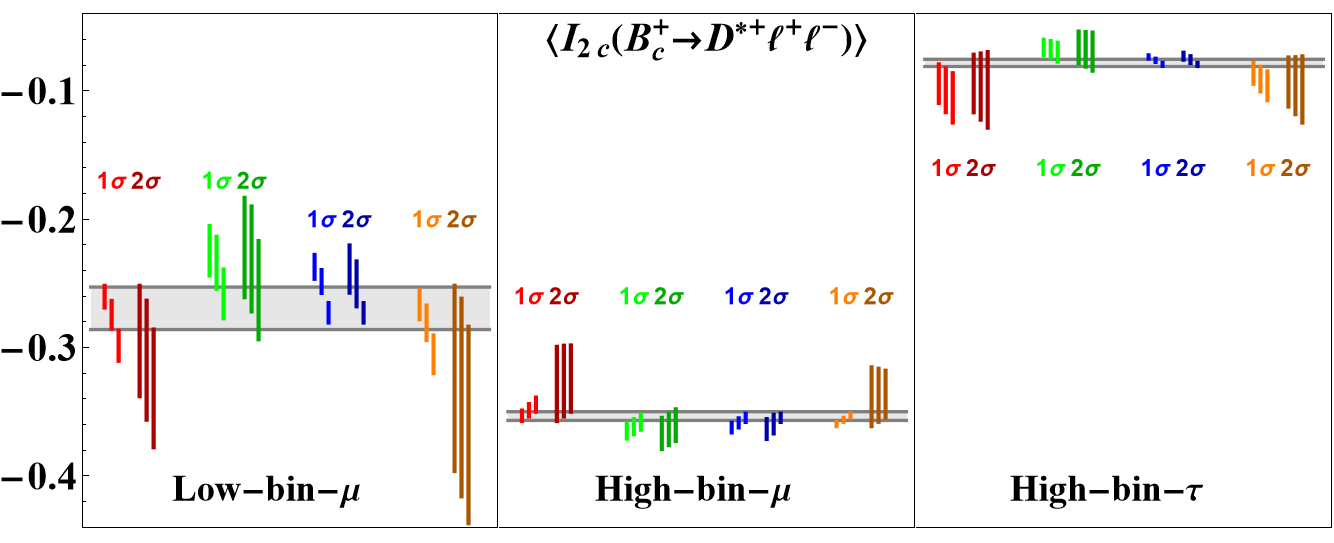}
\includegraphics[width=2.8in,height=1.2in]{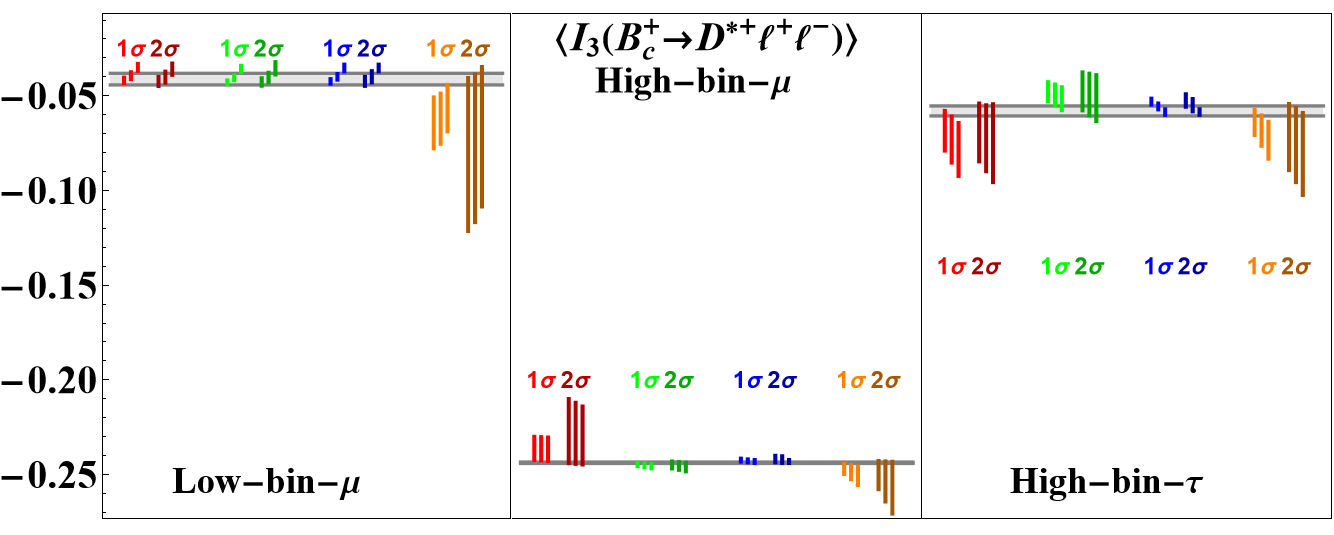}

\includegraphics[width=2.8in,height=1.2in]{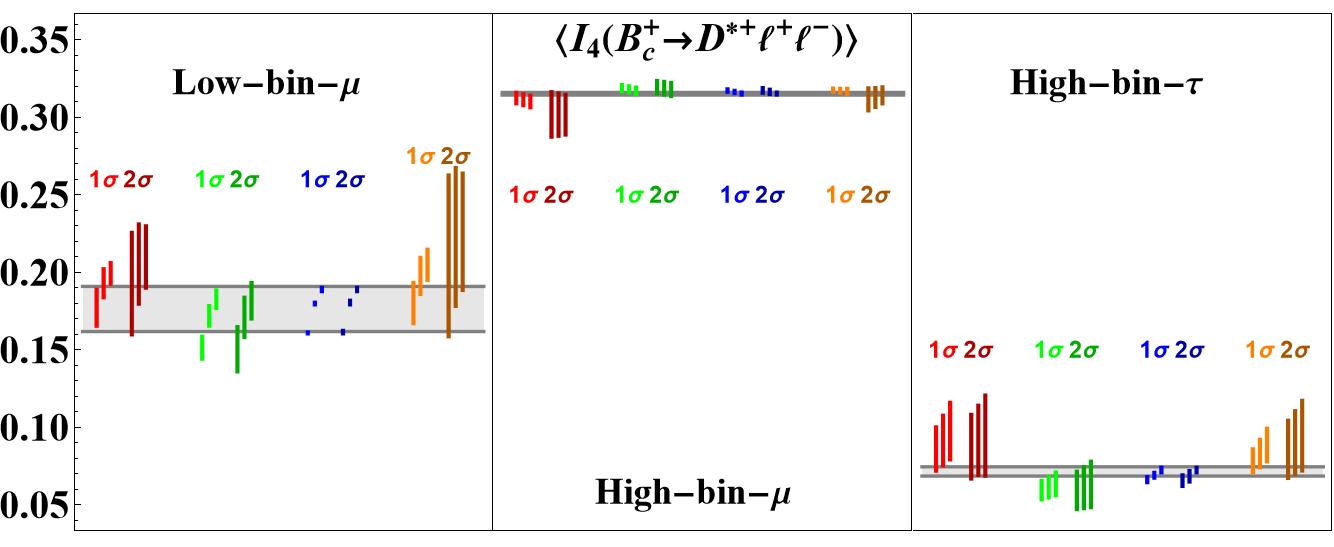}
\includegraphics[width=2.8in,height=1.2in]{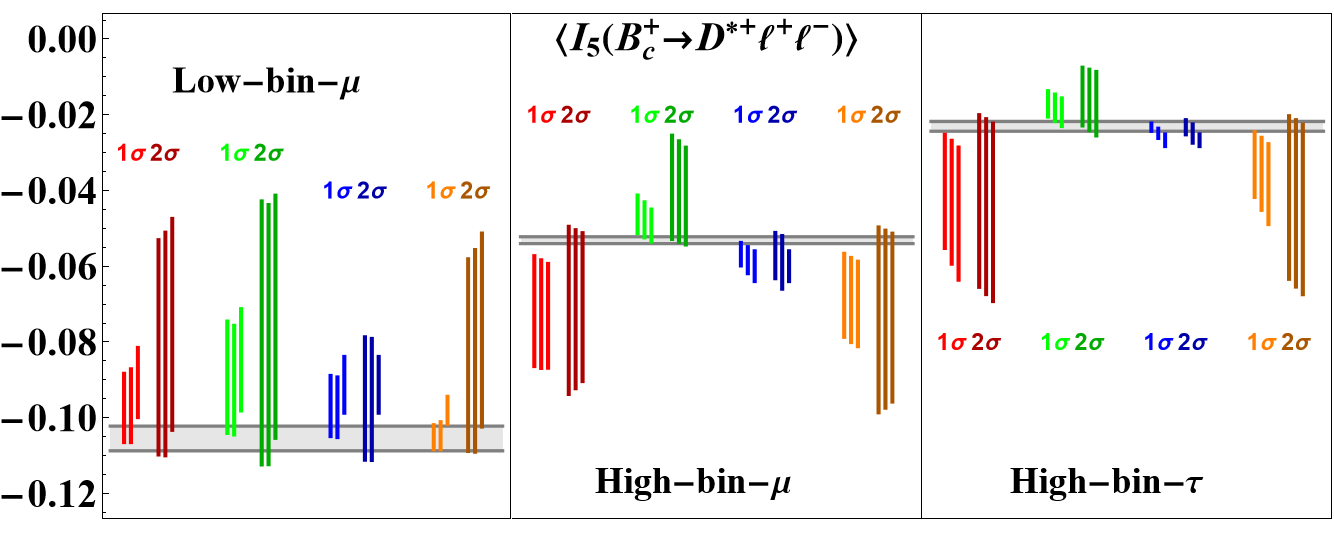}

\includegraphics[width=2.8in,height=1.2in]{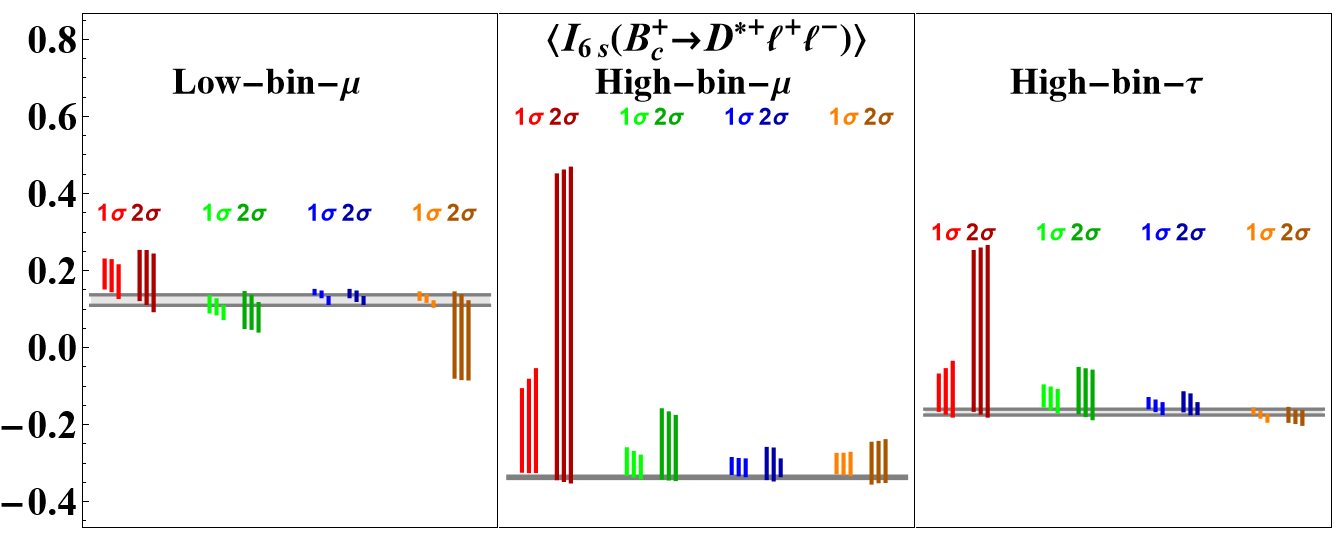}

\caption{Binned predictions for the observables in the 1D NP scenarios SI--SIV. For each observable, the results are shown in the low-$q^{2}$ muon bin, high-$q^{2}$ muon bin, and high-$q^{2}$ tau bin. The colored bars correspond to the allowed $1\sigma$ and $2\sigma$ ranges of the NP WCs, including the form factor uncertainties, while the horizontal gray bands represent the SM predictions.}
\label{1Dbarplots}
\end{figure}
For the angular coefficient $\langle I_{5}\rangle$, it has been observed that for the case of $\mu$ lepton in low $q^{2}$ range, the NP scenarios can be discriminated from the SM predictions. However, some scenarios, such as SII, SIII, and SIV show overlapping with each other for the kinematic range $q^{2}\approx 2-7$ $\text {GeV}^{2}$. In contrast, the scenario SI only overlaps with others for $q^{2}\le 2.5$ $\text{GeV}^{2}$ and $q^{2}\ge 6.5$ $\text{GeV}^{2}$. In the high-$q^2$ $\mu$ range, the angular coefficient $\langle I_{5}\rangle$ is strongly affected by the charmonium resonances, while the NP bands show visible separation away from these ranges. For the $\tau$ as a final state lepton, the hierarchy among the scenarios becomes clearer: SI gives the largest negative magnitude, SII gives the smallest, and SIII and SIV lie in between.
The $\langle I_{6s}\rangle$ shows strong sensitivity to the NP WCs. This behavior is expected because, in the convention used in Eq.~\eqref{AFB1}, $I_{6s}$ enters directly in the numerator of the forward-backward asymmetry. Thus, $\langle I_{6s}\rangle$ follows the main trends seen in $A_{FB}$ and provides complementary information on the helicity structure of the decay. In the high-$q^2$ range, its behavior is also shaped by the resonance contribution and by the interference between WA and PB amplitudes.

Overall, the angular coefficients exhibit substantial sensitivity to the NP WCs and provide complementary information to the branching fraction, forward-backward asymmetry, and longitudinal helicity fraction. In particular, the observables 
$\langle I_{2c}\rangle$, 
$\langle I_{5}\rangle$, and 
$\langle I_{6s}\rangle$ appear to be especially sensitive to the interplay between the weak annihilation amplitudes and the NP-modified short-distance contributions. The effects become more pronounced in the $\tau$ channel due to the enhanced lepton-mass dependence of the helicity amplitudes.

In Fig.~\ref{1Dbarplots}, we present the binned predictions of the observables for the 1D NP scenarios through bar plots, where the complete prediction ranges associated with the NP WCs and form factor uncertainties are displayed. 
For each observable, the results are given for three different bins. The kinematic ranges for the low-$q^2$ $\mu$ bin and high-$q^2$ $\tau$ bin are the same as before,
while for high-$q^2$ $\mu$ bin, we choose the range $[14,q_{\mathrm{max}}^2]~\mathrm{GeV}^2$. In each bin, the gray horizontal band represents the SM prediction together with its theoretical uncertainty originating mainly from the error in the form factors. The colored bars correspond to the different 1D NP scenarios and display the full allowed prediction ranges obtained from the variation of the NP WCs within their corresponding $1\sigma$ (lighter shade) and $2\sigma$ (darker shade) intervals. Furthermore, each set of bars contains three individual predictions corresponding to the allowed NP WCs ranges with lower, central, and upper fixed values of the form factors, thereby incorporating both the hadronic uncertainties and the allowed NP parameter space simultaneously. The previous color convention for the NP scenarios is maintained throughout the analysis. Therefore, the bar plots provide a compact representation of the possible deviations that each NP scenario can generate, with its $1\sigma$ and $2\sigma$ ranges, for a given observable within the selected kinematic bins.

\begin{figure}[H]
\centering
\includegraphics[width=2in,height=1.24in]{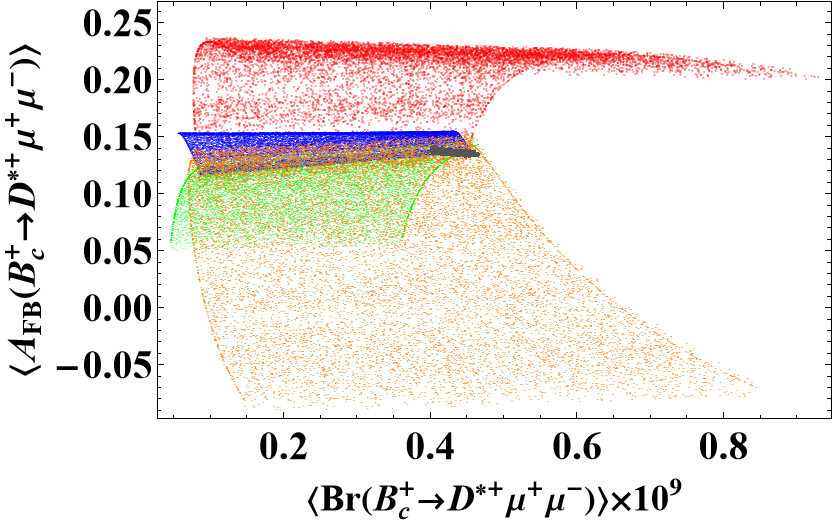}
\includegraphics[width=2in,height=1.24in]{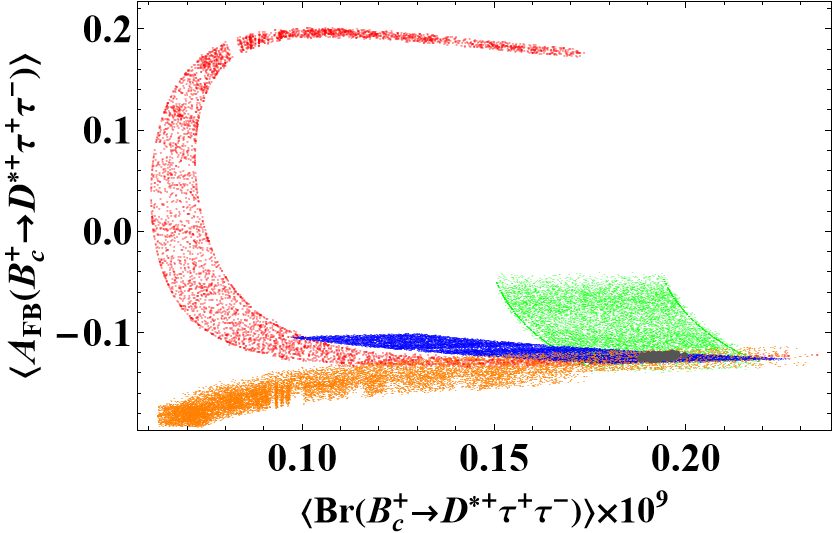}

\includegraphics[width=2in,height=1.24in]{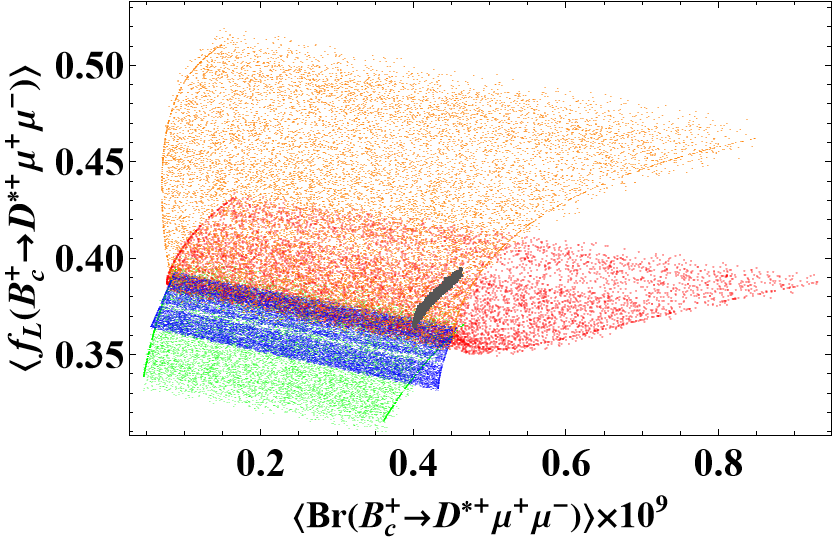}
\includegraphics[width=2in,height=1.24in]{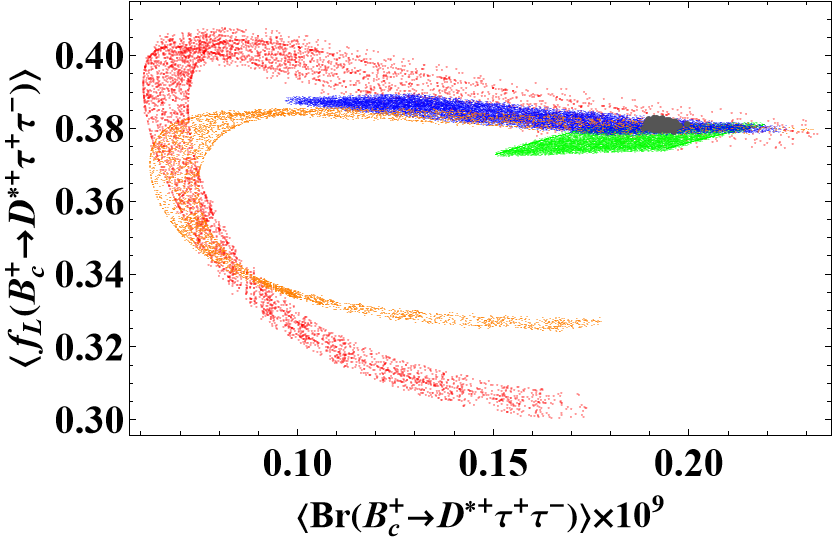}
\caption{Correlations between the integrated branching fraction and the observables $\langle A_{\rm FB}\rangle$ and $\langle f_{L}\rangle$ in the 1D NP scenarios SI--SIV. The left panels correspond to the low-$q^{2}$ bin for muon channel, while the right panels show the high-$q^{2}$ bin for tau channel. The spread of points includes the form factor uncertainties and the WC uncertainties up to $2\sigma$ range.}
\label{Fig1Dcorr}
\end{figure}

A general feature observed in these plots is that the low-$q^{2}$ range possesses comparatively larger SM uncertainty bands due to the enhanced sensitivity to hadronic effects and weak annihilation contributions. In contrast, the high-$q^{2}$ ranges exhibit smaller SM uncertainties but comparatively larger relative deviations induced by the NP scenarios. After excluding the charmonium dominated range, the high-$q^2$ $\mu$ bin $[14,q^2_{\rm max}]~\mathrm{GeV}^2$, provides a comparatively cleaner range for studying NP effects. However, the full kinematic interval $8 \leq q^2 \leq 18.18~\mathrm{GeV}^2$ includes resonance structures, and therefore it is also affected by LD resonance contributions and WA interference. Among the considered scenarios, SI and SIV generally produce the largest deviations from the SM predictions for most observables. 
In particular, sizable shifts are visible in the differential branching ratio, the FB asymmetry, and the angular coefficients. 
The SII and SIII scenarios, on the other hand, tend to remain comparatively closer to the SM predictions in several observables, although noticeable deviations are still present in specific bins.

An important phenomenological implication of these results is that some observables show partially separated prediction ranges among the different NP scenarios, especially in the high-$q^{2}$ $\tau$ range. Although sizable overlap remains once the WCs ranges and form-factor uncertainties are included, this behavior suggests that future precision measurements of these observables at LHCb or other high-luminosity flavor experiments could help discriminate among the different NP scenarios and constrain the underlying WCs associated with the $b\to d\,\ell^{+}\ell^{-}$ transition. The combined possible minimum and maximum predictions obtained from any of the full $1\sigma$ and $2\sigma$ ranges of the NP WCs are summarized in Table~\ref{tab:1Dnums}.

The correlation plots shown in Fig.~\ref{Fig1Dcorr}, illustrate the interplay between the integrated branching fraction and the observables $\langle A_{\rm FB}\rangle$ and $\langle f_{L}\rangle$ for different 1D NP scenarios in the low-bin for muon and in the high-bin for tau channel. In the muon channel, the correlation between $\langle A_{\rm FB}\rangle$ and $\text{Br}(B_c^+\to D^{\ast+}\mu^{+}\mu^{-})$ shows a clear separation among the NP scenarios. The SI scenario predicts the largest positive values of $\langle A_{\rm FB}\rangle$ together with an enhanced branching fraction, while SII produces comparatively smaller asymmetry values. The SIII and SIV scenarios remain closer to the SM-like range with noticeable overlap. For the $\tau$ channel, the SIV scenario generates the largest negative values which increases with increasing branching ratio, whereas SI exhibits the widest spread in the allowed range mostly lying in positive range. In the SII scenario, the allowed points show an anti-correlation between the integrated branching fraction and $\langle A_{FB}\rangle$, such that larger branching-fraction values are associated with smaller values of $\langle A_{FB}\rangle$.

The lower panels display the correlations involving the longitudinal helicity fraction $\langle f_{L}\rangle$. In the muon mode, a positive correlation is observed for NP scenarios, where larger branching fractions correspond to larger values of $\langle f_{L}\rangle$. The SI and SIV scenarios allow comparatively larger longitudinal polarization. In the $\tau$ channel, the allowed ranges for SII and SIII become more compressed because of the restricted phase space. The observable $\langle f_{L}\rangle$ remains relatively large throughout the kinematic range in case of SI and SIV scenarios, indicating enhanced longitudinal polarization contributions in the heavy-lepton case. In short, the correlation plots demonstrate that the combined measurements of the branching fraction, $\langle A_{\rm FB}\rangle$, and $\langle f_{L}\rangle$ can provide useful discrimination among different NP scenarios. The $\tau$ channel is found to be particularly sensitive to NP effects.

\subsection{Probing NP signatures using 2D NP Scenarios}\label{results-2D}

\begin{figure}[H]
\centering
\includegraphics[width=2in,height=1.24in]{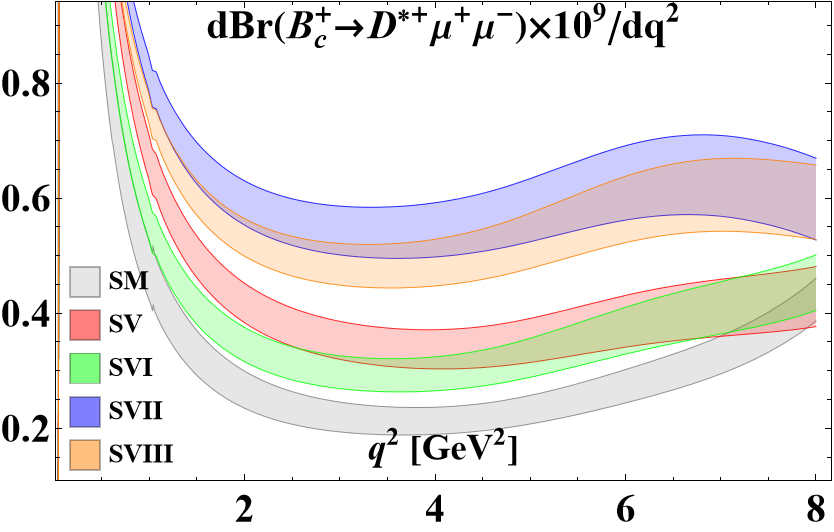}
\includegraphics[width=2in,height=1.24in]{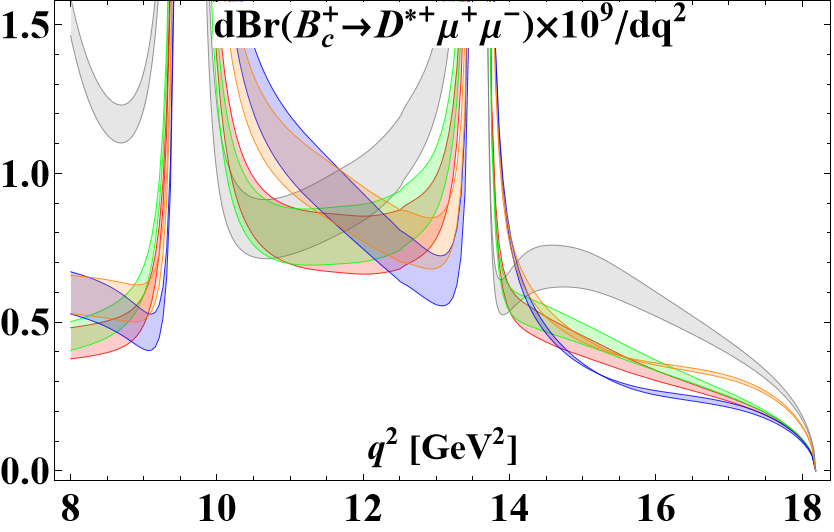}
\raisebox{0.0in}{\includegraphics[width=2in,height=1.24in]{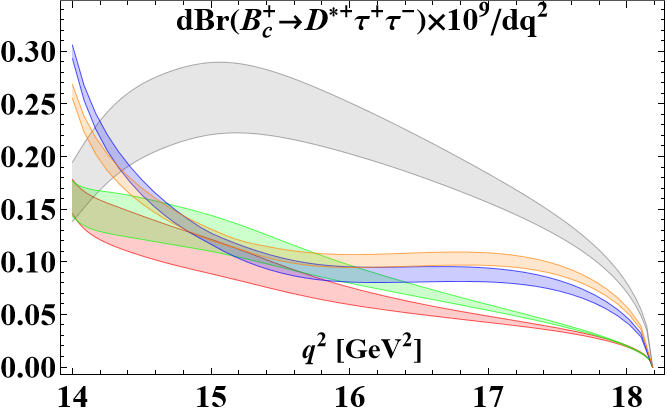}}
\includegraphics[width=2in,height=1.24in]{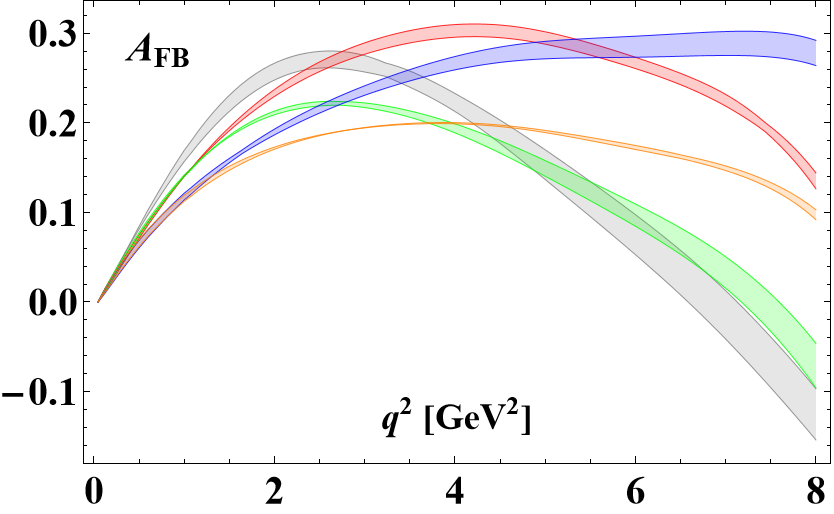}
\includegraphics[width=2in,height=1.24in]{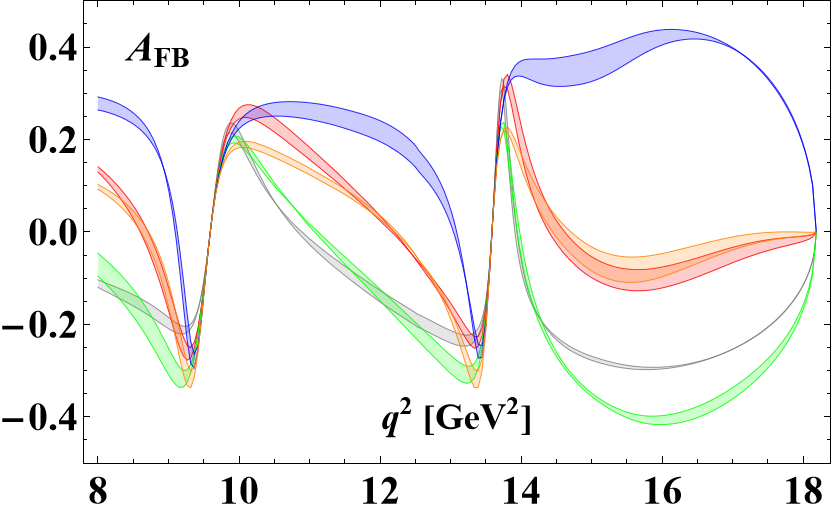}
\raisebox{0.0in}{\includegraphics[width=2in,height=1.24in]{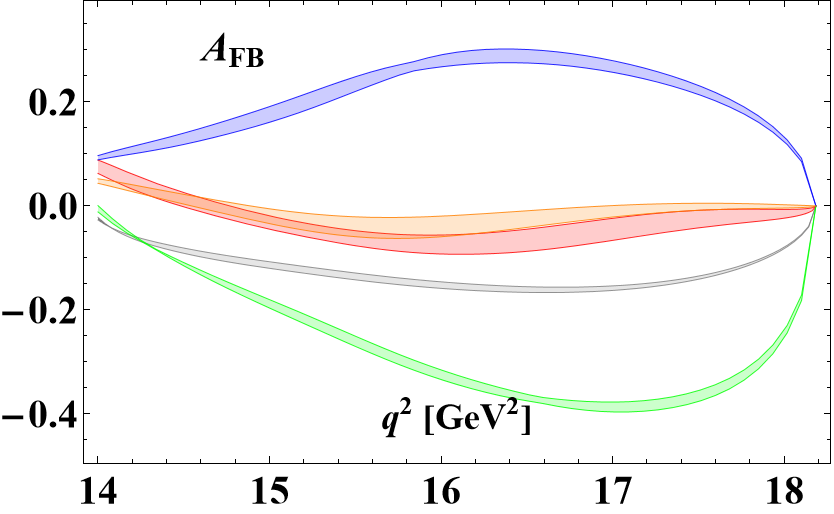}}
\includegraphics[width=2in,height=1.24in]{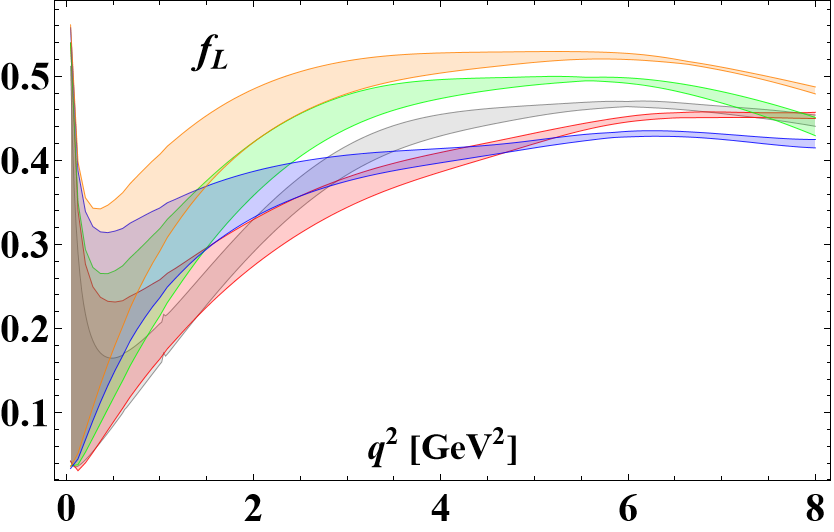}
\includegraphics[width=2in,height=1.24in]{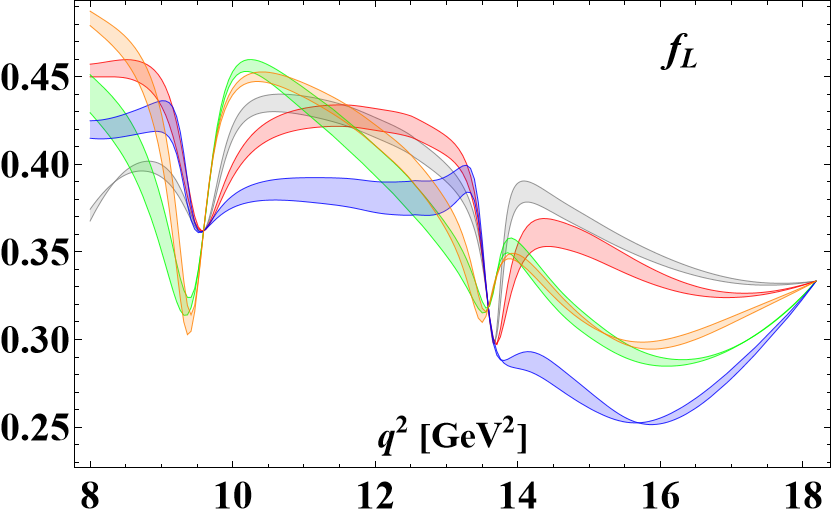}
\raisebox{0.0in}{\includegraphics[width=2in,height=1.24in]{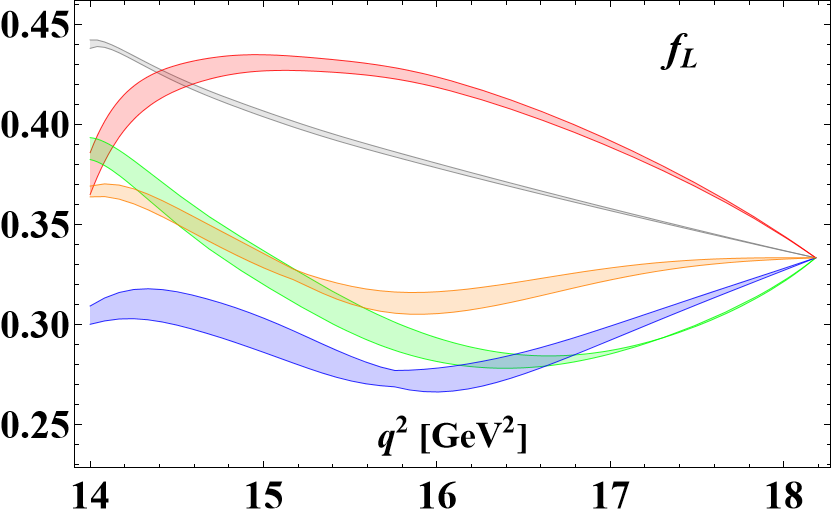}}
\caption{Predictions for the differential branching fraction
$d\text{Br}/dq^{2}$, the FB asymmetry $A_{\rm FB}$, and the longitudinal helicity fraction $f_{L}$ of the decay
$B_c^{+}\to D^{\ast+}\ell^{+}\ell^{-}$ in the SM and the 2D NP scenarios SV--SVIII. The first and second columns display these observables (from top to bottom) as functions of the dilepton invariant mass squared $q^{2}$ for the low- and high-$q^{2}$ ranges of the muon channel, respectively, whereas the third column shows the corresponding predictions for the high-$q^{2}$ range of the tau channel.}
\label{fig:2D_BrAFBfL}
\end{figure}

We now discuss the effects of the 2D NP scenarios SV-SVIII on the observables associated with the decay $B_{c}^+\to D^{\ast+}\ell^{+}\ell^{-}$. 
The corresponding predictions for the differential branching fraction $d\text{Br}/dq^{2}$, the forward-backward asymmetry $A_{\rm FB}$, and the longitudinal helicity fraction $f_{L}$, are shown in Fig.~\ref{fig:2D_BrAFBfL}. To distinguish different NP scenarios, the color bands are plotted using the left side $1\sigma$ value for SV--SVII and the right side $1\sigma$ value for SVIII.

The color coding of the 2D scenarios is given as:

\begin{itemize}
    \item SV $(C_{7}^{NP},\,C_{9}^{NP})$: red,
    \item SVI $(C_{9}^{NP},\,C_{9}^{\prime NP})$: green,
    \item SVII $(C_{9}^{NP},\,C_{10}^{\prime NP})$: blue,
    \item SVIII $(C_{9}^{NP}=-C_{9}^{\prime NP},\,C_{10}^{NP}=+C_{10}^{\prime NP})$: orange.
\end{itemize}

The upper row of Fig.~\ref{fig:2D_BrAFBfL} presents the differential branching fraction. In the low-$q^{2}$ muon range, all 2D NP scenarios tend to enhance the branching fraction relative to the SM prediction, with the SVII scenario producing the largest enhancement over most of the kinematic range. 
The SVI scenario predicts comparatively closer values to the SM case, while the SV and SVIII scenarios lie between these two cases. In the high-$q^{2}$ muon range, the branching fraction exhibits a pronounced sensitivity to the NP WCs near the resonance ranges. 
The SVII scenario again shows the largest enhancement, whereas the SVI scenario remains comparatively suppressed. For the $\tau$ channel, the hierarchy among the NP scenarios changes significantly. The SVII and SVIII scenarios predict comparatively smaller branching fractions than the SM band at higher $q^{2}$, while the SV scenario gives the smallest possible values over most of the allowed range.

The second row of Fig.~\ref{fig:2D_BrAFBfL} displays the forward-backward asymmetry $A_{\rm FB}$. In the low-$q^{2}$ muon range, the SVII scenario produces the largest positive asymmetry, while the SVI scenario predicts a rapid decrease toward negative values near the upper end of the low-$q^{2}$ range. The SV and SVIII scenarios also remain distinct and do not show zero crossing of the $A_{\rm FB}$ in this range. In the high-$q^{2}$ muon range, all scenarios exhibit sizable distortions around the resonance ranges due to the interplay between weak annihilation and PB amplitudes. The SVII scenario produces the largest positive enhancement, whereas the SVI scenario generates large negative deviations at higher $q^{2}$ values. For the $\tau$ mode, the differences among the NP scenarios become particularly pronounced. 
The SVII scenario predicts positive values of $A_{\rm FB}$ over most of the kinematic range, while the SVI scenario develops large negative values. The SV and SVIII scenarios are distinguishable from the SM band but show similar behavior to each other.

The third row of Fig.~\ref{fig:2D_BrAFBfL} shows the longitudinal helicity fraction $f_{L}$. 
In the low-$q^{2}$ muon range, all scenarios predict an increasing behavior with $q^{2}$, where SVI and SVIII scenarios produce larger values. The spread among the scenarios remains moderate in comparison with the branching fraction and forward-backward asymmetry. In the high-$q^{2}$ muon range, visible deviations appear near the resonance ranges, with the SVII scenario predicting comparatively smaller values of $f_{L}$ at larger $q^{2}$. For the $\tau$ channel, the SV scenario gives the largest longitudinal polarization fraction, whereas the SVII scenario predicts the smallest values over most of the allowed kinematic range. SVI and SVIII also predict distinct lower values of $f_{L}$, in comparison to the SM band.

\begin{figure}[H]
\centering
\includegraphics[width=2in,height=1.24in]{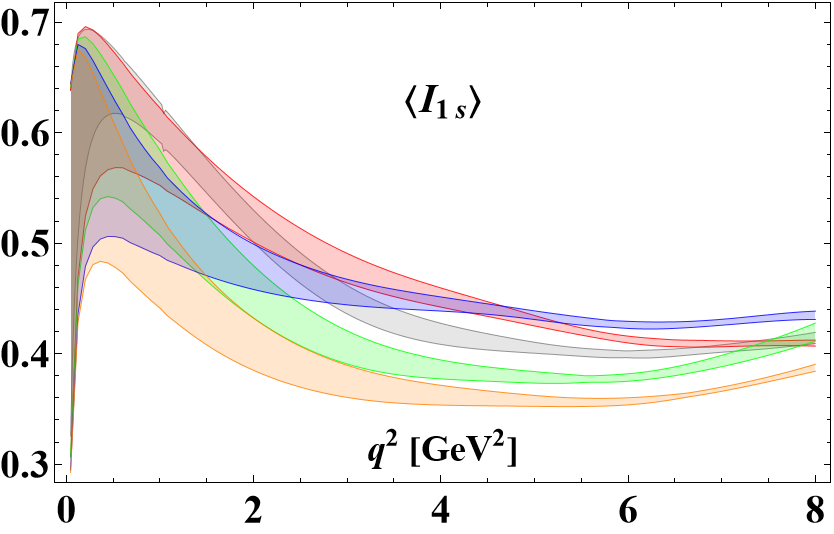}
\includegraphics[width=2in,height=1.24in]{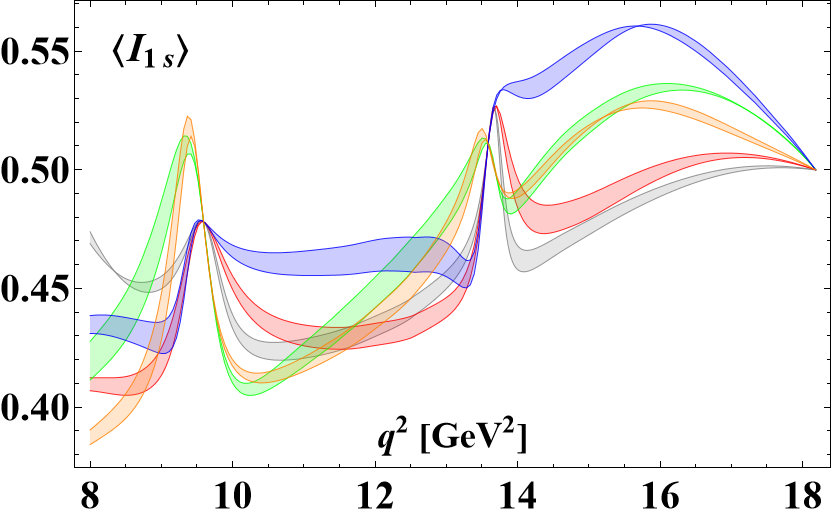}
\raisebox{0.0in}{\includegraphics[width=2in,height=1.24in]{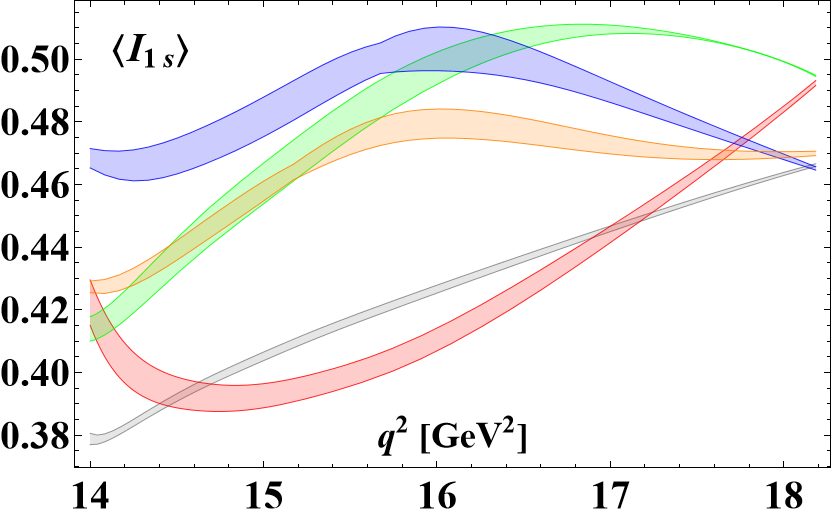}}
\\
\includegraphics[width=2in,height=1.24in]{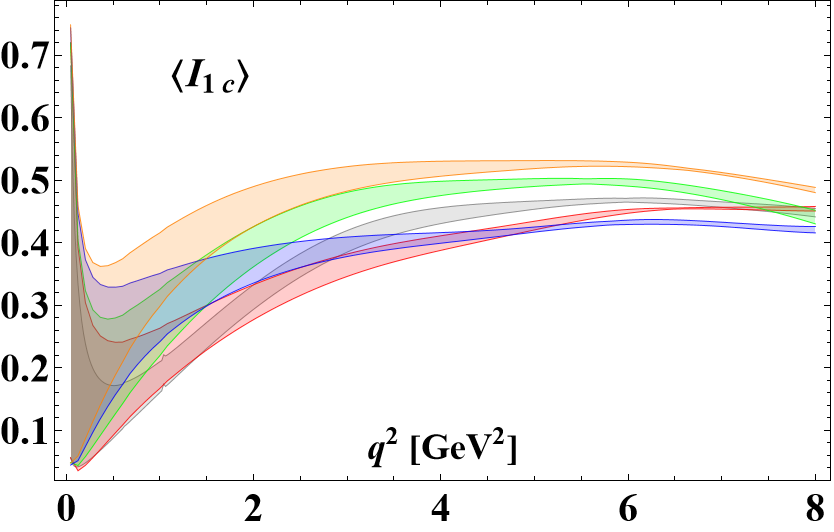}
\includegraphics[width=2in,height=1.24in]{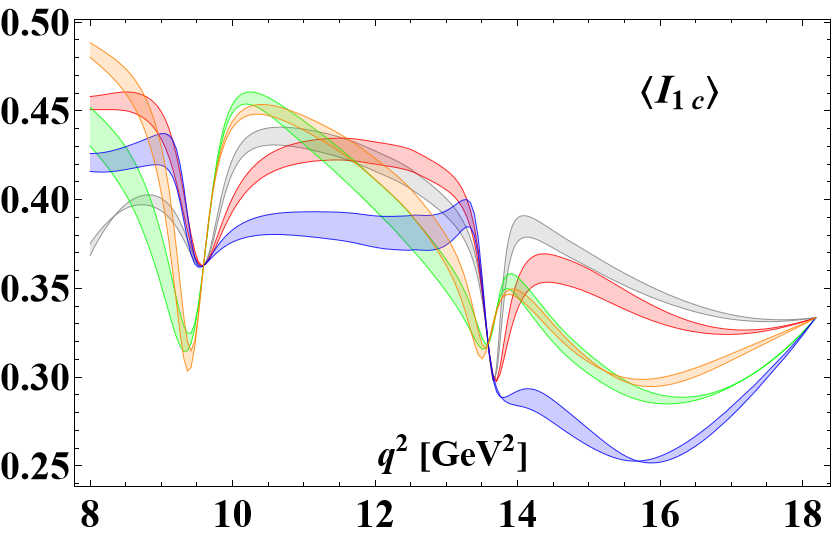}
\raisebox{0.00in}{\includegraphics[width=2in,height=1.24in]{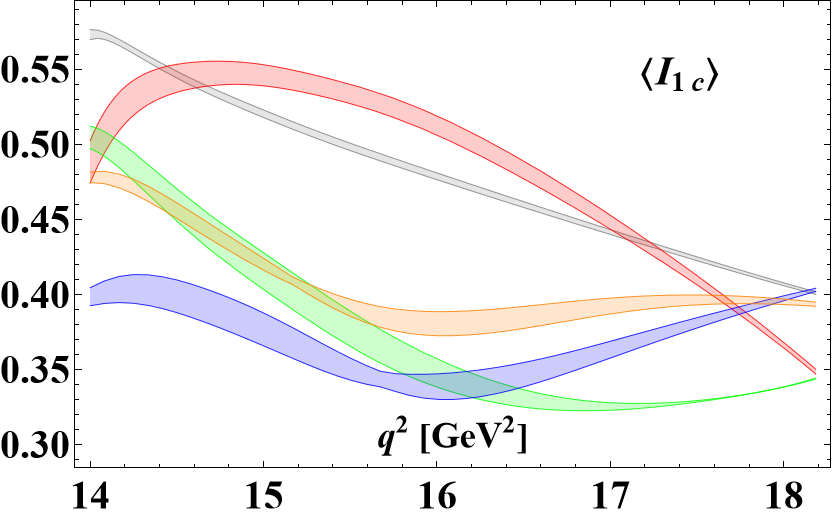}}
\\
\includegraphics[width=2in,height=1.24in]{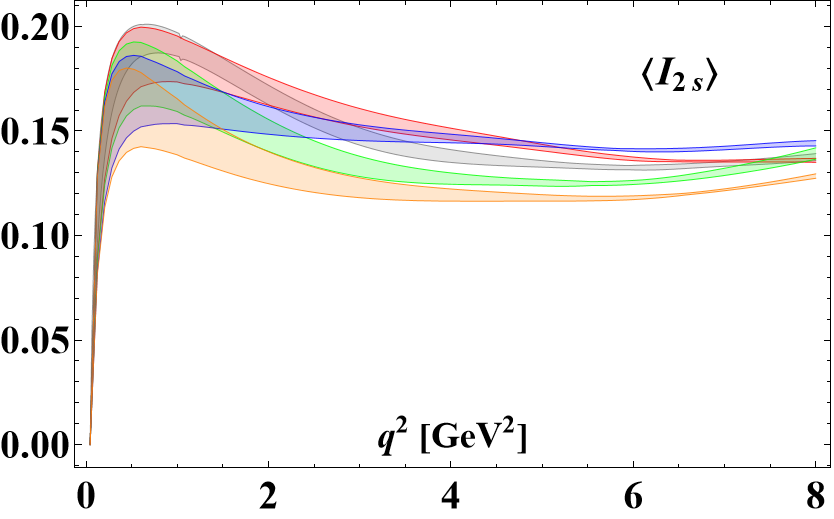}
\includegraphics[width=2in,height=1.24in]{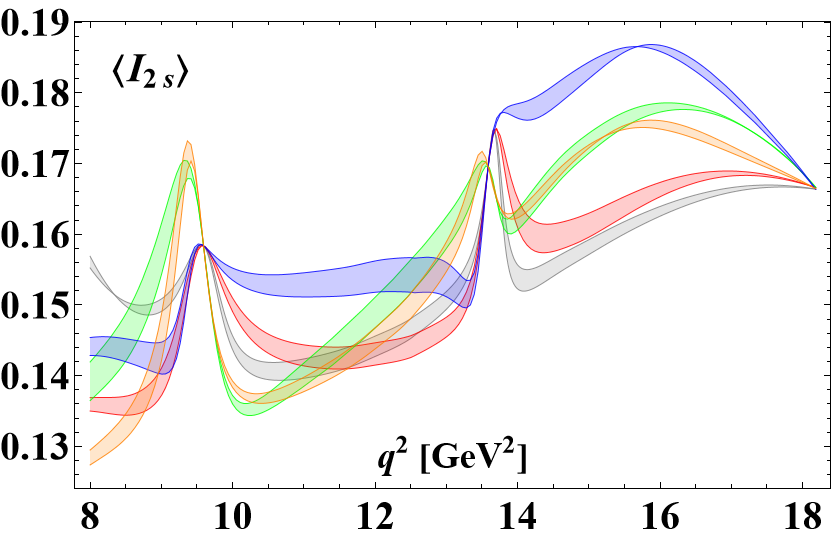}
\raisebox{0.0in}{\includegraphics[width=2in,height=1.24in]{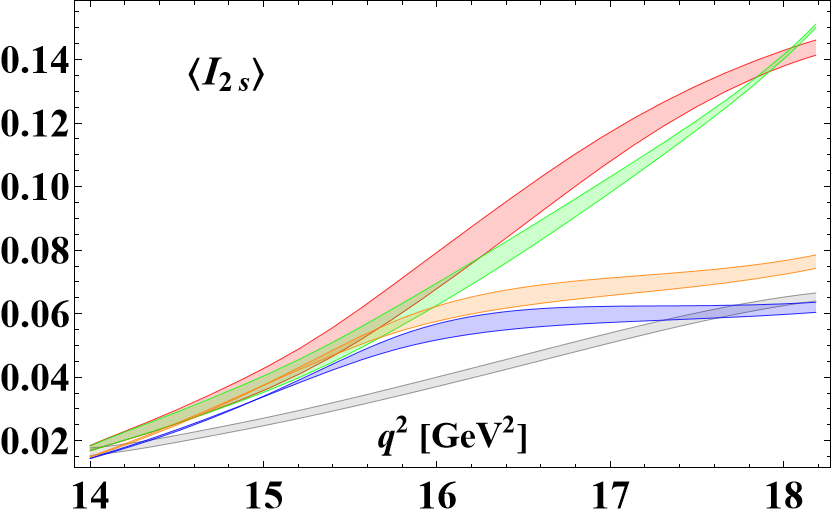}}
\\
\includegraphics[width=2in,height=1.24in]{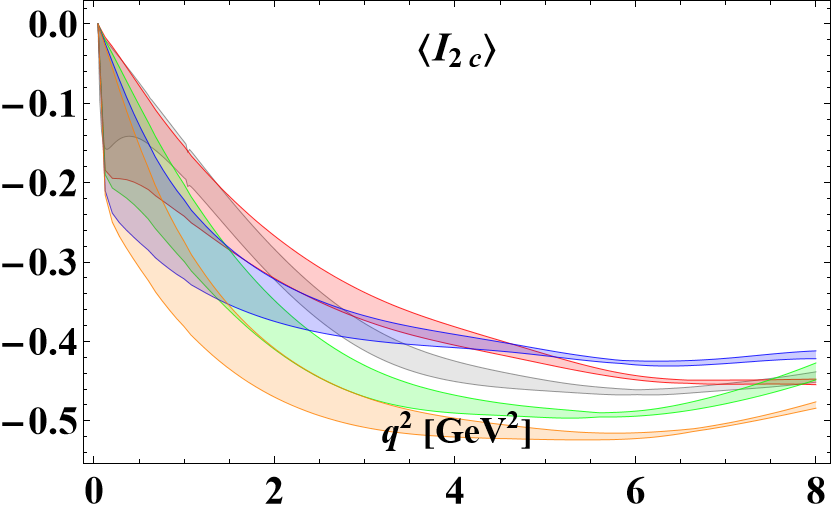}
\includegraphics[width=2in,height=1.24in]{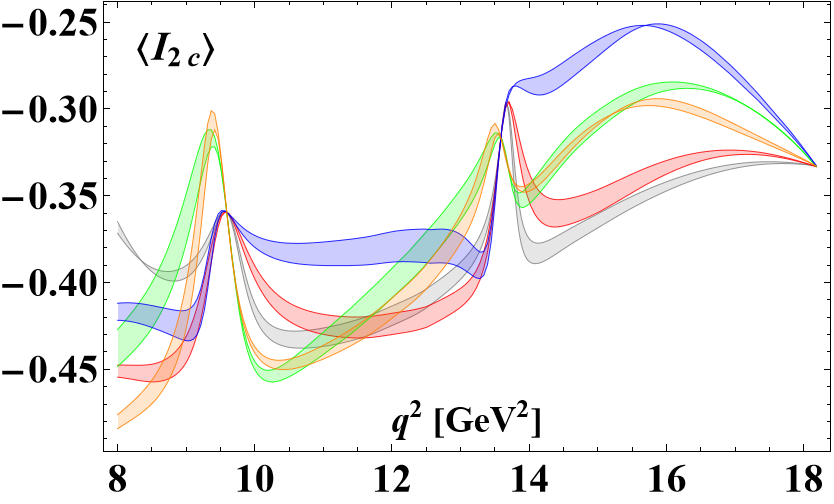}
\raisebox{0.0in}{\includegraphics[width=2in,height=1.24in]{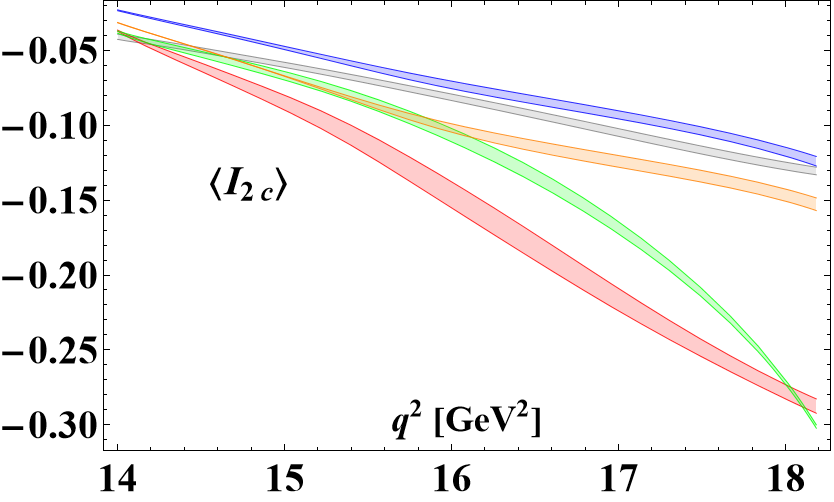}}
\caption{Predictions for the normalized angular coefficients $\langle I_{1s}\rangle$, $\langle I_{1c}\rangle$, $\langle I_{2s}\rangle$, and $\langle I_{2c}\rangle$ of the decay
$B_c^{+}\to D^{\ast+}\ell^{+}\ell^{-}$ in the SM and the 2D NP scenarios SV--SVIII. The arrangement of the panels is the same as in Fig.~\ref{fig:2D_BrAFBfL}.}
\label{fig:2D_I1stoI2c}
\end{figure}

\begin{figure}[H]
\centering
\includegraphics[width=2in,height=1.24in]{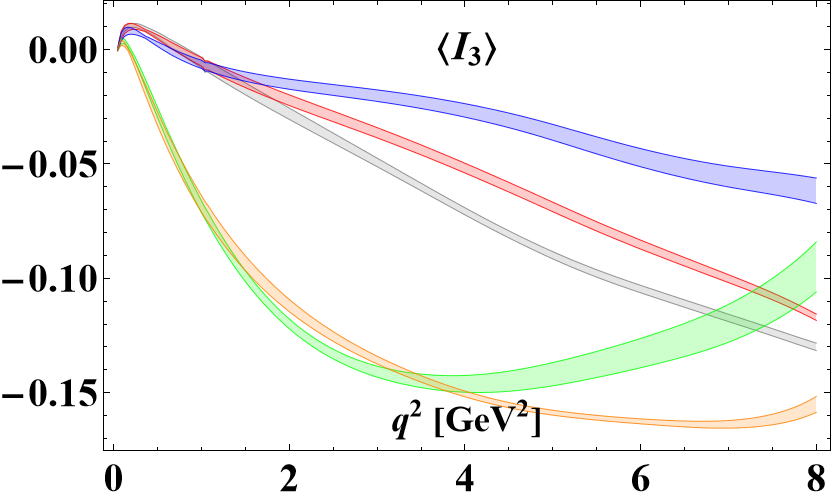}
\includegraphics[width=2in,height=1.24in]{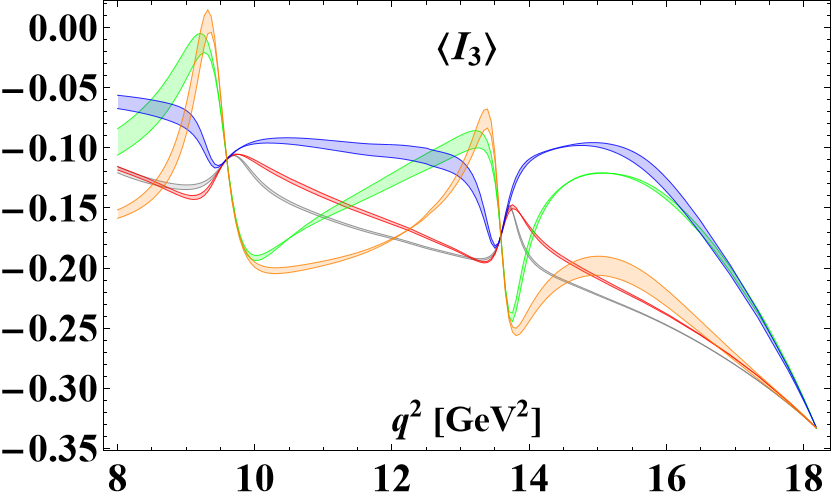}
\raisebox{0.0in}{\includegraphics[width=2in,height=1.24in]{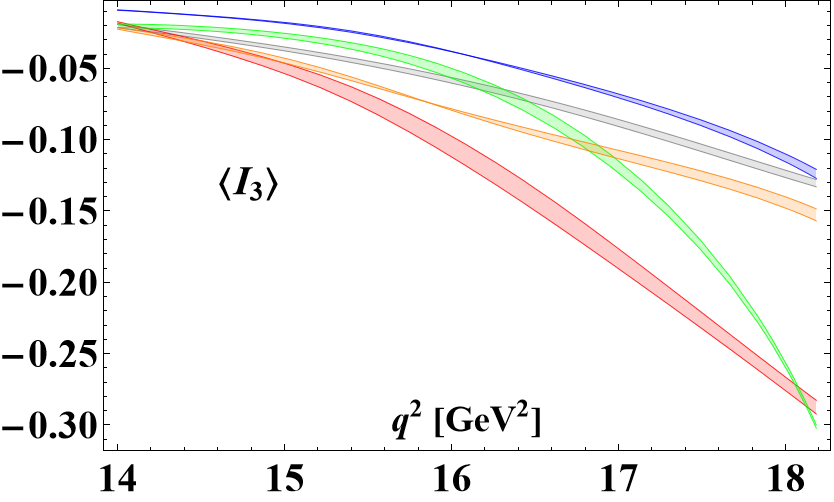}}
\\
\includegraphics[width=2in,height=1.24in]{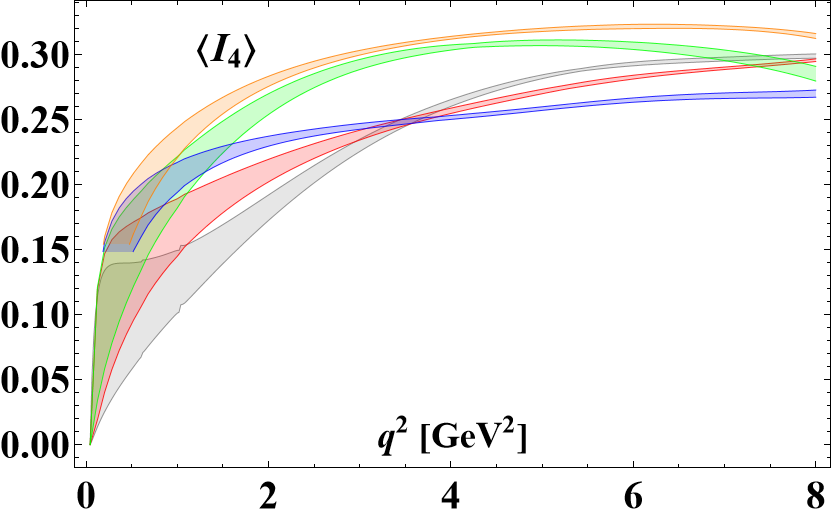}
\includegraphics[width=2in,height=1.24in]{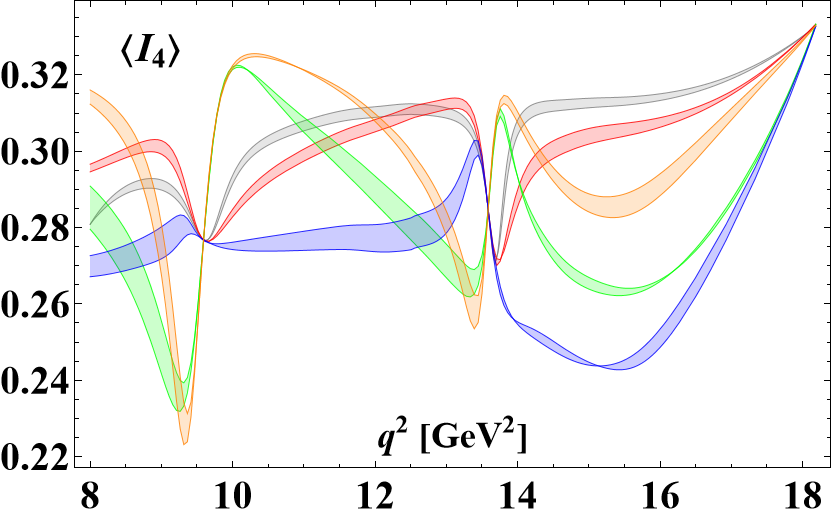}
\raisebox{0.00in}{\includegraphics[width=2in,height=1.24in]{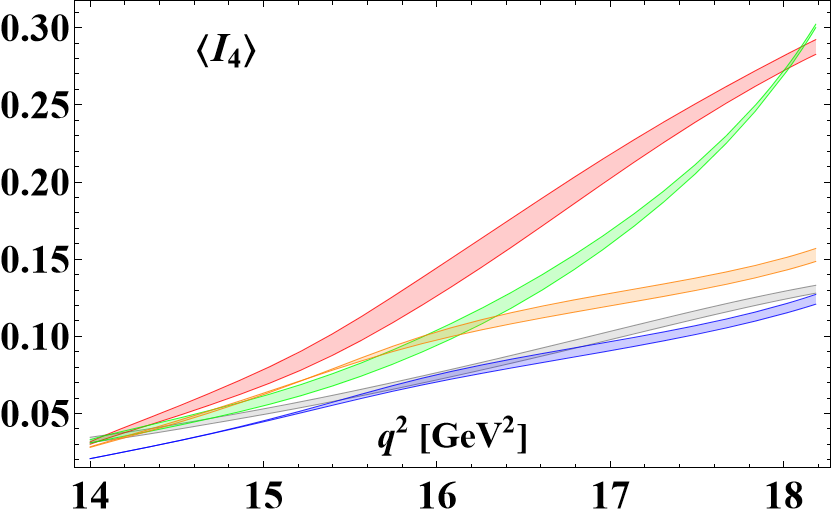}}
\\
\includegraphics[width=2in,height=1.24in]{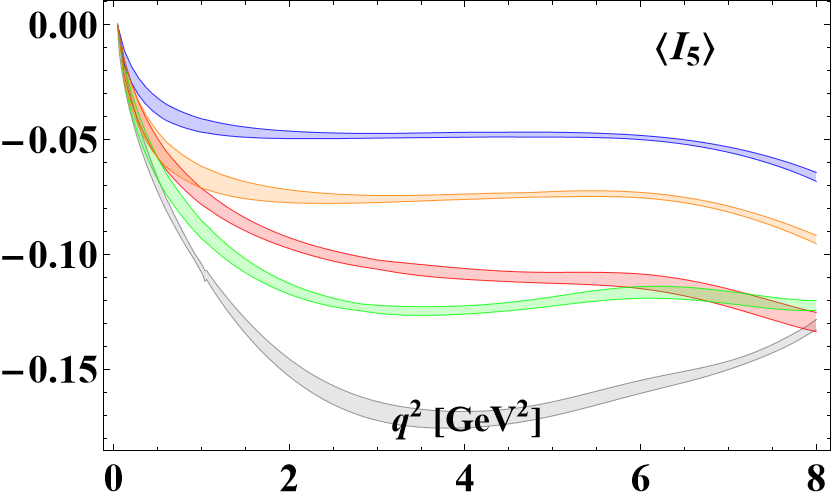}
\includegraphics[width=2in,height=1.24in]{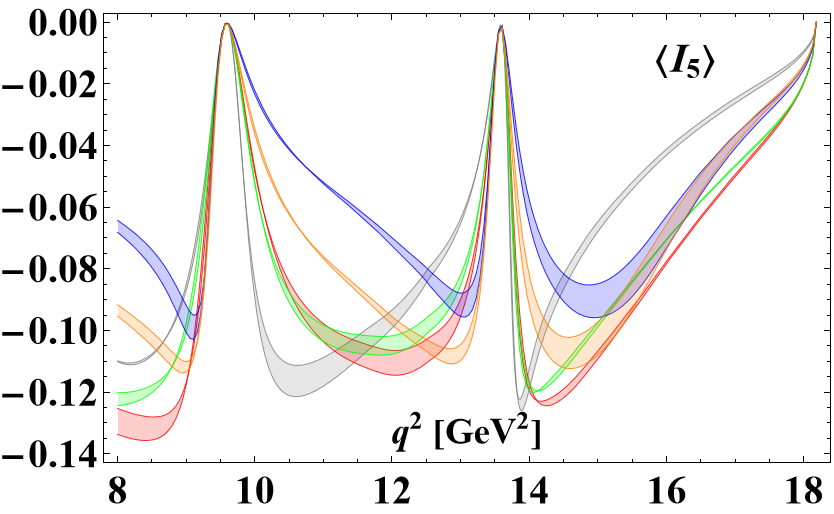}
\raisebox{0.0in}{\includegraphics[width=2in,height=1.24in]{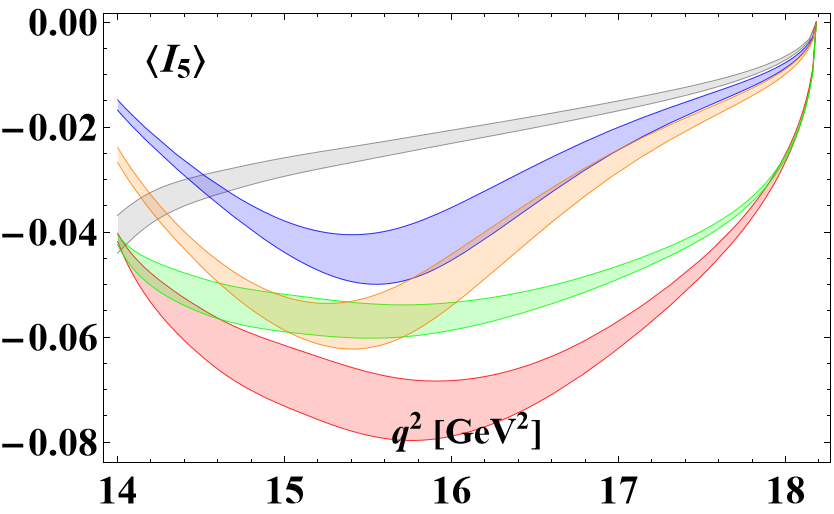}}
\\
\includegraphics[width=2in,height=1.24in]{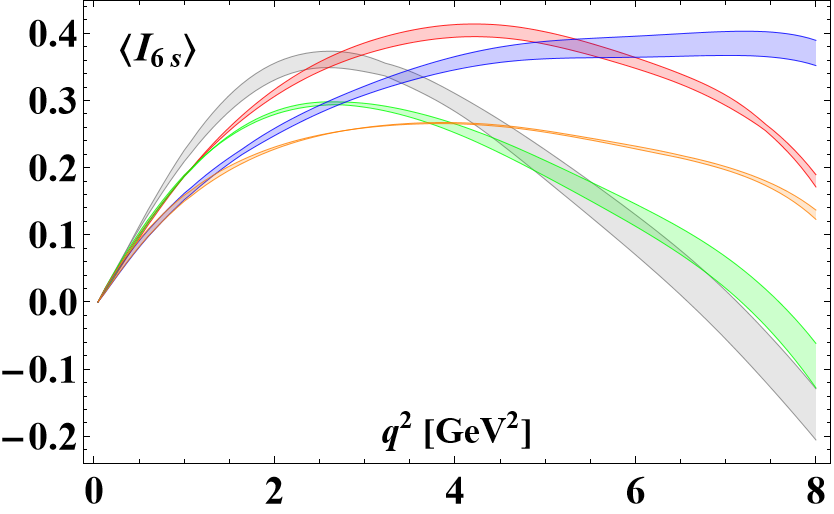}
\includegraphics[width=2in,height=1.24in]{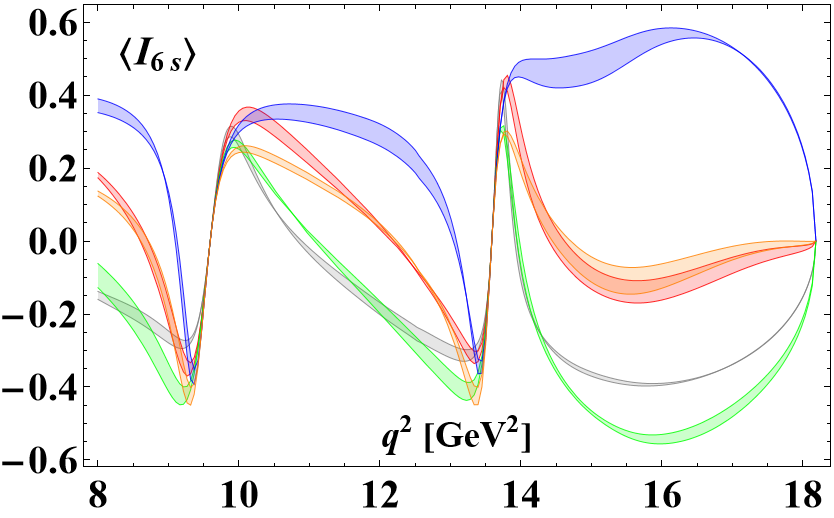}
\raisebox{0.0in}{\includegraphics[width=2in,height=1.24in]{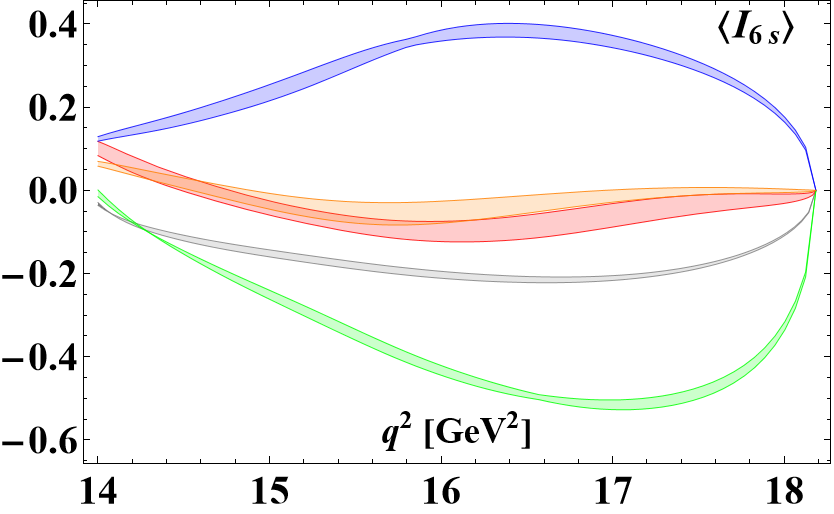}}
\caption{Predictions for the normalized angular coefficients $\langle I_{3}\rangle$, $\langle I_{4}\rangle$, $\langle I_{5}\rangle$, and $\langle I_{6s}\rangle$ of the decay
$B_c^{+}\to D^{\ast+}\ell^{+}\ell^{-}$ in the SM and the 2D NP scenarios SV--SVIII. The arrangement of the panels is the same as in Fig.~\ref{fig:2D_BrAFBfL}.}
\label{fig:2D_I3toI6s}
\end{figure}

The predictions for the normalized angular coefficients 
$\langle I_{1s}\rangle$, 
$\langle I_{1c}\rangle$, 
$\langle I_{2s}\rangle$, and 
$\langle I_{2c}\rangle$ for the 2D NP scenarios are shown in Fig.~\ref{fig:2D_I1stoI2c}, while the corresponding results for 
$\langle I_{3}\rangle$, 
$\langle I_{4}\rangle$, 
$\langle I_{5}\rangle$, and 
$\langle I_{6s}\rangle$ are displayed in Fig.~\ref{fig:2D_I3toI6s}. 
The figures are arranged in the same order as before. The angular coefficients 
$\langle I_{1s}\rangle$ and 
$\langle I_{2s}\rangle$ show moderate sensitivity to the NP WCs in the low-$q^{2}$ muon range, with the SVII scenario generally producing the largest enhancement above $q^2\approx 5$ GeV$^2$, while the SVIII scenario predicts comparatively smaller values. 
In the high-$q^{2}$ muon range, the spread among the NP scenarios becomes more visible near the resonance ranges. For the $\tau$ channel, both $\langle I_{1s}\rangle$ and 
$\langle I_{2s}\rangle$ exhibit a substantial spread among the NP scenarios at higher $q^{2}$ values. The coefficients 
$\langle I_{1c}\rangle$ and 
$\langle I_{2c}\rangle$ display a comparatively stronger dependence on the NP scenarios. In the low-$q^{2}$ range, the SVIII scenario generally produces larger deviations from the SM prediction, whereas the SV and SVII scenarios remain relatively closer to the SM values. For the $\tau$ mode, the hierarchy among the NP scenarios changes significantly. For $\langle I_{1c}\rangle$, scenarios SVI--SVIII deviate largely as compared to the SM predictions in most of the allowed kinematic range, whereas the Scenario SV deviates less from the SM and also gives opposite predictions to SVI--SVIII in the interval $15 \leq q^2 \leq 17~\mathrm{GeV}^2$. For $\langle I_{2c}\rangle$, SV gives the most pronounced deviations, whereas the Scenario SVII gives opposite predictions relative to SV, SVI, and SVIII.    

The observables 
$\langle I_{3}\rangle$ and $\langle I_{4}\rangle$, presented in Fig.~\ref{fig:2D_I3toI6s}, also show sensitivity to the NP WCs. In the low-$q^{2}$ muon range, $\langle I_{3} \rangle$ is a more suitable observable compared to $\langle I_{4} \rangle$ as it is relatively more sensitive to NP scenarios. However, for the $\tau$ mode, both $\langle I_{3} \rangle$ and $\langle I_{4} \rangle$ are equally good observables, as they exhibit decreasing and increasing values for all NP scenarios with almost equivalent magnitudes, respectively. The largest deviations, compared to SM case, are observed for SV and SVI scenarios. For the coefficient $\langle I_5\rangle$, all 2D NP scenarios give negative values over most of the kinematic range. In the low-$q^2$ muon range, $\langle I_5\rangle$ shows a clear dependence on the NP WCs. The SVII scenario gives the least negative values and deviates largely from the SM predictions. Interestingly, all NP scenarios clearly deviate from the SM results and can be discriminated from each other. In the high-$q^2$ muon range, the observable is strongly affected by the charmonium resonance structures. Therefore, the sharp variations in this range may not be interpreted as purely NP effects. For the high-$q^2$ $\tau$ channel, the hierarchy changes: SV gives the largest negative magnitude, while SVII stays closer to the SM band, and SVI and SVIII lie between them. Thus, $\langle I_5\rangle$ is one of the suitable observables for exploring the 2D NP scenarios.

The angular coefficient 
$\langle I_{6s}\rangle$ exhibits one of the strongest sensitivities to the 2D NP scenarios. 
In the low-$q^{2}$ muon range, all scenarios, except SVI, produce sizable positive values, while in the high-$q^{2}$ range the observable develops large deviations. In particular, the SVI scenario generates substantial negative deviations in the $\tau$ channel, whereas the SVII scenario predicts comparatively large positive values over most of the allowed kinematic range. In summary, the angular observables in the 2D NP scenarios exhibit richer phenomenological behavior than in the corresponding 1D scenarios due to the simultaneous variation of two NP WCs. Several observables, especially $\langle I_{1c}\rangle$, $\langle I_{2c}\rangle$, $\langle I_{3}\rangle$, $\langle I_{5}\rangle$, and $\langle I_{6s} \rangle$, show substantial sensitivity to the interference between the weak annihilation amplitudes and the NP-modified short-distance contributions. The effects become particularly pronounced in the $\tau$ channel, making these observables promising probes for distinguishing different NP scenarios in future experimental studies.

\begin{figure}[H]
\centering
\includegraphics[width=2.8in,height=1.2in]{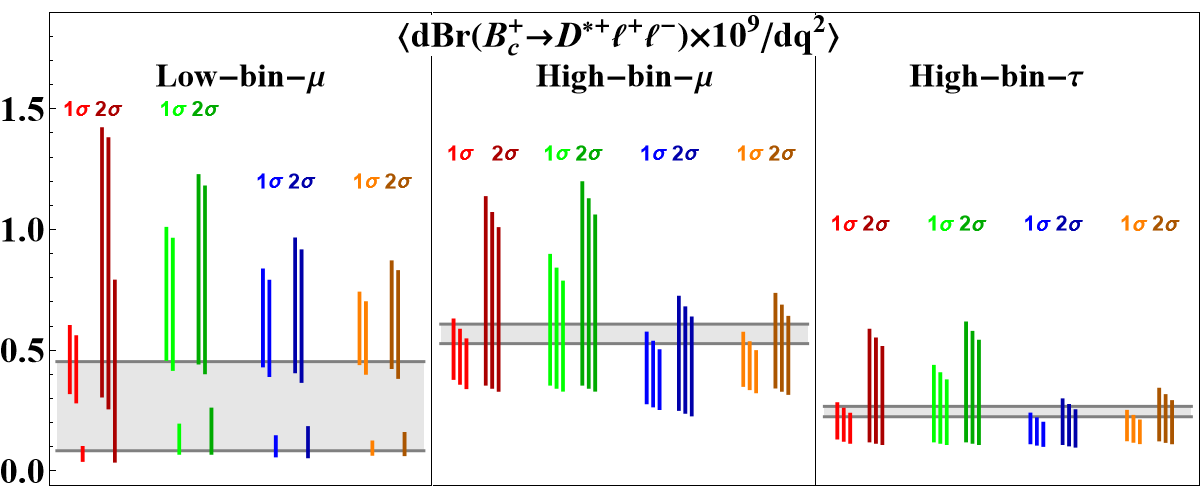}
\includegraphics[width=2.8in,height=1.2in]{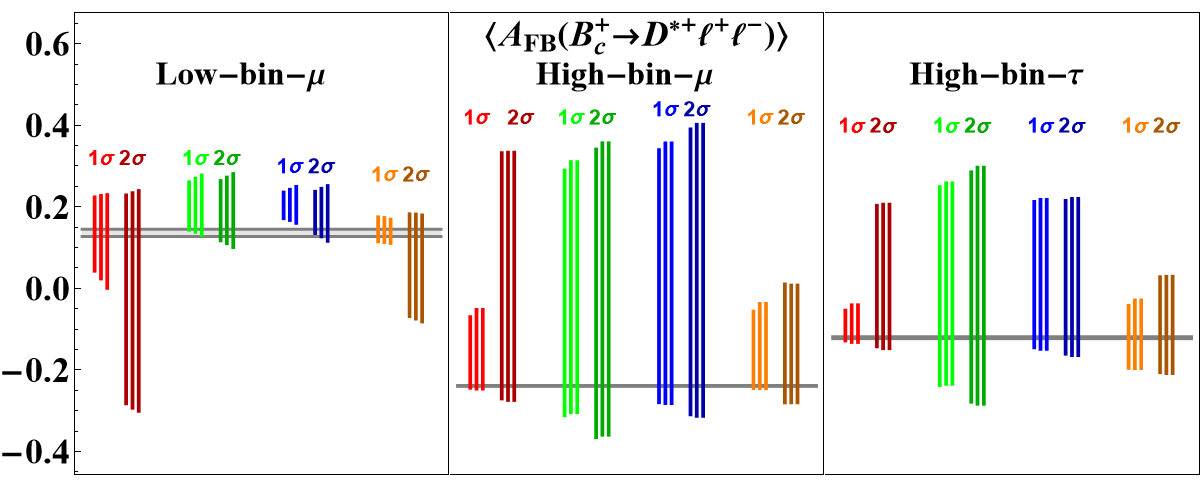}

\includegraphics[width=2.8in,height=1.2in]{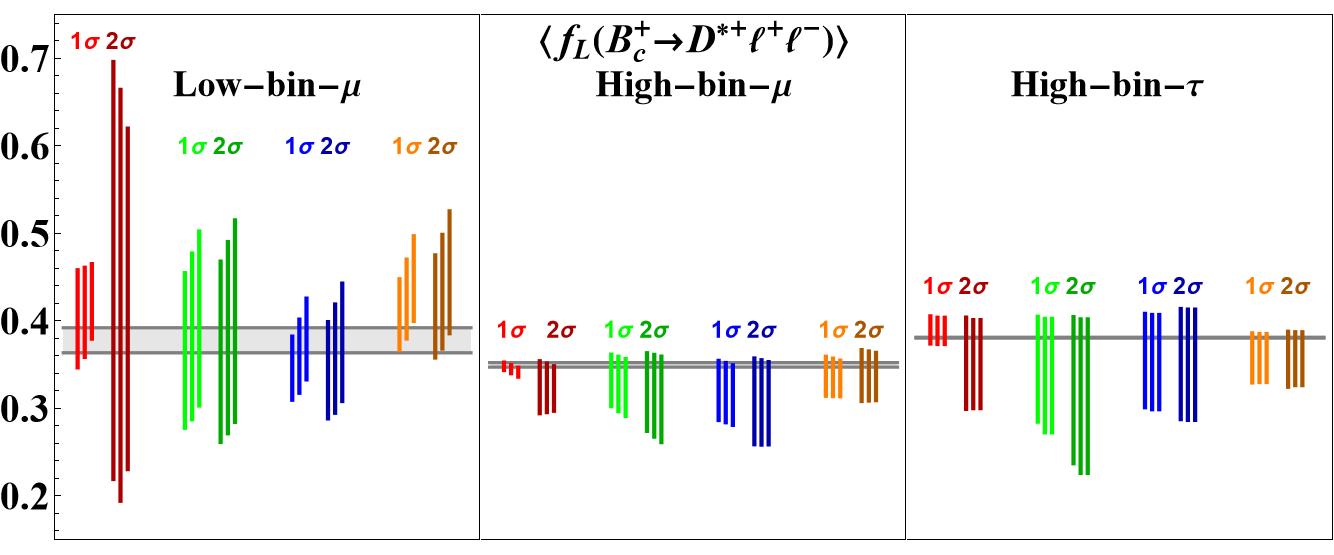}
\includegraphics[width=2.8in,height=1.2in]{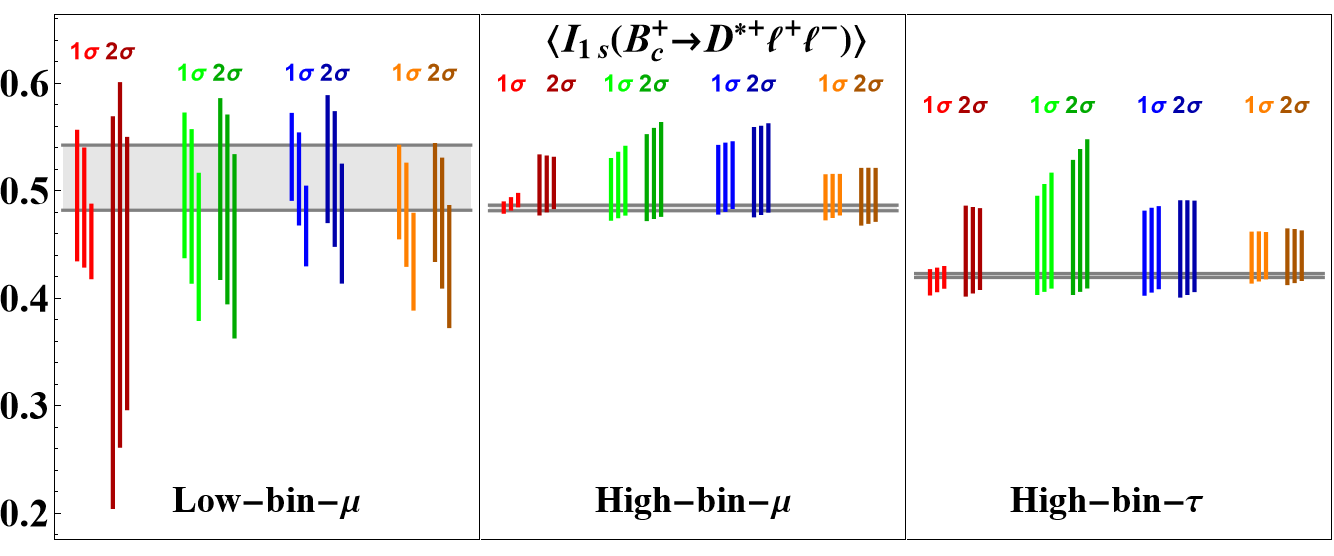}

\includegraphics[width=2.8in,height=1.2in]{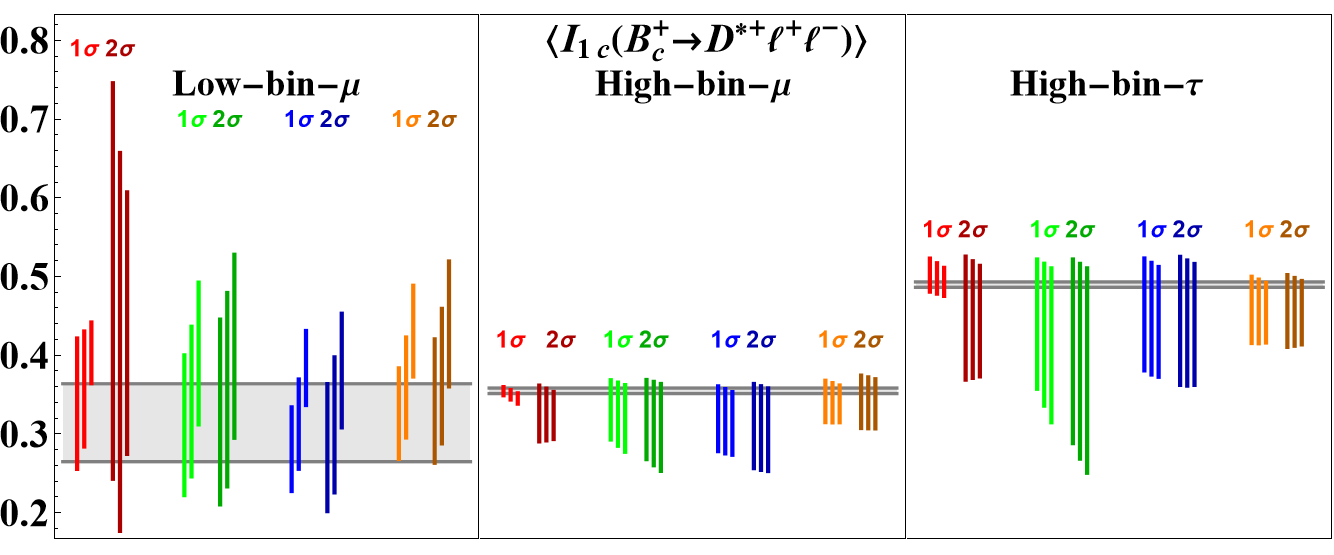}
\includegraphics[width=2.8in,height=1.2in]{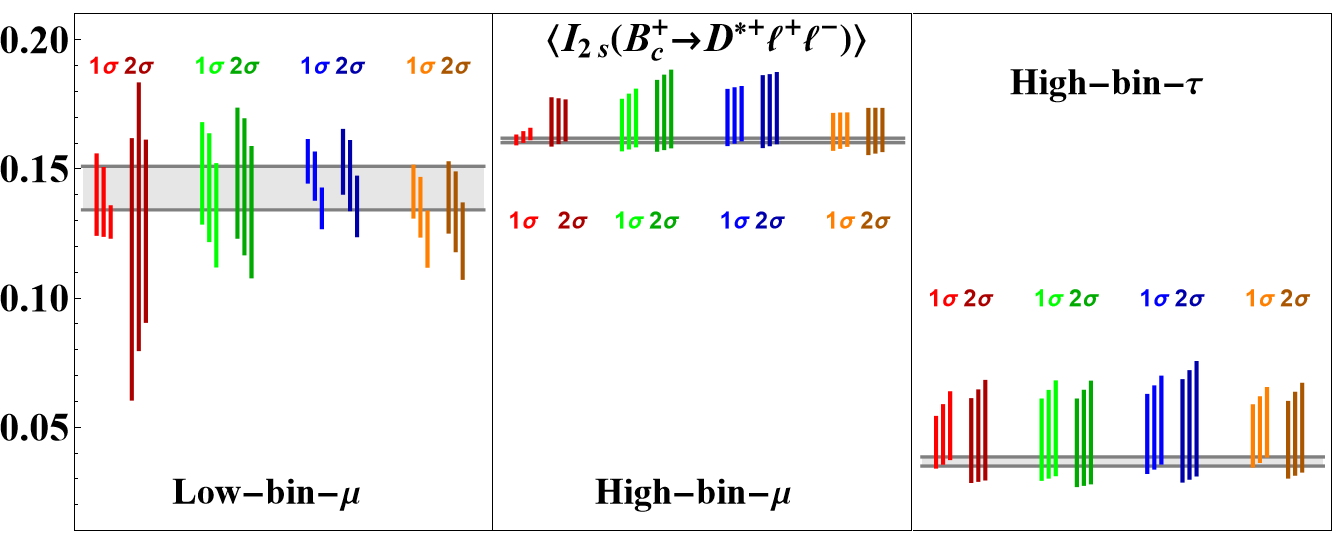}

\includegraphics[width=2.8in,height=1.2in]{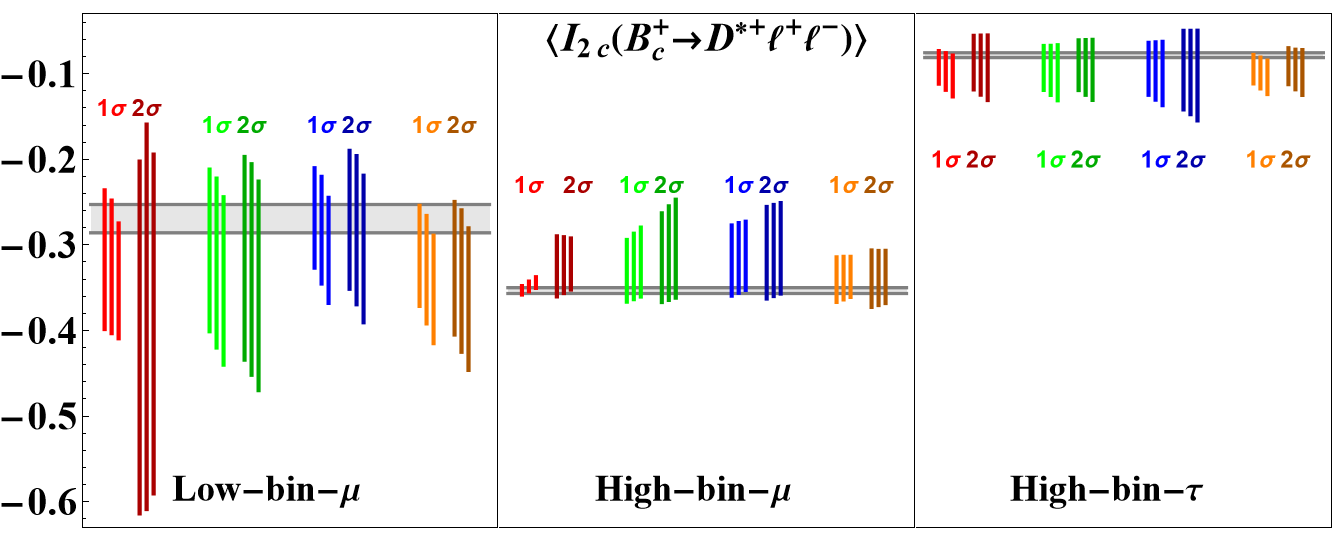}
\includegraphics[width=2.8in,height=1.2in]{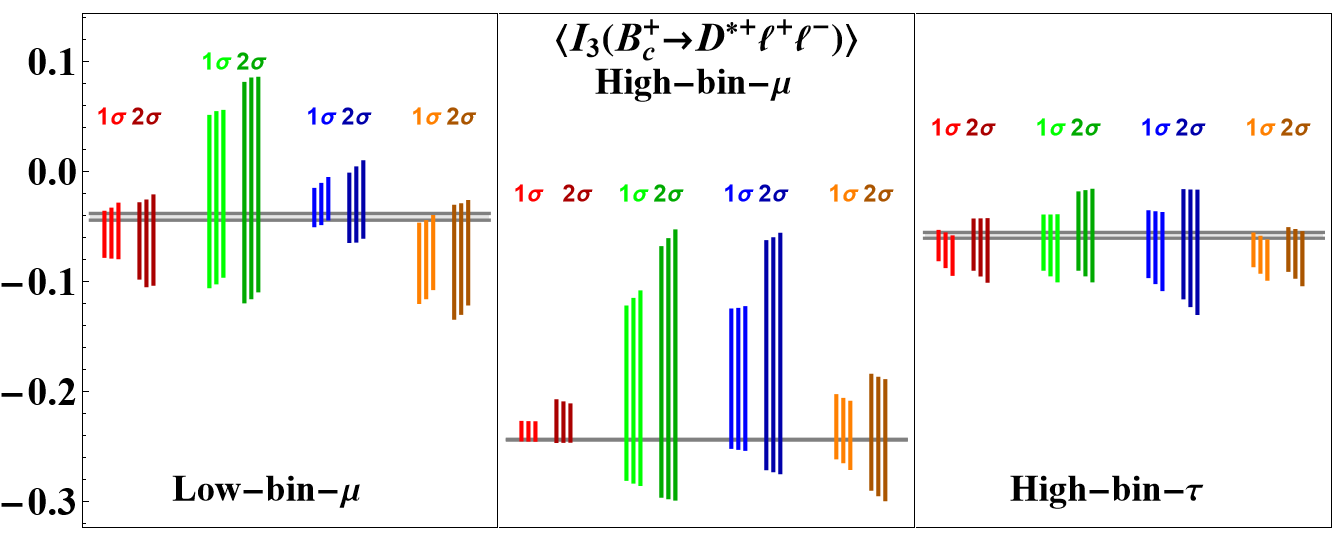}

\includegraphics[width=2.8in,height=1.2in]{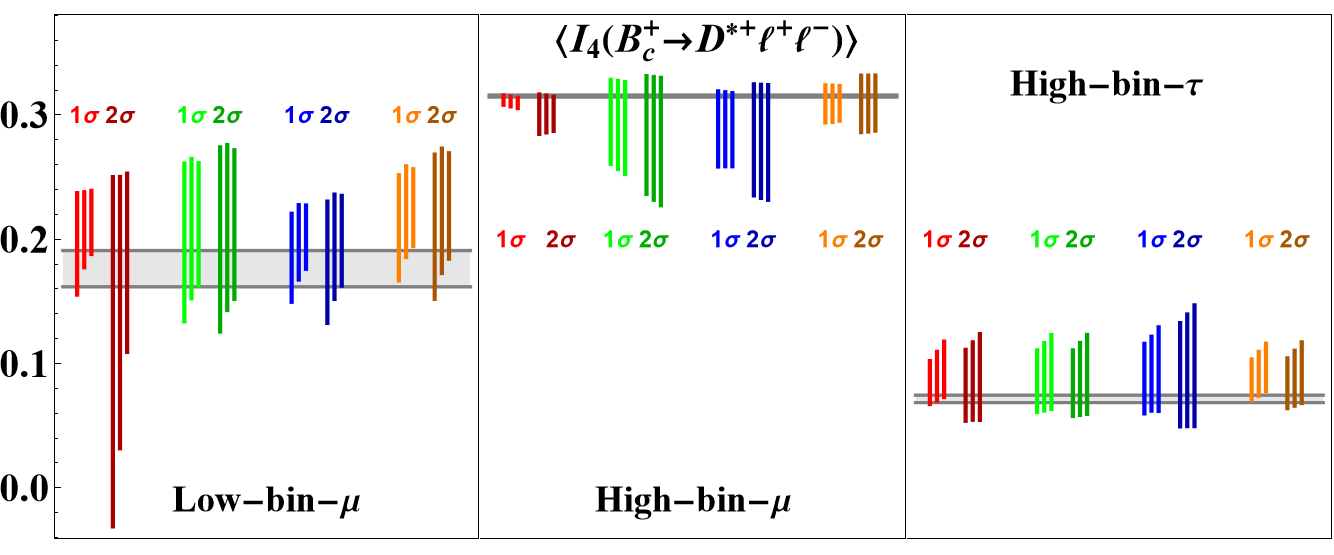}
\includegraphics[width=2.8in,height=1.2in]{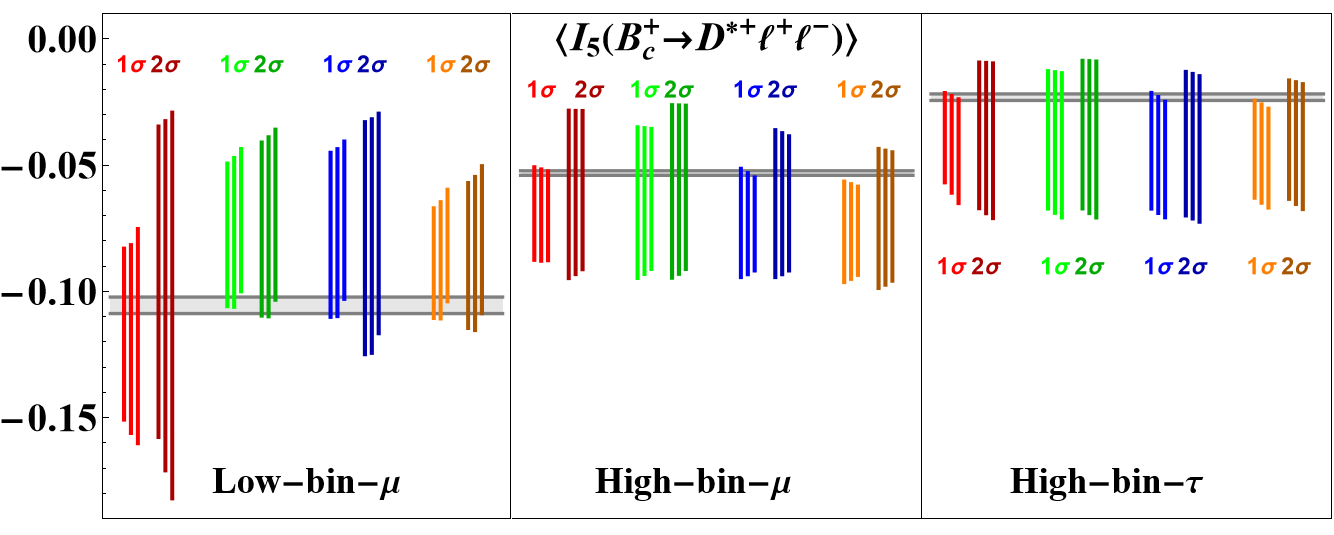}

\includegraphics[width=2.8in,height=1.2in]{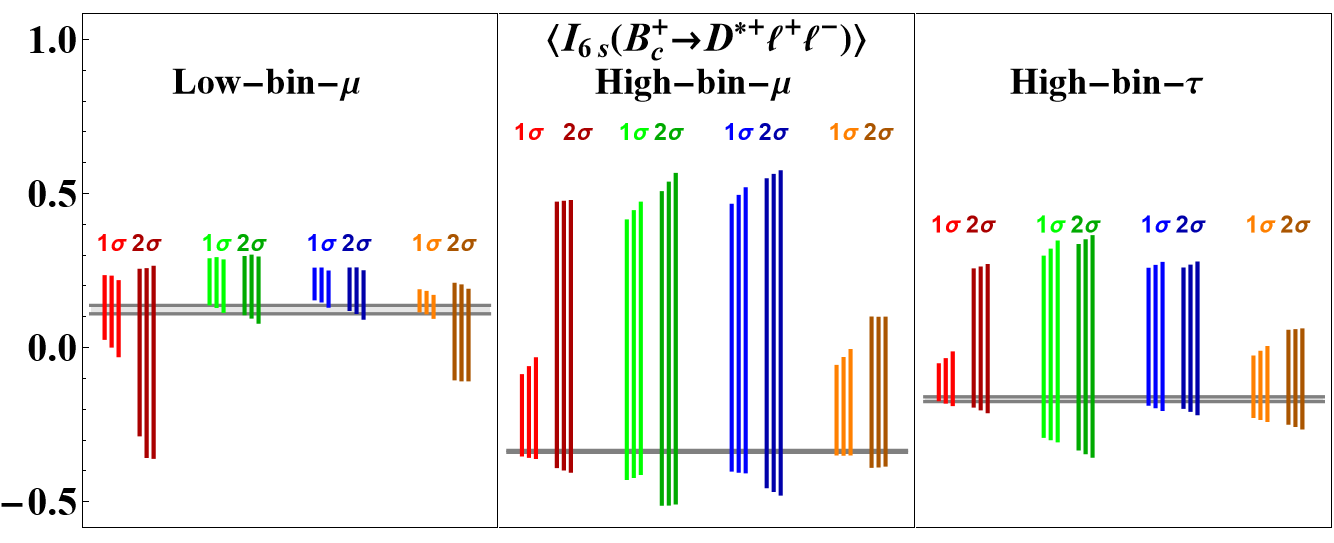}

\caption{Binned predictions for the observables of $B_c^{+}\to D^{\ast+}\ell^{+}\ell^{-}$ in the SM and the 2D NP scenarios SV--SVIII. The panels display the low-$q^{2}$ muon bin, high-$q^{2}$ muon bin, and high-$q^{2}$ tau bin. The colored bars indicate the allowed $1\sigma$ and $2\sigma$ NP ranges including the form factor uncertainties, whereas the gray bands denote the corresponding SM predictions.}
\label{2Dbarplots}
\end{figure}

Figure~\ref{2Dbarplots} shows the binned predictions of the observables in the 2D NP scenarios SV-SVIII. The results are presented for three different bins as discussed previously for Fig.~\ref{1Dbarplots}. The gray horizontal bands represent the SM predictions together with their theoretical uncertainties, while the colored bars correspond to the predictions of the different 2D NP scenarios. As before, the lighter and darker shades indicate the $1\sigma$ and $2\sigma$ ranges of the NP WCs, respectively. Each set of bars describes the effects of the allowed NP parameter space with the choice of lower, central, and upper values of the form factors. Therefore, the bar plots provide a compact illustration of the possible deviations generated by the 2D NP scenarios, in different observables, for the three bins. Compared to the corresponding 1D scenarios, the 2D scenarios generally produce wider prediction ranges due to the simultaneous variation of two NP WCs. The largest deviations are observed in the low- as well as in high-$q^{2}$ ranges, where the SM uncertainties are comparatively smaller and the sensitivity to NP effects becomes more pronounced. The $\tau$ channel continues to display enhanced sensitivity to the NP effects due to the sizable lepton-mass contributions entering the helicity amplitudes. However, with the consideration of the complete $1\sigma$ and $2\sigma$ ranges of the NP WCs, the overlap of the NP predictions with the SM band and among each other appears to be inevitable.

From a phenomenological perspective, the 2D NP scenarios have more wider predictions than the corresponding 1D cases. The simultaneous modification of two WCs enhances the correlations among the observables and allows sizable departures from the SM predictions even in observables that remain comparatively stable in the 1D scenarios. Consequently, a combined analysis of multiple observables in different kinematic ranges may provide strong constraints on the allowed NP parameter space in future experimental measurements. The combined possible minimum and maximum predictions corresponding to any of the full $1\sigma$ and $2\sigma$ ranges of the NP WCs are summarized in Table~\ref{tab:2Dnums}.

\begin{figure}[H]
\centering
\includegraphics[width=2in,height=1.24in]{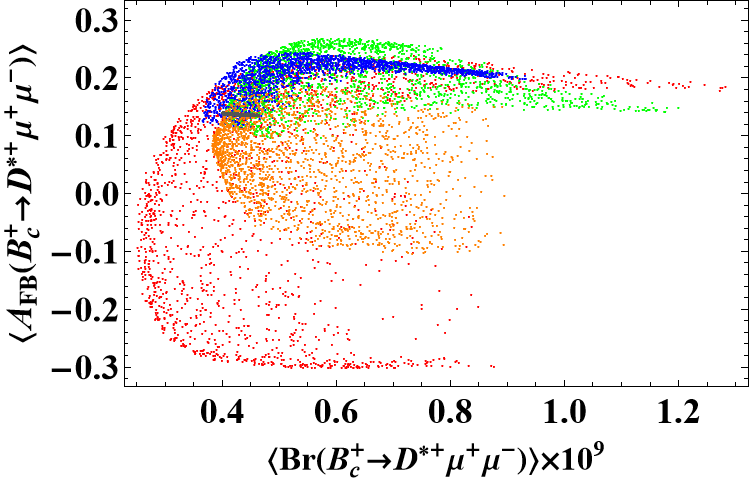}
\includegraphics[width=2in,height=1.24in]{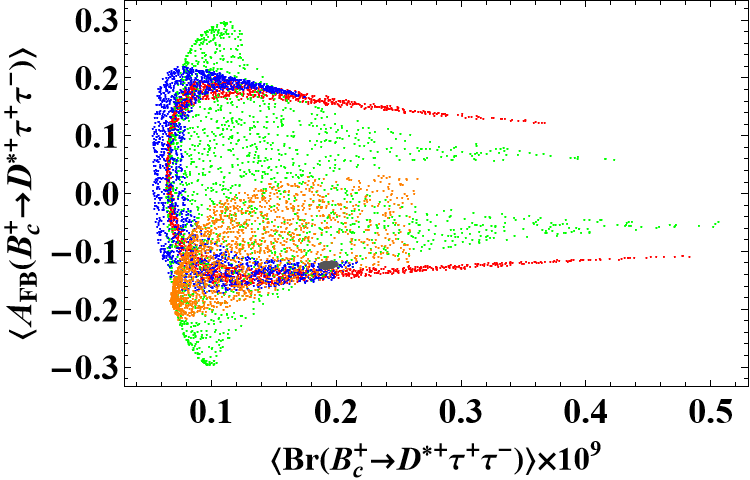}

\includegraphics[width=2in,height=1.24in]{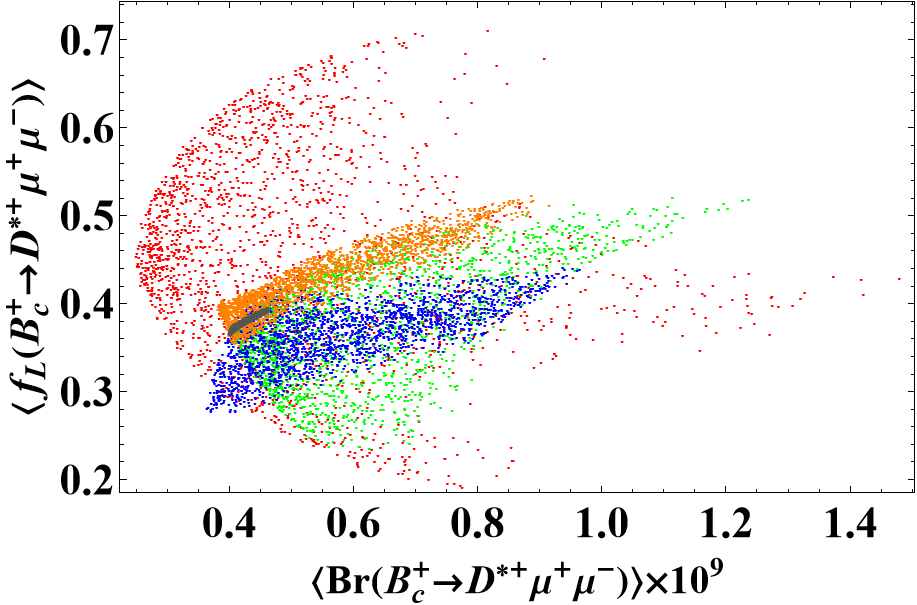}
\includegraphics[width=2in,height=1.24in]{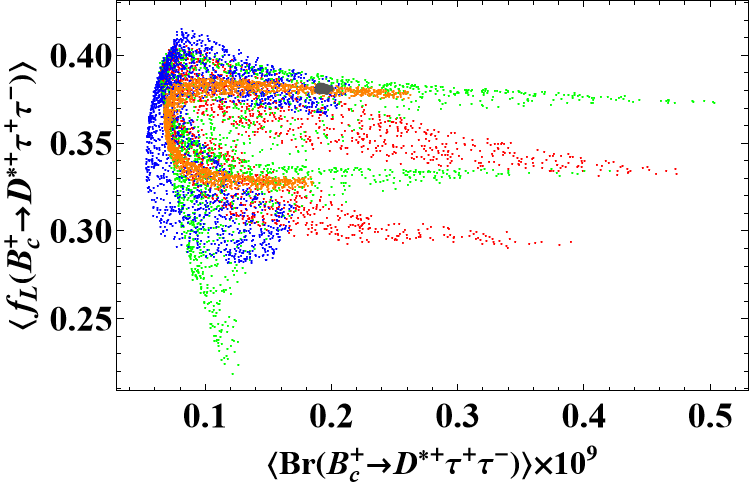}
\caption{Correlation plots between the integrated branching fraction and the observables $\langle A_{\rm FB}\rangle$ and $\langle f_{L}\rangle$ for $B_c^{+}\to D^{\ast+}\ell^{+}\ell^{-}$ obtained from the allowed parameter space of the 2D NP scenarios SV--SVIII. The correlations are plotted in low-$q^2$ bin for muon channel and in high-$q^2$ bin for tau channel. The spread of points includes the form factor uncertainties and the WCs uncertainties up to $2\sigma$ range.}
\label{Fig2Dcorr}
\end{figure}

The correlation plots for the 2D NP scenarios, presented in Fig.~\ref{Fig2Dcorr}, exhibit a richer phenomenology due to the simultaneous variation of two NP WCs. In Fig.~\ref{Fig2Dcorr}, the upper panels show the correlation between the integrated branching fraction and $\langle A_{\rm FB}\rangle$, for different 2D NP scenarios in the low-bin for muon and in the high-bin for tau channel. In the muon channel, the allowed ranges display sizable spreads, particularly for the red scenario, which extends to both large positive and negative values of $\langle A_{\rm FB}\rangle$. The green and blue scenarios predominantly populate the positive asymmetry range and remain comparatively more constrained. The orange scenario occupies an intermediate range with moderate deviations from the SM-like values. In the $\tau$ channel, the allowed parameter space becomes more compressed in the branching fraction, while $\langle A_{\rm FB}\rangle$ develops both positive and negative branches. The green scenario shows the largest spread, whereas the blue scenario remains comparatively narrow and concentrated around moderate asymmetry values.

Similarly, the lower panels, in Fig.~\ref{Fig2Dcorr}, display the correlations involving the longitudinal helicity fraction $\langle f_{L}\rangle$, for low-bin muon and high-bin tau decay. In the muon mode, the red scenario produces the broadest allowed range and extends toward significantly larger values of $\langle f_{L}\rangle$, while the green and blue scenarios prefer comparatively smaller longitudinal polarization fractions. The orange scenario remains localized around the intermediate range. For the $\tau$ channel, the blue and green scenarios exhibit the widest spread of $\langle f_{L}\rangle$, with the former and latter extending toward comparatively lower and higher values of branching fractions, respectively.

The correlation plots for the 2D NP scenarios show that the simultaneous variation of two NP WCs produces broader allowed ranges of the observables than in the corresponding 1D scenarios. The correlation between the integrated branching fraction and $\langle A_{FB}\rangle$ is especially useful, since different 2D scenarios populate different parts of the parameter space in both the $\mu$ and $\tau$ channels. The correlations involving $\langle f_L\rangle$ are comparatively more compressed, but they still provide a complementary check on the decay. Although the allowed ranges of different scenarios partly overlap, the combined use of the branching fraction, $\langle A_{FB}\rangle$, and $\langle f_L\rangle$ can help distinguish among the 2D NP scenarios and constrain the WCs associated with the $b\to d\,\ell^+\ell^-$ transition in future precision measurements.

\begin{table*}[htbp]
\centering
\caption{
Minimum and maximum values of the observables for the SM
and 1D scenarios SI, SII, SIII and SIV.
The three bins correspond respectively to:
low-$q^2$ muon range,
high-$q^2$ muon range,
and high-$q^2$ tauon range.
}
\renewcommand{\arraystretch}{1.2}

\resizebox{\textwidth}{!}{
\begin{tabular}{|c c c c c c c|}
\hline
Observable & range & SM & SI & SII & SIII & SIV \\
\hline

\multirow{3}{*}{$\mathrm{Br}\cross 10^{9}$}

& Low-$q^2$ ($\mu$)
& $(0.083,\,0.452)$
& $(0.077,\,0.930)$
& $(0.046,\,0.465)$
& $(0.062,\,0.456)$
& $(0.070,\,0.847)$ \\

& High-$q^2$ ($\mu$)
& $(0.484,\,0.568)$
& $(0.289,\,0.598)$
& $(0.286,\,0.598)$
& $(0.293,\,0.597)$
& $(0.273,\,0.596)$ \\

& High-$q^2$ ($\tau$)
& $(0.171,\,0.216)$
& $(0.060,\,0.234)$
& $(0.150,\,0.219)$
& $(0.112,\,0.226)$
& $(0.062,\,0.231)$ \\

\hline

\multirow{3}{*}{$A_{\rm FB}$}

& Low-$q^2$ ($\mu$)
& $(0.127,\,0.145)$
& $(0.117,\,0.237)$
& $(0.062,\,0.147)$
& $(0.133,\,0.158)$
& $(-0.077,\,0.147)$ \\

& High-$q^2$ ($\mu$)
& $(-0.240,\,-0.238)$
& $(-0.240,\,0.325)$
& $(-0.241,\,-0.120)$
& $(-0.243,\,-0.197)$
& $(-0.255,\,-0.187)$ \\

& High-$q^2$ ($\tau$)
& $(-0.123,\,-0.118)$
& $(-0.126,\,0.202)$
& $(-0.128,\,-0.043)$
& $(-0.125,\,-0.096)$
& $(-0.178,\,-0.117)$ \\

\hline

\multirow{3}{*}{$f_L$}

& Low-$q^2$ ($\mu$)
& $(0.363,\,0.392)$
& $(0.355,\,0.432)$
& $(0.319,\,0.397)$
& $(0.335,\,0.387)$
& $(0.361,\,0.512)$ \\

& High-$q^2$ ($\mu$)
& $(0.347,\,0.352)$
& $(0.302,\,0.352)$
& $(0.345,\,0.369)$
& $(0.348,\,0.362)$
& $(0.316,\,0.354)$ \\

& High-$q^2$ ($\tau$)
& $(0.380,\,0.381)$
& $(0.304,\,0.404)$
& $(0.373,\,0.381)$
& $(0.379,\,0.389)$
& $(0.325,\,0.385)$ \\

\hline

\multirow{3}{*}{$\langle I_{1s}\rangle$}

& Low-$q^2$ ($\mu$)
& $(0.481,\,0.542)$
& $(0.424,\,0.543)$
& $(0.477,\,0.592)$
& $(0.484,\,0.566)$
& $(0.375,\,0.543)$ \\

& High-$q^2$ ($\mu$)
& $(0.481,\,0.486)$
& $(0.481,\,0.525)$
& $(0.464,\,0.488)$
& $(0.470,\,0.485)$
& $(0.478,\,0.512)$ \\

& High-$q^2$ ($\tau$)
& $(0.419,\,0.422)$
& $(0.405,\,0.478)$
& $(0.418,\,0.422)$
& $(0.408,\,0.421)$
& $(0.415,\,0.461)$ \\

\hline

\multirow{3}{*}{$\langle I_{1c}\rangle$}

& Low-$q^2$ ($\mu$)
& $(0.264,\,0.363)$
& $(0.263,\,0.443)$
& $(0.193,\,0.369)$
& $(0.231,\,0.360)$
& $(0.263,\,0.510)$ \\

& High-$q^2$ ($\mu$)
& $(0.351,\,0.358)$
& $(0.299,\,0.358)$
& $(0.349,\,0.381)$
& $(0.352,\,0.372)$
& $(0.316,\,0.362)$ \\

& High-$q^2$ ($\tau$)
& $(0.486,\,0.493)$
& $(0.379,\,0.521)$
& $(0.486,\,0.493)$
& $(0.488,\,0.512)$
& $(0.412,\,0.499)$ \\

\hline

\multirow{3}{*}{$\langle I_{2s}\rangle$}

& Low-$q^2$ ($\mu$)
& $(0.134,\,0.150)$
& $(0.126,\,0.152)$
& $(0.133,\,0.160)$
& $(0.134,\,0.156)$
& $(0.110,\,0.151)$ \\

& High-$q^2$ ($\mu$)
& $(0.160,\,0.161)$
& $(0.160,\,0.174)$
& $(0.154,\,0.162)$
& $(0.156,\,0.161)$
& $(0.159,\,0.170)$ \\

& High-$q^2$ ($\tau$)
& $(0.034,\,0.038)$
& $(0.034,\,0.066)$
& $(0.022,\,0.040)$
& $(0.030,\,0.038)$
& $(0.034,\,0.059)$ \\

\hline

\multirow{3}{*}{$\langle I_{2c}\rangle$}

& Low-$q^2$ ($\mu$)
& $(-0.285,\,-0.252)$
& $(-0.377,\,-0.251)$
& $(-0.293,\,-0.183)$
& $(-0.280,\,-0.220)$
& $(-0.436,\,-0.251)$ \\

& High-$q^2$ ($\mu$)
& $(-0.356,\,-0.350)$
& $(-0.357,\,-0.298)$
& $(-0.379,\,-0.348)$
& $(-0.371,\,-0.351)$
& $(-0.361,\,-0.315)$ \\

& High-$q^2$ ($\tau$)
& $(-0.080,\,-0.075)$
& $(-0.128,\,-0.069)$
& $(-0.084,\,-0.053)$
& $(-0.080,\,-0.070)$
& $(-0.124,\,-0.073)$ \\

\hline

\multirow{3}{*}{$\langle I_3\rangle$}

& Low-$q^2$ ($\mu$)
& $(-0.044,\,-0.038)$
& $(-0.044,\,-0.032)$
& $(-0.044,\,-0.032)$
& $(-0.044,\,-0.033)$
& $(-0.121,\,-0.034)$ \\

& High-$q^2$ ($\mu$)
& $(-0.244,\,-0.243)$
& $(-0.244,\,-0.210)$
& $(-0.248,\,-0.243)$
& $(-0.243,\,-0.240)$
& $(-0.270,\,-0.242)$ \\

& High-$q^2$ ($\tau$)
& $(-0.060,\,-0.055)$
& $(-0.095,\,-0.054)$
& $(-0.063,\,-0.037)$
& $(-0.060,\,-0.049)$
& $(-0.102,\,-0.054)$ \\

\hline

\multirow{3}{*}{$\langle I_4\rangle$}

& Low-$q^2$ ($\mu$)
& $(0.161,\,0.190)$
& $(0.159,\,0.230)$
& $(0.136,\,0.193)$
& $(0.160,\,0.189)$
& $(0.158,\,0.267)$ \\

& High-$q^2$ ($\mu$)
& $(0.314,\,0.315)$
& $(0.287,\,0.316)$
& $(0.313,\,0.323)$
& $(0.314,\,0.318)$
& $(0.304,\,0.319)$ \\

& High-$q^2$ ($\tau$)
& $(0.068,\,0.074)$
& $(0.066,\,0.120)$
& $(0.047,\,0.077)$
& $(0.062,\,0.073)$
& $(0.067,\,0.116)$ \\

\hline

\multirow{3}{*}{$\langle I_5\rangle$}

& Low-$q^2$ ($\mu$)
& $(-0.108,\,-0.102)$
& $(-0.110,\,-0.047)$
& $(-0.112,\,-0.041)$
& $(-0.111,\,-0.078)$
& $(-0.108,\,-0.051)$ \\

& High-$q^2$ ($\mu$)
& $(-0.054,\,-0.052)$
& $(-0.093,\,-0.049)$
& $(-0.054,\,-0.025)$
& $(-0.065,\,-0.051)$
& $(-0.098,\,-0.049)$ \\

& High-$q^2$ ($\tau$)
& $(-0.024,\,-0.021)$
& $(-0.069,\,-0.020)$
& $(-0.025,\,-0.007)$
& $(-0.028,\,-0.021)$
& $(-0.067,\,-0.020)$ \\

\hline

\multirow{3}{*}{$\langle I_{6s}\rangle$}

& Low-$q^2$ ($\mu$)
& $(0.109,\,0.136)$
& $(0.097,\,0.248)$
& $(0.044,\,0.141)$
& $(0.115,\,0.147)$
& $(-0.080,\,0.140)$ \\

& High-$q^2$ ($\mu$)
& $(-0.340,\,-0.334)$
& $(-0.347,\,0.464)$
& $(-0.341,\,-0.164)$
& $(-0.342,\,-0.263)$
& $(-0.350,\,-0.243)$ \\

& High-$q^2$ ($\tau$)
& $(-0.175,\,-0.160)$
& $(-0.176,\,0.260)$
& $(-0.183,\,-0.055)$
& $(-0.170,\,-0.119)$
& $(-0.198,\,-0.159)$ \\

\hline

\end{tabular}
}
\label{tab:1Dnums}
\end{table*}

\begin{table*}[htbp]
\centering
\caption{
Minimum and maximum values of the observables for the SM and 2D scenarios SV, SVI, SVII and SVIII. The three bins correspond respectively to:
low-$q^2$ muon range,
high-$q^2$ muon range,
and high-$q^2$ tauon range.
}
\renewcommand{\arraystretch}{1.2}

\resizebox{\textwidth}{!}{
\begin{tabular}{|c c c c c c c|}
\hline
Observable & range & SM & SV & SVI & SVII & SVIII \\
\hline

\multirow{3}{*}{$\mathrm{Br}\cross 10^{9}$}

& Low-$q^2$ ($\mu$)
& $(0.083,\,0.452)$
& $(0.042,\,1.417)$
& $(0.074,\,1.232)$
& $(0.059,\,0.957)$
& $(0.068,\,0.861)$ \\

& High-$q^2$ ($\mu$)
& $(0.484,\,0.568)$
& $(0.287,\,1.079)$
& $(0.287,\,1.171)$
& $(0.181,\,0.686)$
& $(0.273,\,0.689)$ \\

& High-$q^2$ ($\tau$)
& $(0.171,\,0.216)$
& $(0.059,\,0.534)$
& $(0.059,\,0.591)$
& $(0.047,\,0.242)$
& $(0.062,\,0.286)$ \\

\hline

\multirow{3}{*}{$A_{\rm FB}$}

& Low-$q^2$ ($\mu$)
& $(0.127,\,0.145)$
& $(-0.300,\,0.238)$
& $(0.101,\,0.280)$
& $(0.116,\,0.251)$
& $(-0.081,\,0.181)$ \\

& High-$q^2$ ($\mu$)
& $(-0.240,\,-0.238)$
& $(-0.274,\,0.332)$
& $(-0.365,\,0.355)$
& $(-0.313,\,0.401)$
& $(-0.280,\,0.009)$ \\

& High-$q^2$ ($\tau$)
& $(-0.123,\,-0.118)$
& $(-0.147,\,0.205)$
& $(-0.283,\,0.295)$
& $(-0.164,\,0.219)$
& $(-0.208,\,0.028)$ \\

\hline

\multirow{3}{*}{$f_L$}

& Low-$q^2$ ($\mu$)
& $(0.363,\,0.392)$
& $(0.194,\,0.695)$
& $(0.261,\,0.514)$
& $(0.288,\,0.442)$
& $(0.358,\,0.525)$ \\

& High-$q^2$ ($\mu$)
& $(0.347,\,0.352)$
& $(0.294,\,0.353)$
& $(0.261,\,0.363)$
& $(0.258,\,0.357)$
& $(0.308,\,0.366)$ \\

& High-$q^2$ ($\tau$)
& $(0.380,\,0.381)$
& $(0.299,\,0.403)$
& $(0.226,\,0.404)$
& $(0.287,\,0.413)$
& $(0.324,\,0.387)$ \\

\hline

\multirow{3}{*}{$\langle I_{1s}\rangle$}

& Low-$q^2$ ($\mu$)
& $(0.481,\,0.542)$
& $(0.206,\,0.611)$
& $(0.370,\,0.580)$
& $(0.416,\,0.585)$
& $(0.369,\,0.543)$ \\

& High-$q^2$ ($\mu$)
& $(0.481,\,0.486)$
& $(0.479,\,0.531)$
& $(0.473,\,0.564)$
& $(0.477,\,0.560)$
& $(0.469,\,0.519)$ \\

& High-$q^2$ ($\tau$)
& $(0.419,\,0.422)$
& $(0.402,\,0.484)$
& $(0.404,\,0.543)$
& $(0.402,\,0.489)$
& $(0.414,\,0.463)$ \\

\hline

\multirow{3}{*}{$\langle I_{1c}\rangle$}

& Low-$q^2$ ($\mu$)
& $(0.264,\,0.363)$
& $(0.169,\,0.744)$
& $(0.205,\,0.510)$
& $(0.204,\,0.451)$
& $(0.263,\,0.518)$ \\

& High-$q^2$ ($\mu$)
& $(0.351,\,0.358)$
& $(0.290,\,0.361)$
& $(0.250,\,0.368)$
& $(0.252,\,0.363)$
& $(0.307,\,0.374)$ \\

& High-$q^2$ ($\tau$)
& $(0.486,\,0.493)$
& $(0.368,\,0.523)$
& $(0.263,\,0.521)$
& $(0.361,\,0.524)$
& $(0.410,\,0.501)$ \\

\hline

\multirow{3}{*}{$\langle I_{2s}\rangle$}

& Low-$q^2$ ($\mu$)
& $(0.134,\,0.150)$
& $(0.062,\,0.178)$
& $(0.108,\,0.171)$
& $(0.124,\,0.164)$
& $(0.108,\,0.151)$ \\

& High-$q^2$ ($\mu$)
& $(0.160,\,0.161)$
& $(0.159,\,0.176)$
& $(0.157,\,0.186)$
& $(0.158,\,0.186)$
& $(0.156,\,0.172)$ \\

& High-$q^2$ ($\tau$)
& $(0.034,\,0.038)$
& $(0.029,\,0.067)$
& $(0.027,\,0.067)$
& $(0.029,\,0.074)$
& $(0.030,\,0.066)$ \\

\hline

\multirow{3}{*}{$\langle I_{2c}\rangle$}

& Low-$q^2$ ($\mu$)
& $(-0.285,\,-0.252)$
& $(-0.614,\,-0.156)$
& $(-0.466,\,-0.196)$
& $(-0.388,\,-0.189)$
& $(-0.446,\,-0.250)$ \\

& High-$q^2$ ($\mu$)
& $(-0.356,\,-0.350)$
& $(-0.359,\,-0.289)$
& $(-0.366,\,-0.252)$
& $(-0.362,\,-0.250)$
& $(-0.372,\,-0.306)$ \\

& High-$q^2$ ($\tau$)
& $(-0.080,\,-0.075)$
& $(-0.130,\,-0.055)$
& $(-0.130,\,-0.060)$
& $(-0.153,\,-0.050)$
& $(-0.124,\,-0.070)$ \\

\hline

\multirow{3}{*}{$\langle I_3\rangle$}

& Low-$q^2$ ($\mu$)
& $(-0.044,\,-0.038)$
& $(-0.103,\,-0.022)$
& $(-0.118,\,0.085)$
& $(-0.063,\,0.007)$
& $(-0.132,\,-0.029)$ \\

& High-$q^2$ ($\mu$)
& $(-0.244,\,-0.243)$
& $(-0.244,\,-0.208)$
& $(-0.297,\,-0.049)$
& $(-0.272,\,-0.064)$
& $(-0.297,\,-0.186)$ \\

& High-$q^2$ ($\tau$)
& $(-0.060,\,-0.055)$
& $(-0.099,\,-0.044)$
& $(-0.098,\,-0.019)$
& $(-0.127,\,-0.018)$
& $(-0.102,\,-0.052)$ \\

\hline

\multirow{3}{*}{$\langle I_4\rangle$}

& Low-$q^2$ ($\mu$)
& $(0.161,\,0.190)$
& $(-0.030,\,0.253)$
& $(0.125,\,0.275)$
& $(0.132,\,0.235)$
& $(0.152,\,0.272)$ \\

& High-$q^2$ ($\mu$)
& $(0.314,\,0.315)$
& $(0.284,\,0.316)$
& $(0.229,\,0.331)$
& $(0.231,\,0.324)$
& $(0.286,\,0.331)$ \\

& High-$q^2$ ($\tau$)
& $(0.068,\,0.074)$
& $(0.053,\,0.123)$
& $(0.057,\,0.122)$
& $(0.049,\,0.145)$
& $(0.064,\,0.116)$ \\

\hline

\multirow{3}{*}{$\langle I_5\rangle$}

& Low-$q^2$ ($\mu$)
& $(-0.108,\,-0.102)$
& $(-0.181,\,-0.029)$
& $(-0.109,\,-0.035)$
& $(-0.123,\,-0.029)$
& $(-0.115,\,-0.050)$ \\

& High-$q^2$ ($\mu$)
& $(-0.054,\,-0.052)$
& $(-0.094,\,-0.026)$
& $(-0.094,\,-0.027)$
& $(-0.094,\,-0.036)$
& $(-0.098,\,-0.043)$ \\

& High-$q^2$ ($\tau$)
& $(-0.024,\,-0.021)$
& $(-0.070,\,-0.009)$
& $(-0.070,\,-0.008)$
& $(-0.072,\,-0.012)$
& $(-0.067,\,-0.016)$ \\

\hline

\multirow{3}{*}{$\langle I_{6s}\rangle$}

& Low-$q^2$ ($\mu$)
& $(0.109,\,0.136)$
& $(-0.355,\,0.258)$
& $(0.082,\,0.294)$
& $(0.095,\,0.253)$
& $(-0.108,\,0.201)$ \\

& High-$q^2$ ($\mu$)
& $(-0.340,\,-0.334)$
& $(-0.400,\,0.472)$
& $(-0.503,\,0.570)$
& $(-0.473,\,0.570)$
& $(-0.382,\,0.113)$ \\

& High-$q^2$ ($\tau$)
& $(-0.175,\,-0.160)$
& $(-0.206,\,0.264)$
& $(-0.351,\,0.361)$
& $(-0.213,\,0.272)$
& $(-0.259,\,0.051)$ \\

\hline

\end{tabular}
}
\label{tab:2Dnums}
\end{table*}

\section{Conclusions}\label{conl}

We have investigated the FCNC semileptonic decay
$B_c^+\to D^{\ast+}\ell^+\ell^-$, with $\ell=\mu,\tau$, in a model-independent NP framework for the underlying $b\to d\ell^+\ell^-$ transition. The analysis includes the PB contribution, LD resonance effects, and the WA amplitude, which plays a particularly important role in rare $B_c$ decays. The $B_c\to D^\ast$ transition form factors are taken from the covariant confined quark model, while the WA form factors are considered within the Bethe-Salpeter approach. Using these inputs, we have studied the differential branching fraction, the FB asymmetry, the longitudinal helicity fraction, and a comprehensive set of normalized angular observables in the selected 1D and 2D NP scenarios involving the WCs $C_7^{(\prime)}$, $C_9^{(\prime)}$, and $C_{10}^{(\prime)}$.

Our results show that the interplay between the PB and WA amplitudes can significantly modify the phenomenology of $B_c^+\to D^{\ast+}\ell^+\ell^-$ decay, particularly in the low-$q^2$ muon range. The WA contribution generates sizable effects in several observables in both the SM and the various NP scenarios and, therefore, should not be neglected in precision studies of this decay. In the high-$q^2$ muon range, the observables are additionally affected by LD resonance contributions, leading to visible structures around the charmonium resonances. Consequently, the ranges above the dominant resonance contributions provide a comparatively cleaner environment for probing NP effects.

Among the observables considered, the differential branching ratio and the FB asymmetry exhibit a strong sensitivity to NP contributions in both the 1D and 2D NP scenarios. Furthermore, in the four-fold decay, several normalized angular observables, particularly $\langle I_{2c}\rangle$, $\langle I_{3}\rangle$, $\langle I_{5}\rangle$, and $\langle I_{6s} \rangle$, show pronounced deviations from their SM predictions and reveal valuable complementary information on the underlying NP structure. The sensitivity of these observables arises from the NP-modified PB amplitudes and their interference with the WA contribution, making them especially useful for constraining the relevant WCs.

The comparison of different NP scenarios indicates that sizable deviations from the SM can occur in both the muon and tau channels. In particular, the $\tau$ mode often exhibits enhanced separation among the NP scenarios due to the larger lepton-mass effects entering the helicity amplitudes. This suggests that measurements involving $\tau$ leptons could provide additional discriminatory power of the observables in future experimental studies.

We also investigated correlations among the integrated observables and found that combined measurements of the branching fraction, $A_{\rm FB}$, and $f_L$ can help distinguish between different NP scenarios and further constrain the allowed parameter space of the NP WCs. Such correlations provide complementary information beyond individual observables and may become particularly useful once experimental precision improves.

In conclusion, our analysis demonstrates that the decay $B_c^+\to D^{\ast+}\ell^+\ell^-$ constitutes a phenomenologically rich and complementary probe of NP in the $b\to d\ell^+\ell^-$ sector. The sizable WA contributions, the sensitivity of several angular observables to NP scenarios, and the distinctive patterns observed in the correlation studies make this decay an attractive channel for future investigations. The observables studied in this work should be explored at future LHCb upgrades and other high-luminosity experiments, providing valuable opportunities to reveal the flavor structure of physics beyond the SM.


\section*{Acknowledgments}
M.A.P and Z.A would like to acknowledge the financial support provided by the Higher Education Commission (HEC) of Pakistan
through Grant no. NRPU/20-15142.

\clearpage
\bibliographystyle{refstyle}
\bibliography{ref}

\providecommand{\href}[2]{#2}\begingroup\raggedright\begin{thebibliography}{10}

\bibitem{Cabibbo:1963yz}
N.~Cabibbo, {\it {Unitary Symmetry and Leptonic Decays}},  {\em Phys. Rev. Lett.} {\bf 10} (1963) 531--533.

\bibitem{Kobayashi:1973fv}
M.~Kobayashi and T.~Maskawa, {\it {CP Violation in the Renormalizable Theory of Weak Interaction}},  {\em Prog. Theor. Phys.} {\bf 49} (1973) 652--657.

\bibitem{LHCb:2022qnv}
{\bf LHCb} Collaboration, R.~Aaij et~al., {\it {Test of lepton universality in $b \rightarrow s \ell^+ \ell^-$ decays}},  {\em Phys. Rev. Lett.} {\bf 131} (2023), no.~5 051803, [\href{https://arxiv.org/abs/2212.09152}{{\tt arXiv:2212.09152}}].

\bibitem{LHCb:2020lmf}
{\bf LHCb} Collaboration, R.~Aaij et~al., {\it {Measurement of $CP$-Averaged Observables in the $B^{0}\rightarrow K^{*0}\mu^{+}\mu^{-}$ Decay}},  {\em Phys. Rev. Lett.} {\bf 125} (2020), no.~1 011802, [\href{https://arxiv.org/abs/2003.04831}{{\tt arXiv:2003.04831}}].

\bibitem{LHCb:2021xxq}
{\bf LHCb} Collaboration, R.~Aaij et~al., {\it {Angular analysis of the rare decay $B_s^0 \to \phi \mu^+ \mu^- $}},  {\em JHEP} {\bf 11} (2021) 043, [\href{https://arxiv.org/abs/2107.13428}{{\tt arXiv:2107.13428}}].

\bibitem{LHCb:2026xvw}
{\bf LHCb} Collaboration, R.~Aaij et~al., {\it {Angular analysis of the $B^+\to \pi^+\mu^+\mu^-$ decay}},  [\href{https://arxiv.org/abs/2604.21987}{{\tt arXiv:2604.21987}}].

\bibitem{LHCb:2015hsa}
{\bf LHCb} Collaboration, R.~Aaij et~al., {\it {First measurement of the differential branching fraction and $C\!P$ asymmetry of the $B^\pm\to\pi^\pm\mu^+\mu^-$ decay}},  {\em JHEP} {\bf 10} (2015) 034, [\href{https://arxiv.org/abs/1509.00414}{{\tt arXiv:1509.00414}}].

\bibitem{BaBar:2013qaj}
{\bf BaBar} Collaboration, J.~P. Lees et~al., {\it {Search for the rare decays $B \to \pi\ell^+\ell^-$ and $B^0 \to \eta\ell^+\ell^-$}},  {\em Phys. Rev. D} {\bf 88} (2013), no.~3 032012, [\href{https://arxiv.org/abs/1303.6010}{{\tt arXiv:1303.6010}}].

\bibitem{Belle:2008tjs}
{\bf Belle} Collaboration, J.~T. Wei et~al., {\it {Search for $B \to \pi l^+ l^-$ Decays at Belle}},  {\em Phys. Rev. D} {\bf 78} (2008) 011101, [\href{https://arxiv.org/abs/0804.3656}{{\tt arXiv:0804.3656}}].

\bibitem{Belle:2024cis}
{\bf Belle, Belle-II} Collaboration, I.~Adachi et~al., {\it {Search for Rare $b\to d\ell^+\ell^-$ Transitions at Belle}},  {\em Phys. Rev. Lett.} {\bf 133} (2024), no.~10 101804, [\href{https://arxiv.org/abs/2404.08133}{{\tt arXiv:2404.08133}}].

\bibitem{LHCb:2018rym}
{\bf LHCb} Collaboration, R.~Aaij et~al., {\it {Evidence for the decay $ {B}_S^0\to {\overline{K}}^{\ast 0}{\mu}^{+}{\mu}^{-} $}},  {\em JHEP} {\bf 07} (2018) 020, [\href{https://arxiv.org/abs/1804.07167}{{\tt arXiv:1804.07167}}].

\bibitem{Belle2WhitePaper2018}
E.~Kou et~al., {\it {The Belle II Physics Book}},  {\em PTEP} {\bf 2019} (2019), no.~12 123C01, [\href{https://arxiv.org/abs/1808.10567}{{\tt arXiv:1808.10567}}].

\bibitem{Ishaq:2013toa}
S.~Ishaq, F.~Munir, and I.~Ahmed, {\it {Lepton polarization asymmetries in $B \to K_{1}l^{+}l^{-}$ decay as a searching tool for new physics}},  {\em JHEP} {\bf 07} (2013) 006.

\bibitem{Huang:2018rys}
Z.-R. Huang, M.~A. Paracha, I.~Ahmed, and C.-D. L\"u, {\it {Testing Leptoquark and $Z^{\prime}$ Models via $B\to K_{1}(1270,1400)\mu^{+}\mu^{-}$ Decays}},  {\em Phys. Rev. D} {\bf 100} (2019), no.~5 055038, [\href{https://arxiv.org/abs/1812.03491}{{\tt arXiv:1812.03491}}].

\bibitem{Munir:2015gsp}
F.~Munir, S.~Ishaq, and I.~Ahmed, {\it {Polarized forward-backward asymmetries of lepton pair in $B\rightarrow K_{1}\ell^{+}\ell^{-}$ decay in the presence of New physics}},  {\em PTEP} {\bf 2016} (2016), no.~1 013B02, [\href{https://arxiv.org/abs/1511.07075}{{\tt arXiv:1511.07075}}].

\bibitem{MunirBhutta:2020ber}
F.~Munir~Bhutta, Z.-R. Huang, C.-D. L\"u, M.~A. Paracha, and W.~Wang, {\it {New physics in $b\to s\ell\ell$ anomalies and its implications for the complementary neutral current decays}},  {\em Nucl. Phys. B} {\bf 979} (2022) 115763, [\href{https://arxiv.org/abs/2009.03588}{{\tt arXiv:2009.03588}}].

\bibitem{Bhutta:2024zwj}
F.~M. Bhutta, A.~Rehman, M.~J. Aslam, I.~Ahmed, and S.~Ishaq, {\it {Angular observables of the four-fold $B \to K_{1}(1270,1400)(\to V P) \ell^{+}\ell^{-}$ decays in and beyond the Standard Model}},  {\em Phys. Rev. D} {\bf 111} (2025), no.~9 095011, [\href{https://arxiv.org/abs/2410.20633}{{\tt arXiv:2410.20633}}].

\bibitem{Das:2018orb}
D.~Das, B.~Kindra, G.~Kumar, and N.~Mahajan, {\it {$B\to K^\ast_2(1430)\ell^+\ell^-$ distributions at large recoil in the Standard Model and beyond}},  {\em Phys. Rev. D} {\bf 99} (2019), no.~9 093012, [\href{https://arxiv.org/abs/1812.11803}{{\tt arXiv:1812.11803}}].

\bibitem{Mohapatra:2021izl}
M.~K. Mohapatra and A.~Giri, {\it {Implications of light $Z^{\prime}$ on semileptonic $B(B_s)\to T\{K_2^*(1430)(f_2^{\prime}(1525))\}\ell^+\ell^-$ decays at large recoil}},  {\em Phys. Rev. D} {\bf 104} (2021), no.~9 095012, [\href{https://arxiv.org/abs/2109.12382}{{\tt arXiv:2109.12382}}].

\bibitem{Rajeev:2020aut}
N.~Rajeev, N.~Sahoo, and R.~Dutta, {\it {Angular analysis of $B_s\, \to\, f_{2}'\,(1525)\,(\to K^+\,K^-)\,\mu^+ \,\mu^-$ decays as a probe to lepton flavor universality violation}},  {\em Phys. Rev. D} {\bf 103} (2021), no.~9 095007, [\href{https://arxiv.org/abs/2009.06213}{{\tt arXiv:2009.06213}}].

\bibitem{LHCb:2012ag}
{\bf LHCb} Collaboration, R.~Aaij et~al., {\it {First observation of the decay $B_c^+ \to J/\psi \pi^+\pi^-\pi^+$}},  {\em Phys. Rev. Lett.} {\bf 108} (2012) 251802, [\href{https://arxiv.org/abs/1204.0079}{{\tt arXiv:1204.0079}}].

\bibitem{Dutta:2019wxo}
R.~Dutta, {\it {Model independent analysis of new physics effects on $B_c \to (D_s,\,D^{\ast}_s)\,\mu^+\mu^-$ decay observables}},  {\em Phys. Rev. D} {\bf 100} (2019), no.~7 075025, [\href{https://arxiv.org/abs/1906.02412}{{\tt arXiv:1906.02412}}].

\bibitem{Mohapatra:2021ynn}
M.~K. Mohapatra, N.~Rajeev, and R.~Dutta, {\it {Combined analysis of $B_c \to D_s^{(*)}\,\mu^+\mu^-$ and $B_c \to D_s^{(*)}\,\nu\bar{\nu}$ decays within $Z'$ and leptoquark new physics models}},  {\em Phys. Rev. D} {\bf 105} (2022), no.~11 115022, [\href{https://arxiv.org/abs/2108.10106}{{\tt arXiv:2108.10106}}].

\bibitem{Zaki:2023mcw}
M.~Zaki, M.~A. Paracha, and F.~M. Bhutta, {\it {Footprints of New Physics in the angular distribution of $B_c \to D_s^{*}(\to D_s \gamma,\,(D_s \pi))\,\ell^{+}\ell^{-}$ decays}},  {\em Nucl. Phys. B} {\bf 992} (2023) 116236, [\href{https://arxiv.org/abs/2303.01145}{{\tt arXiv:2303.01145}}].

\bibitem{Li:2023mrj}
Y.-S. Li and X.~Liu, {\it {Angular distribution of the FCNC process $B_c\to D_s^*(\to D_s\ensuremath{\pi})\ensuremath{\ell}^+\ensuremath{\ell}^-$}},  {\em Phys. Rev. D} {\bf 108} (2023), no.~9 093005, [\href{https://arxiv.org/abs/2309.08191}{{\tt arXiv:2309.08191}}].

\bibitem{Salam:2024nfv}
Q.~M.~U. Salam, I.~Ahmed, R.~Khalid, and I.~U. Rehman, {\it {Exploring new physics in transition $b\to s\ell^+\ell^-$ through different $B_c\to D_s^{(*)}\ell^+\ell^-$ observables}},  {\em J. Phys. G} {\bf 52} (2025), no.~4 045003, [\href{https://arxiv.org/abs/2411.00912}{{\tt arXiv:2411.00912}}].

\bibitem{Geng:2001vy}
C.~Q. Geng, C.-W. Hwang, and C.~C. Liu, {\it {Study of rare $B^+_{c} \to D_{d,s}^{(\ast)+}l\bar l$ decays}},  {\em Phys. Rev. D} {\bf 65} (2002) 094037, [\href{https://arxiv.org/abs/hep-ph/0110376}{{\tt hep-ph/0110376}}].

\bibitem{Azizi:2008vv}
K.~Azizi, F.~Falahati, V.~Bashiry, and S.~M. Zebarjad, {\it {Analysis of the Rare $B_c \rightarrow D_{s,d}^{*}\,\ell^{+}\ell^{-}$ Decays in QCD}},  {\em Phys. Rev. D} {\bf 77} (2008) 114024, [\href{https://arxiv.org/abs/0806.0583}{{\tt arXiv:0806.0583}}].

\bibitem{Azizi:2008vy}
K.~Azizi and R.~Khosravi, {\it {Analysis of the rare semileptonic $B_c \to P(D, D_s ) \ell^{+} \ell^{-} / \nu \bar{\nu}$ decays within QCD sum rules}},  {\em Phys. Rev. D} {\bf 78} (2008) 036005, [\href{https://arxiv.org/abs/0806.0590}{{\tt arXiv:0806.0590}}].

\bibitem{Faessler:2002ut}
A.~Faessler, T.~Gutsche, M.~A. Ivanov, J.~G. Korner, and V.~E. Lyubovitskij, {\it {The Exclusive rare decays $B \to$ K(K*) $\bar{\ell} \ell$ and $B_c \to$ D(D*) $\bar{\ell} \ell$ in a relativistic quark model}},  {\em Eur. Phys. J. direct} {\bf 4} (2002), no.~1 18, [\href{https://arxiv.org/abs/hep-ph/0205287}{{\tt hep-ph/0205287}}].

\bibitem{Choi:2010ha}
H.-M. Choi, {\it {Light-front quark model analysis of the exclusive rare $B_c\to D_{(s)}(\ell^+\ell^-,\nu_{\ell}\bar{\nu}_{\ell})$ decays}},  {\em Phys. Rev. D} {\bf 81} (2010) 054003, [\href{https://arxiv.org/abs/1001.3432}{{\tt arXiv:1001.3432}}].

\bibitem{Wang:2014yia}
W.-F. Wang, X.~Yu, C.-D. L{\"u}, and Z.-J. Xiao, {\it {Semileptonic decays $B_c^+$ {\textrightarrow} $D_{(s)}^{(*)}(l^+\nu_l,l^+l^-,\nu\bar{\nu}$) in the perturbative QCD approach}},  {\em Phys. Rev. D} {\bf 90} (2014), no.~9 094018, [\href{https://arxiv.org/abs/1401.0391}{{\tt arXiv:1401.0391}}].

\bibitem{Yilmaz:2012ah}
U.~O. Yilmaz, {\it {Study of $B_c \to D_s^* \ell^+ \ell^-$ in Single Universal Extra Dimension}},  {\em Phys. Rev. D} {\bf 85} (2012) 115026, [\href{https://arxiv.org/abs/1204.1261}{{\tt arXiv:1204.1261}}].

\bibitem{Ebert:2010dv}
D.~Ebert, R.~N. Faustov, and V.~O. Galkin, {\it {Rare Semileptonic Decays of $B$ and $B_c$ Mesons in the Relativistic Quark Model}},  {\em Phys. Rev. D} {\bf 82} (2010) 034032, [\href{https://arxiv.org/abs/1006.4231}{{\tt arXiv:1006.4231}}].

\bibitem{Maji:2020zlq}
P.~Maji, S.~Mahata, P.~Nayek, S.~Biswas, and S.~Sahoo, {\it {Investigation of rare semileptonic $ B_c \to (D_{s,d}^{(*)} ) \mu^+ \mu^-$ decays with non-universal $ Z'$ effect}},  {\em Chin. Phys. C} {\bf 44} (2020), no.~7 073106, [\href{https://arxiv.org/abs/2003.12272}{{\tt arXiv:2003.12272}}].

\bibitem{Maji:2020wer}
P.~Maji, S.~Biswas, P.~Nayek, and S.~Sahoo, {\it {Charged Higgs contribution on $B_c \rightarrow (D_s, D_s^*) l^+l^-$}},  {\em PTEP} {\bf 2020} (2020), no.~5 053B07, [\href{https://arxiv.org/abs/2003.07041}{{\tt arXiv:2003.07041}}].

\bibitem{Paracha:2011gt}
M.~A. Paracha, I.~Ahmed, and M.~J. Aslam, {\it {Semileptonic charmed $B$ meson decays in Universal Extra Dimension Model}},  {\em Phys. Rev. D} {\bf 84} (2011) 035003, [\href{https://arxiv.org/abs/1101.2323}{{\tt arXiv:1101.2323}}].

\bibitem{Ahmed:2011fy}
I.~Ahmed, M.~A. Paracha, M.~Junaid, A.~Ahmed, A.~Rehman, and M.~J. Aslam, {\it {Analysis of $B_c \to D_s^{*}\,\ell^{+}\ell^{-}$ in the Standard Model Beyond Third Generation}},  [\href{https://arxiv.org/abs/1107.5694}{{\tt arXiv:1107.5694}}].

\bibitem{Ahmed:2011sa}
A.~Ahmed, I.~Ahmed, M.~A. Paracha, M.~Junaid, A.~Rehman, and M.~J. Aslam, {\it {Comparative Study of $B_{c} \to D_{s}^{*}\ell^{+}\ell^{-}$ Decays in Standard Model and Supersymmetric Models}},  8, 2011.
\newblock \href{https://arxiv.org/abs/1108.1058}{{\tt arXiv:1108.1058}}.

\bibitem{Aarfi:2026hgi}
Z.~Aarfi, Q.~M.~U. Salam, I.~Ahmed, F.~M. Bhutta, and M.~A. Paracha, {\it {Weak annihilation contribution to angular observables in $B_{c}^+\rightarrow D^{*+}\ell ^{+}\ell ^{-}$ decays}},  {\em Eur. Phys. J. C} {\bf 86} (2026), no.~5 456, [\href{https://arxiv.org/abs/2602.10903}{{\tt arXiv:2602.10903}}].

\bibitem{Ivanov:2024iat}
M.~A. Ivanov, J.~N. Pandya, P.~Santorelli, and N.~R. Soni, {\it {Decay $B_c^+ \to D_{(s)}^{(*)+} \ell^+\ell^-$ within covariant confined quark model}},  {\em Phys. Rev. D} {\bf 110} (2024) 096003, [\href{https://arxiv.org/abs/2404.15085}{{\tt arXiv:2404.15085}}].

\bibitem{Ju:2013oba}
W.-L. Ju, G.-L. Wang, H.-F. Fu, T.-H. Wang, and Y.~Jiang, {\it {The Study of Rare $B_c\rightarrow D^{(*)}_{s,d}l\bar{l}$ Decays}},  {\em JHEP} {\bf 04} (2014) 065, [\href{https://arxiv.org/abs/1307.5492}{{\tt arXiv:1307.5492}}].

\bibitem{Bause:2022rrs}
R.~Bause, H.~Gisbert, M.~Golz, and G.~Hiller, {\it {Model-independent analysis of $b \rightarrow d$ processes}},  {\em Eur. Phys. J. C} {\bf 83} (2023), no.~5 419, [\href{https://arxiv.org/abs/2209.04457}{{\tt arXiv:2209.04457}}].

\bibitem{Farooq:2024owx}
N.~Farooq, M.~Zaki, M.~A. Paracha, and F.~M. Bhutta, {\it {Effects of family non-universal $Z^{\prime}$ model in angular observables of ${\boldsymbol B {\bf\to} (\boldsymbol\rho, {\boldsymbol a}_{\bf 1})\boldsymbol\mu^+\boldsymbol\mu^- }$ decays}},  {\em Chin. Phys. C} {\bf 48} (2024), no.~10 103107, [\href{https://arxiv.org/abs/2407.00520}{{\tt arXiv:2407.00520}}].

\bibitem{Bobeth:1999mk}
C.~Bobeth, M.~Misiak, and J.~Urban, {\it {Photonic penguins at two loops and $m_t$ dependence of $BR[B \to X_s l^+ l^-]$}},  {\em Nucl. Phys. B} {\bf 574} (2000) 291--330, [\href{https://arxiv.org/abs/hep-ph/9910220}{{\tt hep-ph/9910220}}].

\bibitem{Beneke:2001at}
M.~Beneke, T.~Feldmann, and D.~Seidel, {\it {Systematic approach to exclusive $B \to V l^+ l^-$, $V \gamma$ decays}},  {\em Nucl. Phys.} {\bf B612} (2001) 25--58, [\href{https://arxiv.org/abs/hep-ph/0106067}{{\tt hep-ph/0106067}}].

\bibitem{Asatrian:2001de}
H.~H. Asatrian, H.~M. Asatrian, C.~Greub, and M.~Walker, {\it {Two loop virtual corrections to $B \to X_s l^+ l^-$ in the standard model}},  {\em Phys. Lett. B} {\bf 507} (2001) 162--172, [\href{https://arxiv.org/abs/hep-ph/0103087}{{\tt hep-ph/0103087}}].

\bibitem{Asatryan:2001zw}
H.~H. Asatryan, H.~M. Asatrian, C.~Greub, and M.~Walker, {\it {Calculation of two loop virtual corrections to $b \to s l^+ l^-$ in the standard model}},  {\em Phys. Rev.} {\bf D65} (2002) 074004, [\href{https://arxiv.org/abs/hep-ph/0109140}{{\tt hep-ph/0109140}}].

\bibitem{Greub:2008cy}
C.~Greub, V.~Pilipp, and C.~Schupbach, {\it {Analytic calculation of two-loop QCD corrections to $b \to sl^+ l^-$ in the high $q^2$ region}},  {\em JHEP} {\bf 12} (2008) 040, [\href{https://arxiv.org/abs/0810.4077}{{\tt arXiv:0810.4077}}].

\bibitem{Du:2015tda}
D.~Du, A.~X. El-Khadra, S.~Gottlieb, A.~S. Kronfeld, J.~Laiho, E.~Lunghi, R.~S. Van~de Water, and R.~Zhou, {\it {Phenomenology of semileptonic B-meson decays with form factors from lattice QCD}},  {\em Phys. Rev. D} {\bf 93} (2016), no.~3 034005, [\href{https://arxiv.org/abs/1510.02349}{{\tt arXiv:1510.02349}}].

\bibitem{Aarfi:2025qcp}
Z.~Aarfi, Q.~M.~U. Salam, I.~Ahmed, F.~M. Bhutta, R.~Khalid, and M.~A. Paracha, {\it {Investigating New Physics through the Observables of Semileptonic $B_s \to K^{*}(\to K\pi)\,\mu^{+}\mu^{-}$ Decay}},  {\em PTEP} {\bf 2025} (2025), no.~12 123B10, [\href{https://arxiv.org/abs/2506.20446}{{\tt arXiv:2506.20446}}].

\bibitem{Ju:2014oha}
W.-L. Ju, G.-L. Wang, H.-F. Fu, Z.-H. Wang, and Y.~Li, {\it {The rare semi-leptonic B$_{c}$ decays involving orbitally excited final mesons}},  {\em JHEP} {\bf 09} (2015) 171, [\href{https://arxiv.org/abs/1407.7968}{{\tt arXiv:1407.7968}}].

\bibitem{ParticleDataGroup:2024cfk}
{\bf Particle Data Group} Collaboration, S.~Navas et~al., {\it {Review of particle physics}},  {\em Phys. Rev. D} {\bf 110} (2024), no.~3 030001.

\bibitem{Crivellin:2018yvo}
A.~Crivellin, C.~Greub, D.~M{\"u}ller, and F.~Saturnino, {\it {Importance of Loop Effects in Explaining the Accumulated Evidence for New Physics in B Decays with a Vector Leptoquark}},  {\em Phys. Rev. Lett.} {\bf 122} (2019), no.~1 011805, [\href{https://arxiv.org/abs/1807.02068}{{\tt arXiv:1807.02068}}].

\bibitem{Bobeth:2016llm}
C.~Bobeth, A.~J. Buras, A.~Celis, and M.~Jung, {\it {Patterns of Flavour Violation in Models with Vector-Like Quarks}},  {\em JHEP} {\bf 04} (2017) 079, [\href{https://arxiv.org/abs/1609.04783}{{\tt arXiv:1609.04783}}].

\bibitem{DAlise:2024qmp}
A.~D'Alise, G.~Fabiano, D.~Frattulillo, D.~Iacobacci, F.~Sannino, P.~Santorelli, and N.~Vignaroli, {\it {New physics pathways from B processes}},  {\em Nucl. Phys. B} {\bf 1006} (2024) 116631, [\href{https://arxiv.org/abs/2403.17614}{{\tt arXiv:2403.17614}}].

\end{thebibliography}\endgroup

\end{document}